\pdfoutput=1
\PassOptionsToPackage{pdftex,
pdfencoding=auto,
pdfnewwindow=true,
pdfusetitle=true,
bookmarks=true,
bookmarksnumbered=true,
bookmarksopen=true,
pdfpagemode=UseThumbs,
bookmarksopenlevel=1,
pdfpagelabels=true,
breaklinks=true,
colorlinks=true,
}{hyperref}
\PassOptionsToPackage{usenames,dvipsnames,pdftex}{xcolor}
\documentclass[superscriptaddress,aps,prx,nofootinbib,twocolumn,twoside,reprint,letterpaper,longbibliography]{revtex4-2}
\usepackage[utf8]{inputenx}
\usepackage{fontawesome}
\renewcommand{\bell}{\text{\faBellO}}
\renewcommand{\triangle}{\bigtriangleup}
\newcommand{\biloc}{\text{\scalebox{2}[1]{\rotatebox{90}{\faShareAlt}}}}

\usepackage[letterpaper,centering,pdftex]{geometry}
\AtBeginDocument{\newgeometry{
  ignoreall=true,
  heightrounded=true,
  margin=0.9in,
}
}

\usepackage{amssymb,amsfonts}
\usepackage{mathtools}
\usepackage[usenames,dvipsnames]{xcolor}
\definecolor{purple}{RGB}{128,0,128}
\definecolor{ultramarine}{RGB}{63, 0, 255}
\definecolor{medblue}{RGB}{0, 0, 100}
\definecolor{googleblue}{RGB}{34, 0, 204}
\definecolor{panblue}{RGB}{0,24,150}
\definecolor{carmine}{RGB}{150, 0, 24}
\definecolor{gray}{RGB}{150, 150, 150}
\definecolor{darkgreen}{RGB}{0, 80, 0}

\definecolor{jflyOrange}{RGB}{230,159,0}
\definecolor{jflySkyBlue}{RGB}{86,180,233}
\definecolor{jflyBluishGreen}{RGB}{0,158,115}
\definecolor{jflyYellow}{RGB}{240,228,66}
\definecolor{jflyBlue}{RGB}{0,114,178}
\definecolor{jflyVermillion}{RGB}{213,94,0}
\definecolor{jflyReddishPurple}{RGB}{204,121,167}

\newcommand{\term}[1]{\textcolor{medblue}{\textbf{#1}}}
\DeclarePairedDelimiter{\expec}{\langle}{\rangle}
\DeclarePairedDelimiter{\braces}{\lbrace}{\rbrace}
\DeclarePairedDelimiter{\abs}{\lvert}{\rvert}
\DeclarePairedDelimiter{\braks}{\lbrack}{\rbrack}
\usepackage[]{hyperref}

\hypersetup{linkcolor=carmine,citecolor=darkgreen,urlcolor=googleblue,anchorcolor=OliveGreen}
\usepackage{bbold}
\usepackage{graphicx}
\graphicspath{{figs/}}
\usepackage{subfigure}
\usepackage{changes}

\usepackage{microtype}
\microtypecontext{spacing=nonfrench}
\microtypesetup{
expansion={true,nocompatibility},
protrusion={true,nocompatibility},
activate={true,nocompatibility},
tracking=true,
kerning=true,
spacing={true}
}
\usepackage[all=normal,floats=tight,mathspacing=tight,wordspacing=tight,paragraphs=normal,tracking=tight,charwidths=tight,mathdisplays=normal,sections=normal,margins=normal]{savetrees}
\everypar=\expandafter{\the\everypar\loosness=-1 }
\usepackage{nicefrac}
\usepackage{overpic}
\usepackage{ulem}
\usepackage[style=american,autopunct=true]{csquotes} 

\def\id{\boldsymbol{\mathbb{1}}}
\DeclareMathOperator{\Tr}{Tr}
\def\H{{\mathcal H}}
\def\be{\begin{equation}}

\def\ee{\end{equation}}
\usepackage{paralist} 

\newcommand{\shortto}[1][3pt]{\mathrel{%
   \hbox{\rule[\dimexpr\fontdimen22\textfont2-.2pt\relax]{#1}{.4pt}}%
   \mkern-4mu\hbox{\usefont{U}{lasy}{m}{n}\symbol{41}}}}

\makeatletter

\setbox0\hbox{$\xdef\scriptratio{\strip@pt\dimexpr
    \numexpr(\sf@size*65536)/\f@size sp}$}

\newcommand{\scriptveryshortarrow}[1][3pt]{{%
    \hbox{\rule[\scriptratio\dimexpr\fontdimen22\textfont2-.2pt\relax]
               {\scriptratio\dimexpr#1\relax}{\scriptratio\dimexpr.4pt\relax}}%
   \mkern-4mu\hbox{\let\f@size\sf@size\usefont{U}{lasy}{m}{n}\symbol{41}}}}

\makeatother

\usepackage{braket}

\begin{document}
\title{Quantum Inflation: A General Approach to Quantum Causal Compatibility}

\author{Elie Wolfe}
\affiliation{Perimeter Institute for Theoretical Physics, Waterloo, N2L 2Y5 Ontario, Canada}

\author{Alejandro Pozas-Kerstjens}
\affiliation{Departamento de An\'alisis Matem\'atico, Universidad Complutense de Madrid, 28040 Madrid, Spain}
\affiliation{ICFO-Institut de Ci\`encies Fot\`oniques, The Barcelona Institute of Science and Technology, 08860 Castelldefels (Barcelona), Spain}

\author{Matan Grinberg}
\affiliation{Perimeter Institute for Theoretical Physics, Waterloo, N2L 2Y5 Ontario, Canada}
\affiliation{Princeton University, Princeton, New Jersey 08544, USA}

\author{Denis Rosset}
\affiliation{Perimeter Institute for Theoretical Physics, Waterloo, N2L 2Y5 Ontario, Canada}

\author{Antonio Ac\'in}
\affiliation{ICFO-Institut de Cienci\`es Fot\`oniques, The Barcelona Institute of Science and Technology, 08860 Castelldefels (Barcelona), Spain}
\affiliation{ICREA, Passeig Lluis Companys 23, 08010 Barcelona, Spain}

\author{Miguel Navascu\'es}
\affiliation{Institute for Quantum Optics and Quantum Information (IQOQI), Boltzmanngasse 3 1090 Vienna, Austria}

\begin{abstract}
Causality is a seminal concept in science: Any research discipline, from sociology and medicine to physics and chemistry, aims at understanding the causes that could explain the correlations observed among some measured variables.
While several methods exist to characterize classical causal models, no general construction is known for the quantum case.
In this work, we present quantum inflation, a systematic technique to falsify if a given quantum causal model is compatible with some observed correlations.
We demonstrate the power of the technique by reproducing known results and solving open problems for some paradigmatic examples of causal networks.
Our results may find applications in many fields: from the characterization of correlations in quantum networks to the study of quantum effects in thermodynamic and biological processes.
\end{abstract}

\maketitle

\section{INTRODUCTION}
It can be argued that one of the main challenges in any scientific discipline is to identify which causes are behind the correlations observed among some measured variables, encapsulated by their joint probability distribution.
Understanding this problem is crucial in many situations, such as, for example, the development of medical treatments, taking data-based social policy decisions, the design of new materials or the theoretical modeling of experiments. More precise characterizations of causal correlations enable better decision among competing explanations for given statistics, a task known as \term{causal discovery}. Advances in causal understanding also enable quantification of causal effects from purely observational data, thus extracting counterfactual conclusions even in instances where randomized or controlled trials are not feasible, a task known as \term{causal inference}~\cite{pearl,morgan2007counterfactuals,ShpitserPearlIdentification}.

\begin{figure}[!b]
  \begin{center}
    \subfigure[\label{fig:Bell}]
    {\centering
      \begin{minipage}[t]{0.20\textwidth}
      \centering\includegraphics[scale=0.45]{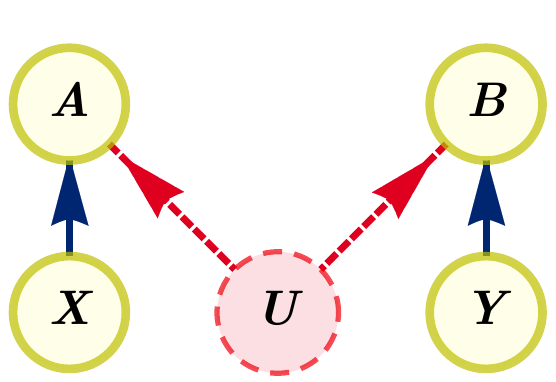}\end{minipage}
    }
    \subfigure[\label{fig:TriangleNoSettings}]
    {\centering
      \begin{minipage}[t]{0.23\textwidth}
      \centering\includegraphics[scale=0.45]{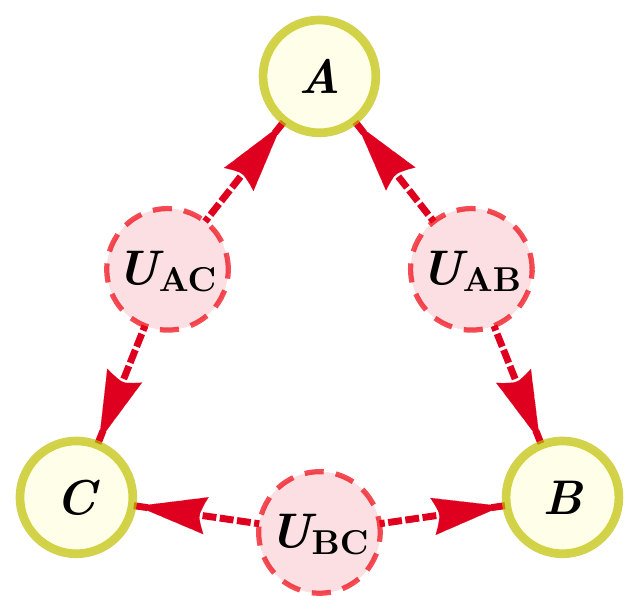}\end{minipage}
    }
  \end{center}
  \caption[]{
  DAG representation of different causal scenarios. The red, dashed circles are latent nodes, and the yellow, single-lined circles denote observed variables. %
  (a)~The Bell scenario is one of the simplest causal structures exhibiting a classical-quantum gap, that is, where there exist distributions that can be realized upon taking the latent nodes to represent quantum states (subsequently denoted as $U\,{=}\,\rho$), but not when the latent nodes are taken to represent distributions over classical hidden variables (subsequently denoted as $U\,{=}\,\Lambda$). %
  (b)~The triangle scenario, with three observed and three latent nodes, also presents a classical-quantum gap.
  \label{fig:dag}
  }
\end{figure}

Bayesian causal networks, in the form of directed acyclic graphs (DAGs), provide the tools to formalize such problems.
These graphs, examples of which are shown in Fig.~\ref{fig:dag}, encode the causal relations between the various variables in the problem, which could be either observed or non-observed.
The latter, also known as latent, are required in many relevant situations in order to explain correlations among the observed.
The fundamental task addressed in this work underlies both causal discovery and causal inference, and is known as the \term{causal compatibility} problem. It consists of deciding whether a given joint probability distribution over some observed variables can be explained by a given candidate Bayesian causal network. Equivalently, the objective of causal compatibility can be viewed as characterizing the set of distributions compatible with a given Bayesian network.

In all cases, the measured variables in a causal network are, by definition, classical. However, causal networks may be classical or quantum depending on whether correlations are established by means of classical or quantum information. Because of its importance and broad range of applications, there is a vast literature devoted to understanding the problem of causal compatibility for classical causal networks; see for instance Ref.~\cite{pearl}.
On the contrary, very little is known for the quantum case despite the fact that nature is ultimately quantum and quantum effects are expected to be crucial for the understanding of many relevant phenomena in many scientific disciplines.
Moreover the two problems are known to be different, as one of the consequences of Bell's theorem~\cite{Bell66,BellReview} is that quantum causal networks can explain correlations for which the analogous classical network fails~\cite{fritz2012bell,HLP,Wood2015,Chaves2015relaxing,WolfeBellQuantified}.
Our work addresses these issues and provides a systematic construction to tackle the problem of causal compatibility for quantum causal networks.

As mentioned, several results already exist in the classical case.
Whenever the network does not contain any latent variable the solution is rather simple and it suffices to check whether all the conditional independences associated to the network topology are satisfied~\cite{pearl}. The problem, however, becomes much more difficult as soon as the network also includes latent variables, as their presence generally implies nontrivial
inequalities on the observed probabilities.
A general method to tackle the causal compatibility problem, known as the \term{inflation technique}, is obtained in~\cite{wolfe2016inflation}. It consists of a hierarchy of conditions, organized according to their computational cost, that are necessary for a Bayesian network to be able to explain the observed correlations. Moreover, the hierarchy is asymptotically sufficient, in the sense that the candidate Bayesian network is compatible if, and only if, all conditions in the hierarchy are satisfied~\cite{navascues2017inflation}.

When moving to quantum causal scenarios, the problem of causal compatibility presents several new features.
In the classical case, the cardinality of the latent variables can be upper bounded~\cite{rosset2016finite} and, therefore, the problem is decidable.
In the quantum case, however, a similar upper bound cannot exist because the problem of quantum causal compatibility is undecidable, as implied by recent results on quantum correlations~\cite{Slofstra,Ji2020Connes}.
Yet, this fact does not preclude the existence of a method similar to inflation to tackle the question.
Unfortunately, the inflation technique cannot be straightforwardly adapted to the quantum case because it relies on information broadcasting, a primitive that is not plausible with quantum information~\cite{BroadcastingMixed,NoCloningGeneral2006}.
Other causal analysis techniques which are fundamentally quantum have been proposed.
Notable among these is the quantum entropy vector approach in Ref.~\cite{Chaves2015}, which is applicable to all causal structures but uses only those constraints on entropies imposed by the causal structure, the scalar extension in Ref.~\cite{Pozas2019}, which imposes stronger constraints but cannot be applied to causal structures in which all observed nodes are causally connected, such as the triangle scenario in Fig.~\ref{fig:TriangleNoSettings} or the covariance matrix approach in Ref.~\cite{aberg2020covariance}, which provides low-degree polynomial (and, thus, rough) approximations of the sets of quantum correlations in networks.

The main result of our work is the construction of \term{quantum inflation}, a systematic technique to study causal compatibility in any quantum Bayesian network.
It can be seen as a quantum analog of the classical inflation technique which avoids the latter's reliance on information broadcast.
In Sec.~\ref{sec:structures} we introduce the graphical notation to be used throughout this work and define the causal compatibility problem.
We next explain quantum inflation by means of a simple example in Sec.~\ref{sec:simple}, deferring its detailed construction for arbitrary two-layer DAGs to Sec.~\ref{sec:details}.
Section~\ref{sec:arbitrary} generalizes the technique to apply to \emph{every} possible multilayer causal structure involving unobserved quantum systems, utilizing the two-layer construction as its elementary constituent. Section~\ref{sec:classical} then considers a slight modification to quantum inflation to obtain a new tool for assessing causal compatibility with classical models.
The next three sections illustrate the power of quantum inflation through different applications.
First, in Sec.~\ref{sec:results}, we focus on the study of quantum correlations and use quantum inflation to characterize correlations achievable in various tripartite quantum causal networks, including the derivation of quantum causal incompatibility witnesses.
Second, we show in Sec.~\ref{sec:cryptography} how quantum inflation can be used for cryptography purposes in quantum networks, in particular, to bound the eavesdropper's predictability on the observed measurement outcomes.
Third, in Sec.~\ref{sec:mediationanalysis}, we consider a standard application in causality and show how to use quantum inflation to bound the strength of causal effects (this is, to perform do-conditional estimation) in the presence of quantum common causes.
Finally, we conclude in Sec.~\ref{sec:conclusions}.

We expect quantum inflation to find an application in many different contexts. The most immediate one is probably the characterization of correlations in quantum networks and its use for quantum information protocols. However, tools to characterize classical causal connections are commonly used in many scientific disciplines; see, for instance, Refs.~\cite{biomolecular,genomics} for examples in the context of biomolecular investigations or genomics. Progress in our understanding of these and other processes at smaller scales will eventually have to face quantum phenomena and understand how causal explanations are affected by them. In fact, some recent works advocate, admittedly not always in fully convincing ways, that quantum effects may play a role in biological processes~\cite{qbio1,qbio2}. We expect quantum inflation to be a fundamental method to address all these scientific challenges that are to come.

\section{DEVICE-INDEPENDENT QUANTUM CAUSAL STRUCTURES}\label{sec:structures}
In this work we consider physical scenarios where an experimenter can observe a number of random variables. Some of such observed variables can be influenced by other observed variables and also by unknown quantum processes. Our goal is to determine whether the postulated causal relations between the observed variables and such hidden processes are compatible with the observed statistics.

\begin{figure}[t]
    \centering
   \includegraphics[width=0.9\columnwidth]{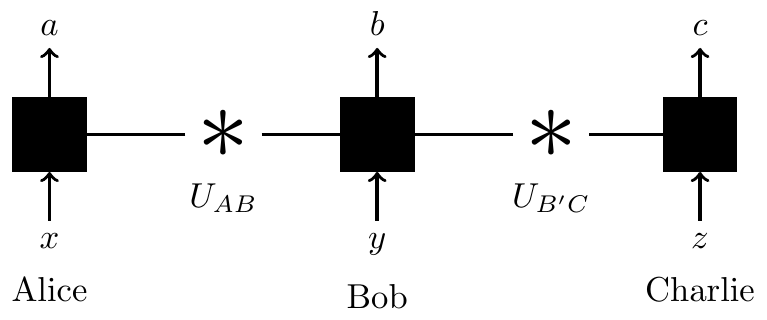}
  \caption{The quantum tripartite line in the device-independent framework.}
  \label{fig:SWAP}
\end{figure}

Think, for instance, of the scenario depicted in Fig.~\ref{fig:SWAP}. The quantum states $U_{AB}$ and $U_{B'C}$ are, respectively, distributed to the parties Alice and Bob, and Bob and Charlie. Alice (respectively, Charlie) probes her (respectively, his) subsystem with a measurement labeled by $X$ (respectively, $Z$), obtaining the result $A$ (respectively, $C$). Bob probes both of his parts of states $U_{AB}$ and $U_{B'C}$ via a collective measurement labeled $Y$, obtaining the outcome $B$. If Alice, Bob, and Charlie repeat this experiment several times, then they can estimate the probabilities ${\{P_{\text{ABCXYZ}}(a,b,c,x,y,z)\}}$, that, from now on, we just refer to as ${P(a,b,c,x,y,z)}$. We further assume that, for each party, the choice of measurement setting is done independently of the other two parties.

This scenario is known as the \emph{tripartite-line causal scenario}~\cite{branciard2010bilocality,branciard2012bilocality}, and it is meant to model the action of a quantum repeater in a device-independent way, where we assume no knowledge of the state preparations or measurement devices involved or, for that matter, the Hilbert spaces $\H_A$, $\H_B$, $\H_{B'}$, and $\H_C$ where those objects act.

For any Hilbert space $\H$, let ${\cal B}(\H)$ denote the set of linear bounded operators mapping $\H$ to itself. The distribution ${P_{\text{ABCXYZ}}}$ is realizable in the tripartite-line causal scenario if and only if there exist Hilbert spaces ${\H_A}$, ${\H_B}$, ${\H_{B'}}$, and ${\H_C}$, quantum states
\begin{align}\begin{aligned}
&U_{AB}\in{\cal B}(\H_A\otimes\H_B), &&U_{AB}\succeq 0,&&\Tr\braks{U_{AB}}=1,\\
&U_{B'C}\in{\cal B}(\H_{B'}\otimes\H_C), &&U_{B'C}\succeq 0,&&\Tr\braks{U_{B'C}}=1,
\label{states}
\end{aligned}\end{align}
\noindent and positive operator valued measures (POVMs)
\begin{align}\begin{aligned}
&E_{a|x}\in {\cal B}(\H_A), && E_{a|x}\succeq 0,\;\;\;\; \sum_a E_{a|x}=\id_A,\\
&F_{b|y}\in {\cal B}(\H_B\otimes \H_{B'}), && F_{b|y}\succeq 0,\;\;\;\; \sum_b F_{b|y}=\id_{BB'},\\
&G_{c|z}\in {\cal B}(\H_C), && G_{c|z}\succeq 0,\;\;\;\; \sum_c G_{c|z}=\id_C,
\label{meas}
\end{aligned}\end{align}
such that
\begin{align}\begin{split}
&P_{\text{ABCXYZ}}(a,b,c,x,y,z)=\\
&\quad P_\text{X}(x)P_\text{Y}(y)P_\text{Z}(z)\times\\
&\quad\; \Tr\braks*{(U_{AB}\otimes U_{B'C})(E_{a|x}\otimes F_{b|y}\otimes G_{c|z})}.
\label{repre}
\end{split}\end{align}

For simplicity, rather than sketching the corresponding quantum experiment, we prefer to represent the relations between the variables $X$, $Y$, $Z$, $A$, $B$, $C$, $U_{AB}$, and $U_{B'C}$, by means of what we hereby call \emph{device-independent quantum causal structures}, or \term{quantum causal structures}, for short. This structure fully determines the form of the observed statistics, as in Eq.~\eqref{repre}.

A quantum causal structure is a Directed Acyclic Graph (\term{DAG}) where each node represents a classical or a quantum system. All quantum nodes are assumed to be latent: This assumption means that we can extract information from them only through their classical successors. Conversely, all classical variables are regarded as directly observable. The outgoing edges of a quantum node represent different subsystems or sectors of the corresponding quantum state, whereas the outgoing edges of a classical node denote copies of the node variable. The interpretation of the DAG is that each node is generated by applying an unknown deterministic quantum operation to all of its incoming edges. A \emph{root node} (a node without incoming edges) can be regarded as the result of applying a transformation to a system with no information content: The corresponding random variable or quantum system is, therefore, assumed uncorrelated with any other root nodes.

\begin{figure}[b]
  \begin{center}
    \subfigure[\label{fig:Bilocality}]
      {\centering
        \begin{minipage}[t]{0.5\linewidth}
        \includegraphics[scale=0.45]{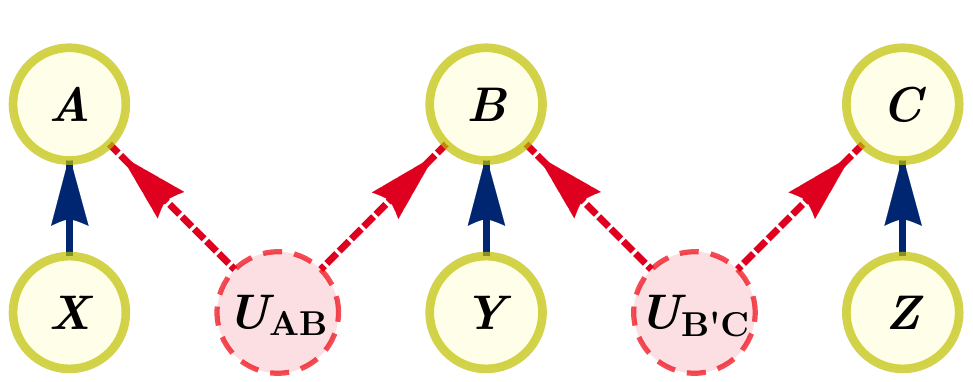}
        \end{minipage}
        }
    \hfill
    \subfigure[\label{fig:ArbitraryStructure}]
    {\centering
      \begin{minipage}[t]{0.46\linewidth}
      \centering\includegraphics[scale=0.45]{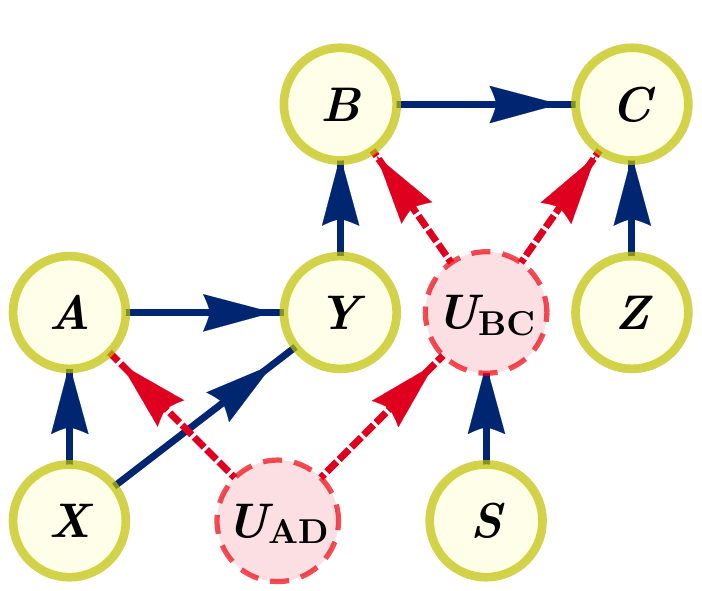}\end{minipage}
    }
  \end{center}
  \caption[]{
  DAG representation of two example causal scenarios with quantum latent nodes.
  (a)~The tripartite-line causal scenario, where two causally independent parties $A$ and $C$ each share some quantum entanglement with a central party $B$, later denoted $\mathcal{Q}^{{\protect\biloc}}$ for brevity.
  (b)~Arbitrary quantum-latent causal structures contain directed edges beyond the traditional network connections of latent-to-terminal edges and root-to-terminal edges. A method for analyzing correlations in general structures with quantum latent nodes is given in Sec.~\ref{sec:arbitrary}.
  \label{fig:qdags}
  }
\end{figure}

The quantum causal structure corresponding to the tripartite-line causal scenario (Fig.~\ref{fig:SWAP}) is depicted in Fig.~\ref{fig:Bilocality}. Since the variables $X$, $Y$, and $Z$ and the quantum states $U_{AB}$ and $U_{B'C}$ are assumed independent, they are represented by root nodes. The random variable $A$ is the result of applying a quantum operation from the random variable $X$ and one part of state $U_{AB}$ to $A$'s sample space. Since the outcome of the map is a classical variable, any such map corresponds to a quantum measurement on $\H_A$, labeled by $X$, i.e., to the POVMs (\ref{meas}).

When we specify the quantum operations mapping incoming edges to nodes in a quantum causal structure ${\cal G}$, we speak of a \emph{quantum realization} of ${\cal G}$. A distribution ${P_\text{obs}}$ over the observed nodes ${\pmb{V}\hspace{-1ex}}_{O}$ of ${\cal G}$ is said to be compatible with ${\cal G}$ if there exists a quantum realization that generates the distribution ${P_\text{obs}}({\pmb{V}\hspace{-1ex}}_{O})$.

Coming back to our example, ${P_{\text{ABCXYZ}}}$ is compatible with the quantum causal structure in Fig.~\ref{fig:Bilocality} if and only if $P$ admits a decomposition of the form~\eqref{repre}, for some Hilbert spaces $\H_A$, $\H_B$, $\H_{B'}$, and $\H_C$ and operators satisfying Eqs. \eqref{states} and \eqref{meas}.

As another example, consider the DAG in Fig.~\ref{fig:ArbitraryStructure}, with quantum nodes $U_{AD}$ and $U_{BC}$.
A distribution ${\{P_{\text{ABCXYZS}}\}}$ is compatible with this causal structure if and only if there exists a state ${U_{AD}}$, quantum channels ${\Omega_s\in {\cal B}(\H_D)\to{\cal B}(\H_{B}\otimes \H_C) }$, and POVMs ${E_{a|x}\in {\cal B}(\H_A)}$, ${F_{b|y}\in {\cal B}(\H_{B'})}$, and ${G_{c|bz}\in{\cal B}(\H_{C})}$ such that
\begin{align}\label{eq:arbitrarycompat}
&P_{\text{ABCXYZS}}(a,b,c,x,y,z,s)=\nonumber\\
&\quad P_\text{X}(x)P_{\text{Y{\textbar}AX}}(y|a,x)P_\text{Z}(z)P_\text{S}(s)\times\\\nonumber
&\quad \;\Tr\braks*{(\id_A\otimes\Omega_s)(U_{AD})(E_{a|x}\otimes F_{b|y}\otimes G_{c|bz})}.
\end{align}
In general terms, the quantum causal structures we consider are arbitrary combinations of the following elements:

\begin{itemize}
    \item \textit{preparations}, controlled or not, of multipartite quantum states of arbitrary dimension (as an extension to the formalism, we also add sources of global shared randomness in Sec.~\ref{sec:results:optim:linear}),
    \item \textit{transformations} of (multipartite) quantum systems, described by quantum channels which may be classically controlled, and
    \item \textit{measurements}, i.e., operations which generate classical random variables from (a set of) quantum states.
\end{itemize}

Without loss of generality we can assume that all measurements are projective and that all transformations are unitary, since we can dilate all such operations and then subsume all the local ancilla states into the causal structure's preparations.

When we depict such a process as a directed acyclic graph, the vertices in the DAG represent the outcomes of these process elements, not the process elements themselves. That is, DAGs are distinct from circuit diagrams. For instance, in Fig.~\ref{fig:ArbitraryStructure} the node $B$ indicates the postmeasurement random variable arising from a measurement on some subsystem of $U_{BC}$ classically controlled by $Y$. The node $U_{BC}$ there indicates the (bipartite) posttransformation state arising from a transformation (classically controlled by $S$) applied to a fraction of the subsystems of $U_{AD}$.

Note that combinations of these elements can describe quite complex operations.
For instance, an operation with both classical and quantum inputs and outputs is captured in this formalism by a controlled transformation followed by a projective measurement on a subspace of the resulting state.

Given a quantum causal structure ${\cal G}$, the purpose of this work is to tackle the following problem.

\vspace{10pt}
\noindent \textbf{Causal Compatibility Problem}\\
Input: a device-independent causal structure ${\cal G}$, and a probability distribution ${P_\text{obs}}({\pmb{V}\hspace{-1ex}}_{O})$ over its observed nodes ${\pmb{V}\hspace{-1ex}}_{O}$.\\
Output: if ${P_\text{obs}}({\pmb{V}\hspace{-1ex}}_{O})$ is compatible with ${\cal G}$, output \texttt{COMPATIBLE}. Otherwise, output \texttt{INCOMPATIBLE}.
\vspace{10pt}

The main challenge when addressing the causal compatibility problem comes from the fact that the Hilbert spaces $\H^{(i)}_Q$ required to reproduce the observed statistics ${P_\text{obs}}({\pmb{V}\hspace{-1ex}}_{O})$ are not known \emph{a priori}, and they might well be infinite dimensional. In fact, for the causal structure depicted in Fig.~\ref{fig:Bell} (the quantum correlations scenario), the causal compatibility problem is known to be undecidable~\cite{Slofstra}. Moreover, even the weaker problem of deciding if ${P_\text{obs}}({\pmb{V}\hspace{-1ex}}_{O})$ can be roughly approximated by a compatible distribution cannot be solved, in general, by a Turing machine~\cite{Ji2020Connes}.

The quantum inflation technique, introduced in the next sections, is a tool to prove that certain distributions ${P_\text{obs}}({\pmb{V}\hspace{-1ex}}_{O})$ are incompatible with the considered causal structure ${\cal G}$. That is, for some instances, the inflation technique allows us to solve the causal compatibility problem in the negative. Even though the no-go results of Refs.~\cite{Slofstra,Ji2020Connes} imply that this tool, as any other tool designed to tackle the quantum causal compatibility problem, cannot be complete,\footnote{Namely, that for many quantum-latent causal structures ${\cal G}$, there will exist incompatible distributions ${P_\text{obs}}({\pmb{V}\hspace{-1ex}}_{O})$ whose incompatibility cannot be detected through the quantum inflation technique. It is important to note that the undecidability proofs of Refs.~\cite{Slofstra,Ji2020Connes} imply a limitation not specific to quantum inflation, however: Any further method for quantum causal compatibility must also be incomplete.} in the next sections we illustrate its applicability by proving the incompatibility of large sets of observed distributions for different quantum causal structures.

\section{QUANTUM INFLATION BY EXAMPLE}\label{sec:simple}
\begin{figure}[b]
  \centering
  \hfill
  \subfigure[\label{fig:TriangleSubsystems}]{
    \begin{overpic}[scale=0.43]{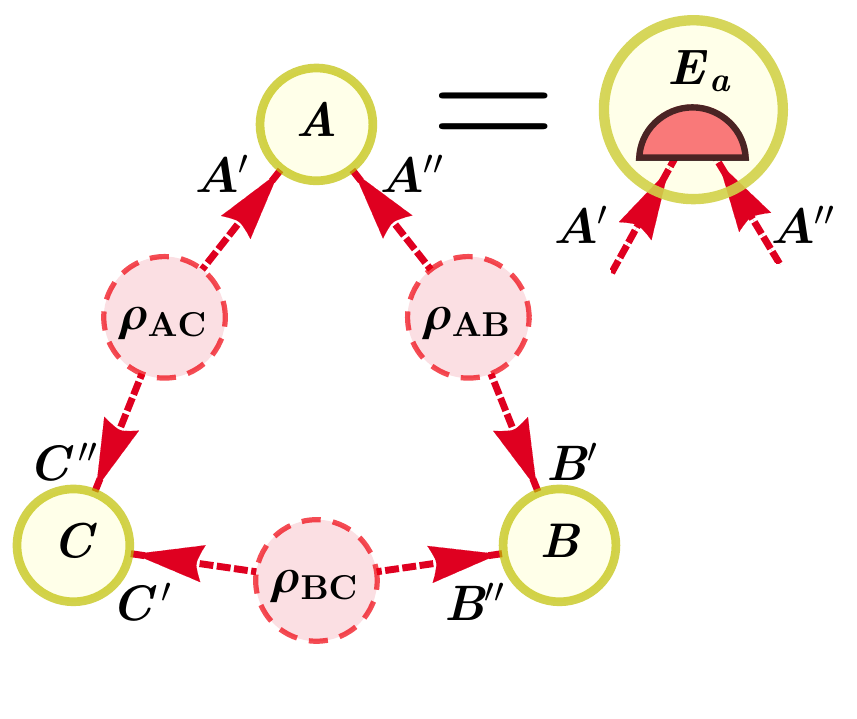}
        \put(90,35){\huge\faArrowRight}
    \end{overpic}
  }
  \hfill
  \subfigure[\label{fig:TriangleQInf_implicit}]{
    \includegraphics[scale=0.43]{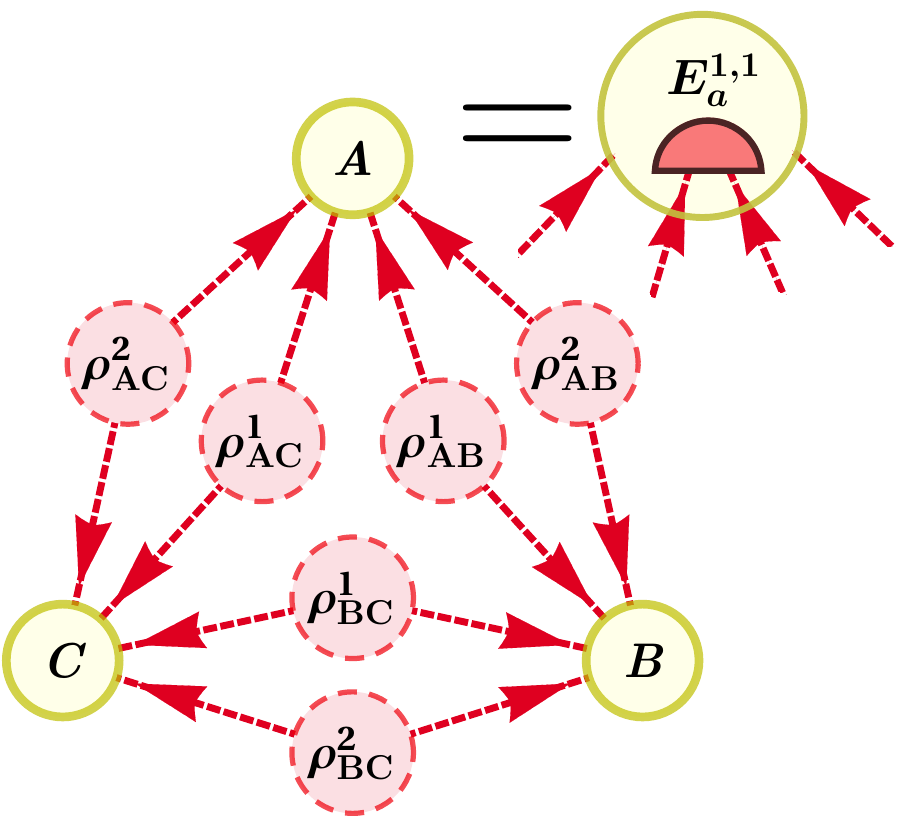}
  }
  \hfill
  \caption[]{Quantum inflation in the triangle scenario. (a) In the original scenario, by probing systems $A'$ and $A''$ with the quantum measurement $\{E_a\}_a$, a value $a$ for the random variable $A$ is generated. The values $b$ and $c$ for the random variables $B$ and $C$ are produced similarly.
  (b) In quantum inflation, we distribute $n$ (in the case shown, $n\,{=}\,2$) independent copies of the same states to the parties, which now use the original measurement operators on different pairs of copies of the states they receive. For instance, the measurement operators $\{E_a^{1,1}\}_a$ act on the states corresponding to copies $\rho_{AB}^1$ and $\rho_{AC}^1$, and the measurements with other superindices are defined in an analogous way.}
  \label{fig:quantum_inflation}
\end{figure}

For the sake of clarity, it is convenient to start the presentation of quantum inflation through an example.
Consider the quantum causal network depicted in Fig.~\ref{fig:TriangleSubsystems}, whereby three random variables $A$, $B$, and $C$ (taking the values $a$, $b$, and $c$, respectively) are generated by conducting bipartite measurements over the ends of three bipartite quantum states $\rho_{AB}$, $\rho_{BC}$, and $\rho_{AC}$ (from now on, quantum latent variables are denoted with $\rho$, instead of $U$).
We are handed the distribution $P_\text{obs}(a,b,c)$ of observed variables and asked if it is compatible with this model.
How to proceed?

Suppose that there existed indeed bipartite states $\rho_{AB}$, $\rho_{BC}$, and $\rho_{AC}$ of systems $A''B'$, $B''C'$, and $A'C''$ and commuting measurement operators $E_a$, $F_b$, and $G_c$, acting on systems $A'A''$, $B'B''$, and $C'C''$, respectively, which are able to reproduce the correlations $P_\text{obs}(a,b,c)$.
Now imagine how the scenario would change if $n$ independent copies $\rho_{AB}^i$, $\rho_{BC}^i$, and $\rho_{AC}^i$, \mbox{$i=1,...,n$} of each of the original states are distributed instead, as depicted in Fig.~\ref{fig:TriangleQInf_implicit}.
Call $\rho$ the overall quantum state before any measurement is carried out.
For any $i,j=1,...,n$ we can, in principle, implement measurement $\{E_a\}_a$ on the $i^\text{th}$ copy of $\rho_{AC}$ and the $j^\text{th}$ copy of $\rho_{AB}$: We denote by $\{E^{i,j}_a\}_a$ the corresponding measurement operators.
Similarly, call $\{F^{i,j}_b\}_b$, (respectively, $\{G^{i,j}_c\}_c$) the measurement $\{F_b\}_b$ (respectively, $\{G_c\}_c$) over the states $\rho_{AB}^i$ and $\rho_{BC}^j$ (respectively, $\rho_{BC}^i$ and $\rho_{AC}^j$).

classical

The newly defined operators and their averages, ${\left< X \right>_\rho \coloneqq \Tr\braks{\rho X}}$, satisfy nontrivial relations.
For example, for ${H=E,F,G}$ and ${i\not=k, j\not=l}$ the operators $H^{i,j}_{\_}$ and $H^{k,l}_{\_}$ act on different Hilbert spaces and, hence, commute, \mbox{$\left[H^{i,j}_{\_},H^{k,l}_{\_} \right]=0$}.
Similarly, expressions such as ${\expec*{ E^{1,1}_a E^{1,2}_{a'}F^{2,2}_b}_\rho}$ and ${\expec*{ E^{1,2}_aE^{1,1}_{a'}F^{1,2}_b}_\rho}$ can be shown identical, since one can arrive at the second one from the first one just by exchanging $\rho_{AB}^1$ with $\rho_{AB}^2$.
More generally, for any function ${Q(\{E^{i,j}_a,F^{k,l}_b,G^{m,n}_c\})}$ of the measurement operators and any three permutations $\pi$, $\pi'$, and $\pi''$ of the indices ${1,...,n}$, one should have
\begin{align}\begin{split}
  \big\langle &Q(\{E^{i,j}_a,F^{k,l}_b,G^{m,n}_c\})\big\rangle_\rho\\
  &=\expec*{ Q(\{E^{\pi(i),\pi'(j)}_a,F^{\pi'(k),\pi''(l)}_b,G^{\pi''(m),\pi(n)}_c\})}_\rho.
  \label{permut}
\end{split}\end{align}
Finally, note that, if we conduct the measurements $\{E^{i,i},F^{i,i}, G^{i,i}\}_{i=1}^n$ at the same time (something we can do, as they all commute with each other), then the measurement outcomes ${a^1,...,a^n,b^1,..., b^n,c^1,...,c^n}$ are distributed according to
\begin{align}
  \expec*{\prod_{i=1}^nE^{i,i}_{a^i}F^{i,i}_{b^i}G^{i,i}_{c^i}}_\rho =\prod_{i=1}^n P_\text{obs}(a^i,b^i,c^i).
  \label{stat}
\end{align}
That is, if the original distribution $P_\text{obs}(a,b,c)$ is compatible with the network in Fig.~\ref{fig:TriangleSubsystems}, then there should exist a Hilbert space $\mathcal{H}$, a state $\rho: \mathcal{H}\to \mathcal{H}$ and operators $\{E^{i,j}_a\}_{i,j,a}$, $\{F^{k,l}_b\}_{k,l,b}$, and $\{G^{m,n}_c\}_{m,n,c}$ satisfying the above relations.
If such is the case, we say that $P_\text{obs}(a,b,c)$ admits an \term{${\boldsymbol{n}^\text{th}\text{-order quantum inflation}}$}.
By increasing the index of $n$, we arrive at a hierarchy of conditions, each of which must be satisfied by any compatible distribution $P_\text{obs}(a,b,c)$.

At first glance, disproving the existence of a quantum inflation looks as difficult as the original feasibility problem.
However, the former task can be tackled via noncommutative polynomial optimization (NPO) theory~\cite{npo}.
Originally developed to characterize quantum correlations in Bell scenarios through the Navascues-Pironio-Acin (NPA) hierarchy~\cite{npa,npa2}, the general goal of NPO theory is to optimize the expectation value of a polynomial over operators subject to a number of polynomial operator and statistical constraints.
This optimization is achieved by means of a hierarchy of semidefinite programming (SDP) tests~\cite{sdp}; see also Appendix~\ref{app_npo}.
In our particular case, we are dealing with a feasibility problem. The polynomial operator constraints we wish to enforce on $E^{i,j}_a$, $F^{k,l}_b$, and $G^{m,n}_c$ are that they define complete families of projectors, which commute when acting on different quantum systems.
The statistical constraints are given by Eqs.~\eqref{permut} and \eqref{stat}.
If for some $n$ we are able to certify, via NPO theory, that $P_\text{obs}(a,b,c)$ does not admit an $n^\text{th}$-order quantum inflation, then we would prove that $P_\text{obs}(a,b,c)$ does not admit a realization in the quantum network in Fig.~\ref{fig:TriangleSubsystems}. An application of this method for this precise scenario is given in Sec.~\ref{sec:SimpleTriangle}.

The method just described can be easily adapted to bound the statistics of any network in which the observed variables are defined by measurements on the quantum latent variables, the triangle scenario described in this section being only an example of such networks.
To test the incompatibility of a distribution $P_\text{obs}$, one would
\begin{itemize}
\item consider a modified network with $n$ copies of each of the latent variables;
\item extend the original measurement operators to act on all possible copies of each system;
\item work out how operator averages relate to $P_\text{obs}$ and to each other; and
\item use NPO theory to disprove the existence of a state and operators satisfying the inferred constraints.
\end{itemize}
In Sec.~\ref{sec:arbitrary} we further show how to extend the notion of quantum inflation to prove infeasibility in general quantum causal structures, where there might be causal connections among observed variables, as well as from observed to latent variables.

\section{DETAILED DESCRIPTION}\label{sec:details}
To illustrate the details of the construction, we first consider a subset of causal scenarios in which single measurements are applied to different quantum states. They correspond to two-layer DAGs in which arrows coming from a first layer, consisting in both observed and latent variables, go to a second layer of observed variables. Each of the variables in the second layer is regarded as an \emph{outcome variable}, since it is the result of conducting a measurement on a quantum state. The tuple of values of all the classical observed parents of such a variable can be understood as the \emph{measurement setting} used to produce this outcome; this case is, for instance, variables $X$, $Y$, and $Z$ in the tripartite-line scenario in Fig.~\ref{fig:Bilocality}.

The essential premise of quantum inflation is to ask what would happen if multiple copies of the original (unspecified) quantum states are simultaneously available to each party.
In this gedanken experiment the parties use copies of their original measurement apparatus to perform $n$ simultaneous measurements on the $n$ copies of the original quantum states now available to them.
There are different ways in which a party can align her measurements to act on the states now available; thus, we must explicitly specify \emph{upon which unique set} of Hilbert spaces a given measurement operator acts nontrivially.
Let us therefore denote measurement operators by
\begin{align*}
\hat{O}^{\boldsymbol{s}|k}_{i|m}\equiv\hat{O}\big(&\text{outcome variable}{=}k,\,\text{ spaces}{=}\boldsymbol{s},\\
&\;\;\text{setting}{=}m,\,\text{ outcome}{=}i\big),
\end{align*}
where the four indices specify
\begin{compactenum}
\item \term{$k$}, the index or name of the \emph{outcome variable} in the original causal graph,
\item \term{$\boldsymbol{s}$}, the Hilbert spaces the given operator acts on,
\item \term{$m$}, the measurement \emph{setting} being used, and
\item \term{$i$}, the \emph{outcome} associated with the operator.
\end{compactenum}

In the example in Fig.~\ref{fig:quantum_inflation}, we have
\begin{align}
  \begin{split}\label{eq:notationtranslation}
  &E_a^{i,j} = E^{\{A_i',A_j''\}}_{a} = \hat{O}^{\{A_i',A_j''\}|A}_{a|\emptyset} ,
  \\ &F_b^{i,j} = F^{\{B'_i, B''_j\}}_{b} = \hat{O}^{\{B'_i, B''_j\}|B}_{b|\emptyset} ,
  \\ &G_c^{i,j} = G^{\{C'_i,C''_j\}}_{c} = \hat{O}^{\{C'_i,C''_j\}|C}_{c|\emptyset} .
  \end{split}\end{align}
Now, using an $n\,{=}\,2$ quantum inflation as in Fig.~\ref{fig:TriangleQInf_implicit}, one finds that $\boldsymbol{s}$ for outcome variable $k\,{=}\,A$ may be sampled from precisely four possibilities, each value being a different \emph{tuple}:
\begin{equation*}
  \quad\boldsymbol{s}\in\bigg\{
  \{A'_{1},A''_{1}\},\{A'_{1},A''_{2}\},\{A'_{2},A''_{1}\},\{A'_{2},A''_{2}\}
  \bigg\},
\end{equation*}
\noindent where $A'_i$ (respectively, $A''_i$) denotes the factor $A$ of the Hilbert space where $\rho^i_{AC}$ (respectively, $\rho^i_{AB}$) acts.

These operators are regarded as the noncommuting variables of an NPO problem where the polynomial constraints are derived according to rules pertaining to the operators' projective nature as well as a number of commutation rules. The statistical constraints are then enforced by imposing symmetry under permutations of the state indices, and also equating certain expectation values with products of observed probabilities.

\subsubsection*{Projection rules}
For fixed $\boldsymbol{s},k,m$, the noncommuting variables $\{\hat{O}^{\boldsymbol{s}|k}_{i|m}\}_i$ must correspond to a complete set of measurement operators. Since we do not restrict the dimensionality of the Hilbert space where they act, we can take them to be a complete set of projectors. That is, they must obey the relations
\begin{subequations}\begin{align}
\label{rule1}
&\hat{O}^{\boldsymbol{s}|k}_{i|m}=(\hat{O}^{\boldsymbol{s}|k}_{i|m})^\dagger,\hat{O}^{\boldsymbol{s}|k}_{i|m}\hat{O}^{\boldsymbol{s}|k}_{i'|m}=\delta_{ii'}\hat{O}^{\boldsymbol{s}|k}_{i|m},\forall\boldsymbol{s},k,i,i',m \\
\label{rule2}
&\sum\nolimits_{i}\hat{O}^{\boldsymbol{s}|k}_{i|m}=\id, \mbox{ for all }\boldsymbol{s},k,m.
\end{align}
\label{rules}
\end{subequations}
These relations imply, in turn, that each of the noncommuting variables is a bounded operator. Hence, by Ref.~\cite{npo}, the hierarchy of SDP programs provided by NPO is complete; i.e., if said distribution does not admit an $n^\text{th}$-order inflation, then one of the NPO SDP relaxations will detect its infeasibility.

\subsubsection*{Commutation rules}
Operators acting on different Hilbert spaces must commute. More formally,
\begin{align}
\left[\hat{O}^{\boldsymbol{s}_1|k_1}_{i_1|m_1},\hat{O}^{\boldsymbol{s}_2|k_2}_{i_2|m_2}\right] =0
\qquad\text{ if }\boldsymbol{s}_1{\cap}\boldsymbol{s}_2{=}\emptyset.
\label{commutation:copies}
\end{align}
As shown in Sec.~\ref{sec:classical}, it is possible to construct an alternative SDP for constraining the correlations of \emph{classical} causal structures by imposing that all measurement operators commute.

\subsubsection*{Symmetry under permutations of the indices}
Because of the way inflated networks are constructed from the original network, all averages of products of the noncommuting variables must be invariant under any permutation $\pi$ of the source indices. Call $\rho$ the overall quantum state of the inflated network (since we do not cap the Hilbert space dimension, we can assume that all state preparations in the original network are pure). Then we have that
\begin{align}\begin{split}
  \label{eq:symconstraint}
  \big\langle\hat{O}^{\boldsymbol{s}_1|k_1}_{i_1|m_1} & \!\!\cdot \hat{O}^{\boldsymbol{s}_2|k_2}_{i_2|m_2}\!\!\cdot\dots\cdot\hat{O}^{\boldsymbol{s}_n|k_n}_{i_n|m_n}\big\rangle_\rho \\
  = &
  \expec*{\hat{O}^{\pi(\boldsymbol{s}_1)|k_1}_{i_1|m_1}  \!\!\cdot \hat{O}^{\pi(\boldsymbol{s}_2)|k_2}_{i_2|m_2}\!\!\cdot\dots\cdot\hat{O}^{\pi(\boldsymbol{s}_n)|k_n}_{i_n|m_n}}_\rho.
\end{split}\end{align}
An example of such statistical constraints imposed in the triangle scenario is given in Eq.~\eqref{permut}.
Another example, for an inflation level $n\,{=}\,3$, is the following:
\begin{subequations}\begin{align}\label{eq:relabelstatesexample}
  \nonumber&\expec*{E_0^{\{A'_1,A''_1\}}E_1^{\{A'_2,A''_2\}}F_0^{\{B'_1,B''_3\}}G_0^{\{C'_1,C''_1\}}}_\rho\\
  \nonumber&\qquad\quad^{\text{apply }\rho_{AB}^1\leftrightarrow\rho_{AB}^2}\\
  &=\expec*{E_0^{\{A'_1,A''_2\}}E_1^{\{A'_2,A''_1\}}F_0^{\{B'_2,B''_3\}}G_0^{\{C'_1,C''_1\}}}_\rho\\
  \nonumber&\qquad\quad^{\text{apply }\rho_{AB}^1\leftrightarrow\rho_{AB}^3}\\
  &=\expec*{E_0^{\{A'_1,A''_2\}}E_1^{\{A'_2,A''_3\}}F_0^{\{B'_2,B''_3\}}G_0^{\{C'_1,C''_1\}}}_\rho\\
  \nonumber&\qquad\quad^{\text{apply }\rho_{BC}^1\leftrightarrow\rho_{BC}^3}\\
  &=\expec*{E_0^{\{A'_1,A''_2\}}E_1^{\{A'_2,A''_3\}}F_0^{\{B'_2,B''_1\}}G_0^{\{C'_3,C''_1\}}}_\rho,
\end{align}\end{subequations}
\noindent where, for readability, we identify ${\hat{O}^{\boldsymbol{s}|A}_{a|\emptyset}}$ ${(\hat{O}^{\boldsymbol{s}|B}_{b|\emptyset})}$ ${[\hat{O}^{\boldsymbol{s}|C}_{c|\emptyset}]}$ with $E^{\boldsymbol{s}}_a$ ${(F^{\boldsymbol{s}}_b)}$ ${[G^{\boldsymbol{s}}_c]}$ per Eq.~\eqref{eq:notationtranslation}.

\subsubsection*{Consistency with identifiable monomials}
Finally, the crucial ingredient that connects an inflated network to the observed correlations is that, as described by Eq.~\eqref{stat} in Sec.~\ref{sec:simple}, certain expectation values pertaining to the inflated network can be related to products of the probabilities of $P_\text{obs}$. We refer to those products of probabilities as the \term{identifiable monomials}. Let $n$ be the order of the considered inflation. We first single out the $j^\text{th}$ copy of each state, for $j=1,\ldots,n$. For each random variable $k$, we write $\vec{j}_k$ for the set of Hilbert spaces on which $k$ acts with the copy labels of all Hilbert spaces set to $j$ (e.g., $\vec{1}_A=\{A'_1,A''_1\}$ and $\vec{2}_A=\{A'_2,A''_2\}$ in Fig.~\ref{fig:quantum_inflation}); we write $i_{j,k}$ for the measurement outcome of $k$ and $m_{j,k}$ for its measurement setting, so we end up describing the operator $\hat{O}^{\vec{j}_k|k}_{i_{j,k}|m_{j,k}}$. For fixed $j$, the product of these operators over all random variables $k$ has an expectation value that reproduces the observed probability ${P_\text{obs}}{\left(\bigcap_k (i_{j,k}|m_{j,k},k)\right)}$. Now, if we furthermore take the product of $j$ over all $n$ copies, the expectation value of the resulting operator reproduces a degree-$n$ monomial comprised of observable probabilities. Formally, for any set of indices $\{i_{j,k},m_{j,k}\}_{j,k}$, we have that
\begin{align}
&\expec*{\prod_{j=1}^n\prod_{k}\hat{O}^{\vec{\boldsymbol{j}}_{k}|k}_{i_{j,k}|m_{j,k}}\hspace{-0.3ex}}_{\hspace{-1ex}\rho}=\prod_{j=1}^n {P_\text{obs}}{\left(\bigcap_k (i_{j,k}|m_{j,k},k)\hspace{-0.5ex}\right)},
\label{consistent}
\end{align}
where $(i|m,k)$ denotes the event of probing $k$ with setting $m$ and obtaining the result $i$, and where $\vec{\boldsymbol{j}}_{k}$ indicates that this operator acts nontrivially on only the Hilbert spaces associated with copy index $\boldsymbol{\sigma}{=}\{jj\dots\}$ of those original sources which are parents of party $k$. See Eq.~\eqref{stat} for an example of this consistency rule applied to the triangle scenario.

The constraints~\eqref{rules}, \eqref{commutation:copies}, \eqref{eq:symconstraint} and \eqref{consistent} are satisfied by the overall state and operators involved in an $n^\text{th}$-order quantum inflation of a particular causal process with observed distribution $P_\text{obs}$. All have a form suitable to assess their physical realizability through NPO via a feasibility problem.

The same machinery can easily be adapted to optimization problems beyond feasibility. Crucially, the left-hand side of Eq.~\eqref{consistent}, under the constraints \eqref{rules}-\eqref{eq:symconstraint}, can be interpreted as a relaxation of the convex hull of $n$ independent copies of a distribution $P_\text{obs}$ compatible with the considered causal structure.
Therefore, quantum inflation can be exploited not only for assessing whether an observed probability distribution can be generated in a particular quantum causal scenario, but also to optimize Bell-like polynomial expressions $B(P_\text{obs})$ over all quantum $P_\text{obs}$ admitting a given causal explanation, so long as the polynomial  $B(P_\text{obs})$ makes reference only to identifiable monomials.
The key idea is to express any Bell-like polynomial $B$ as a linear function over the expectation values $\expec*{\prod_j O_j}$, via the correspondence \eqref{consistent}. Minimizing or maximizing such a linear combination of expectation values can also be cast as a noncommutative polynomial optimization problem, that can be similarly tackled with NPO.
In many polynomial optimization problems, no information about the distribution $P_\text{obs}$ is specified in advance; in Sec.~\ref{sec:opt} we show an explicit example of how Eq.~\eqref{consistent} is exploited in the optimization of polynomial Bell operators. In other cases, such as the quantum network security analysis in Sec.~\ref{sec:cryptography}, we consider polynomial optimization subject to partial specification of the identifiable monomials, for instance, if one has constraints pertaining to marginals of $P_\text{obs}$.

In fact, this idea can be generalized even further to carry optimizations over polynomials of arbitrary operator averages, i.e., not necessarily those averages related to products of observed probabilities (we show a simple example in Sec.~\ref{sec:results:nonlinear}). Moreover, the quantum inflation technique can also be applied for studying classical causal inference. We explain how to do so in Sec.~\ref{sec:classical}, but, before that, we describe how to extend the ideas above to arbitrary quantum causal networks.

\section{ARBITRARY CAUSAL SCENARIOS}\label{sec:arbitrary}
In the previous section, we provide a systematic method to characterize the correlations achievable in a subset of quantum causal networks, namely, two-layer DAGs, where no node has both parents and children and where every observable node has at most one observable parent. Let us denote such DAGs to have a \term{network form}.
In this section, we generalize all previous methodology to characterize correlations in arbitrary causal structures composed of the elements described in Sec.~\ref{sec:structures}.
Namely, we explain how an arbitrary causal structure composed of these elements can be reduced to network form cases, which we can already solve.

\subsection{Visible nodes with parents and children: Maximal interruption}\label{sec:interruption}
The first generalization we consider extends our techniques for network form DAGs to arbitrary so-called \term{latent exogenous} causal structures, where all unobserved nodes are root nodes but otherwise classical variables can have both parents and children.

\begin{figure}[b]
  \begin{center}
    \hfill
    \subfigure[\label{fig:instrumental}]
     {\centering
      \includegraphics[scale=0.45]{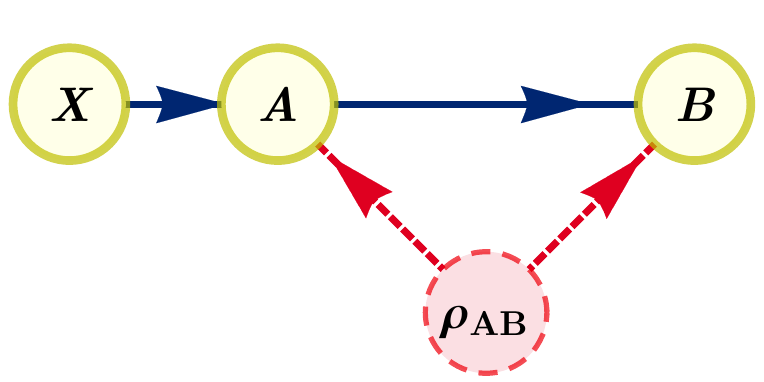}
     }
    \hfill
    \subfigure[\label{fig:instinterrupted}]
     {\centering
      \includegraphics[scale=0.45]{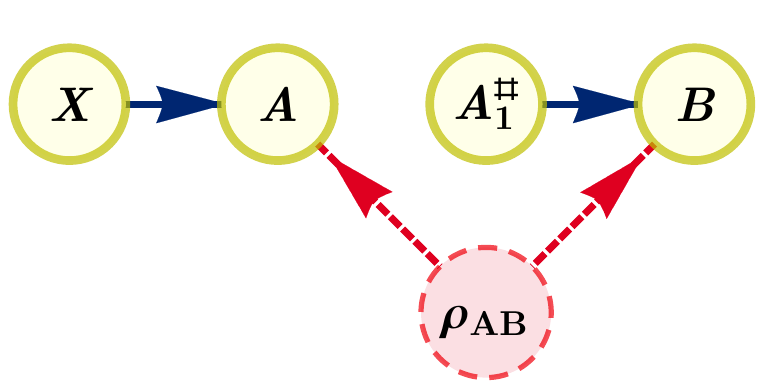}
     }
     \hfill
  \end{center}
  \caption[]{(a) The instrumental causal structure and (b) its maximal interruption, which is isomorphic to the standard causal structure of a Bell experiment. Tsirelson-type inequalities pertaining to the interrupted graph are translated into constraints for the quantum instrumental scenario~\cite{Himbeeck2018instrumental,Agresti2019,Quantifying2020} via the postselection relation ${P_{\text{\ref{fig:instrumental}}}(A{=}a,B{=}b|X{=}x)} = {P_{\text{\ref{fig:instinterrupted}}}(A{=}a,B{=}b|X{=}x,A^{\#}_1{=}a)}$.
  }
  \label{fig:instinterruptionexample}
\end{figure}

\begin{figure}[b]
  \begin{center}
    \hfill
    \subfigure[\label{fig:BellPlus}]
     {\centering
      \includegraphics[scale=0.45]{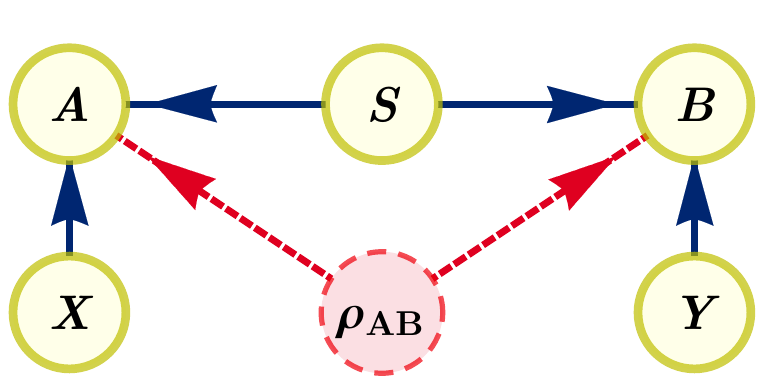}
     }
    \hfill
    \subfigure[\label{fig:BellPlusinterrupted}]
     {\centering
      \includegraphics[scale=0.45]{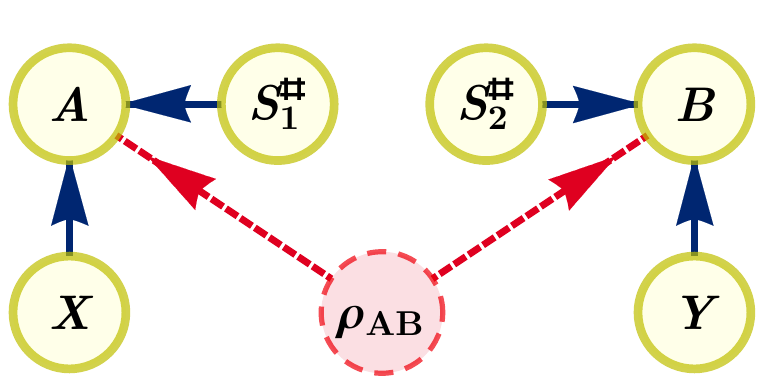}
     }
     \hfill
  \end{center}
  \caption[]{(a) A Bell-like causal structure with a setting common to both parties and (b) its maximal interruption, which is a Bell causal structure assigning distinct outcome variables wholly distinct input variables. Quantum correlations in the common-setting scenario (a) can be explored by applying the standard NPA semidefinite programming hierarchy~\cite{npa,npa2} to the scenario in (b) and then translating back to (a) according to the postselection relation ${P_{\text{\ref{fig:BellPlus}}}(A{=}a,B{=}b|X{=}x,Y{=}y,S{=}s)} = {P_{\text{\ref{fig:BellPlusinterrupted}}}(A{=}a,B{=}b|X{=}x,Y{=}y,S^{\#}_1{=}S^{\#}_2{=}s)}$.
  }
  \label{fig:Bellinterruptionexample}
\end{figure}

We use the term \term{maximal interruption} to refer to our procedure for mapping the correlations of any latent-exogenous causal structure to the correlations of a unique network form structure. The procedure is based on recursive application of two simple rules.

First, suppose a directed edge is found to originate from an \emph{endogenous} (nonroot) observable node, such as node $A$ in Fig.~\ref{fig:instrumental}. Such a node is playing two distinct roles simultaneously: On the one hand, it acts like a setting variable relative to its children. On the other hand, it acts like an outcome variable relative to its parents. The standard NPA semidefinite programming hierarchy~\cite{npa,npa2} presumes that every variable is exclusively either a setting or an outcome, however. Interruption splits such dual-role nodes in two, yielding a gedanken experiment graph with an effective outcome node as well as an effective setting node; see Fig.~\ref{fig:instinterrupted}. Postselecting on the effective outcome variable, for example, $A$ in Fig.~\ref{fig:instinterrupted}, taking a value that matches that of the effective setting variable, $A^{\#}_1$ in Fig.~\ref{fig:instinterrupted}, allows us to characterize the set of distributions compatible with the original graph as a projection of the space of distributions compatible with the interrupted graph. This idea has precedent in classical causal networks in the single-world intervention graphs pioneered by Ref.~\cite{Richardson2013SingleWI} (see also the node-splitting procedure of~Ref.~\cite{BarrettQCM}), as well as the $e$-separation technique introduced in Ref.~\cite{evans2012graphicalmethods}. Interruption of the sort illustrated in Fig.~\ref{fig:instinterruptionexample} is already exploited in quantum information research in Refs.~\cite{Himbeeck2018instrumental,Agresti2019,Quantifying2020}.

\begin{figure}[b]
  \begin{center}
    \hfill
    \subfigure[\label{fig:tointerrupt}]
     {\centering
      \includegraphics[scale=0.45]{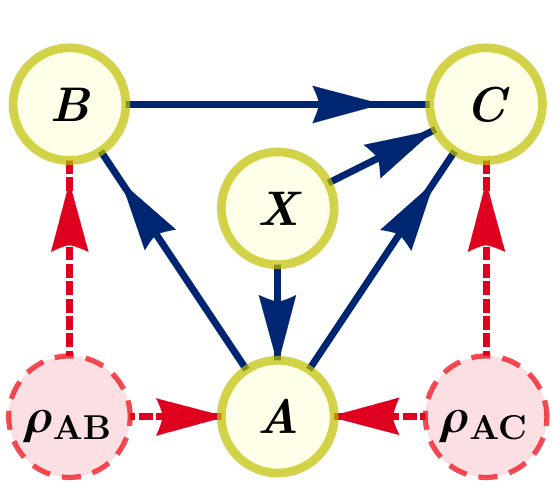}
     }
    \hfill
    \subfigure[\label{fig:interrupted}]
     {\centering
      \includegraphics[scale=0.45]{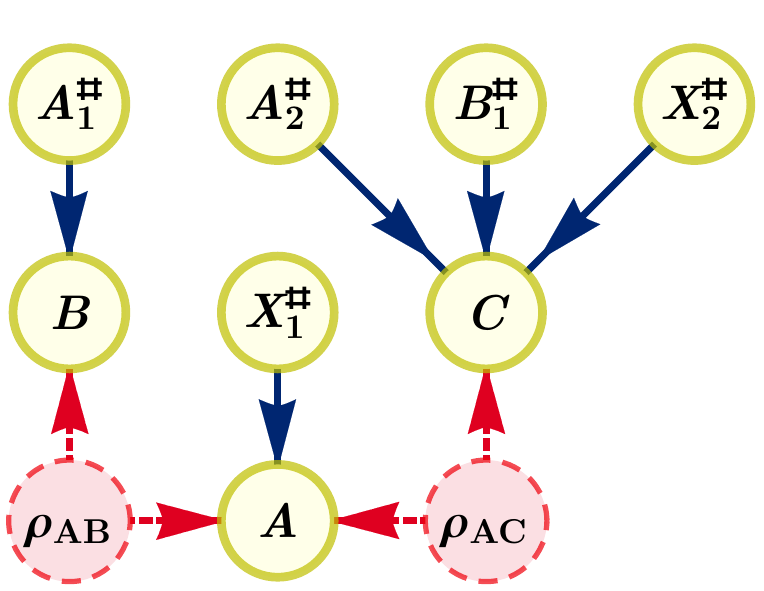}
     }
     \hfill
  \end{center}
  \caption[]{An example causal structure (a) and its maximal interruption (b). Note that in (a) the observable nodes $A$ and $B$ both possess at least one parent node as well as at least one child node. Additionally, the observable nodes $A$ and $X$ have multiple child nodes. Constraints on distributions extended to the interrupted graph (such as those obtained via quantum inflation) are translated into constraints on the original distribution using the postselection relation ${P_{\text{\ref{fig:tointerrupt}}}(A{=}a,B{=}b,C{=}c|X{=}x)} = {P_{\text{\ref{fig:interrupted}}}(A{=}a,B{=}b,C{=}c|X^\#_1{=}X^\#_2{=}x,A^{\#}_1{=}A^{\#}_2{=}a,B^{\#}_1{=}b)}$. The maximally interrupted DAG in (b) has the two-layer network structure considered in Sec.~\ref{sec:details}.
  }
  \label{fig:interruptionexample}
\end{figure}

The other rule relevant to maximal interruption involves the case where one observable node has multiple children, such as node $S$ in Fig.~\ref{fig:BellPlus}. If every outcome node is considered a ``party'' in the interpretation of Bell scenario causal structures, then such nodes can be thought of setting variables which are common to more than one party. Think of Fig.~\ref{fig:BellPlus} as a Bell experiment in which Alice's (respectively, Bob's) setting consists of one bit determined by a source of private randomness $X$ (respectively, $Y$), as well as one bit determined by a shared random number $S$. Without loss of generality, we can characterize the set of correlations that Alice and Bob can observe in this scenario by instead characterizing compatibility relative to Fig.~\ref{fig:BellPlusinterrupted} and then postselecting on $S^{\#}_1{=}S^{\#}_2$. That is, we can characterize the set of correlations that Alice and Bob \emph{would} observe if the $S$ bit for Alice and the $S$ bit for Bob were independent---i.e., Fig.~\ref{fig:BellPlusinterrupted}---such that the set of distributions compatible with Fig.~\ref{fig:BellPlus} is the projection of the set of distributions compatible with the independent-settings scenario onto the subspace where $S^{\#}_1{=}S^{\#}_2$.

Maximal interruption, then, is the combination of both these two forms of node splitting. Graphically, maximal interruption modifies a graph $\mathcal{G}$ as follows: For every observed variable $V$ which is neither a root node nor a terminal node, as well as for any $V$ having multiple children, we introduce $k$ new variables $\{V^{\#}_i\}_i$ where $k$ denotes the number of edges outgoing from $V$. We then replace each edge formerly originating from $V$ by an edge originating from the corresponding $V^{\#}_i$.  In the resulting (partially) \emph{interrupted} graph $\mathcal{G}'$ , $V$ is a terminal node and every $V^{\#}_i$ is a root node. The correlations in $\mathcal{G}'$ are related to those in the original graph $\mathcal{G}$ by postselection, namely
\begin{align}\label{eq:interruptioncondition}
P_{\mathcal{G}}(\dots,V{=}v) = P_{\mathcal{G}'}(\dots,V{=}v|V^{\#}_1{=}\dots{=}V^{\#}_k{=}v).
\end{align}
Proceeding in this fashion, any latent exogenous causal structure can be converted into a network form.
A graphical example of this conversion is shown in Fig.~\ref{fig:interruptionexample}.

Distributions over the nodes of the interrupted graph $\mathcal{G}'$ constitute \emph{extensions} of distributions pertaining to the original graph. That is, observed statistics $P$ pertaining to $\mathcal{G}$ only partially specify an extension $P'$ to $\mathcal{G}'$. Critically, a distribution $P$ is compatible with a graph $\mathcal{G}$ if and only if there exists some \term{valid extension} $P'$ of $P$ which is compatible with the interrupted graph  $\mathcal{G}'$. $P'_{\boldsymbol{AB}\vert\boldsymbol{A}^\#}$ is said to constitute a valid extension of the original distribution $P_{\boldsymbol{AB}}$ when $P'_{\boldsymbol{AB}\vert\boldsymbol{A}^\#}$ is compatible with the interruption graph  $\mathcal{G}'$ and recovers the original distribution $P_{\boldsymbol{AB}}$  under postselection. Formally, we have the following lemma.\par
\medskip
\begin{samepage}
  \noindent\textbf{Fundamental Lemma of Interruption}
  \begin{align}\begin{split}\label{eq:interruption}
    P_{\boldsymbol{A}\boldsymbol{B}}\in & \textsf{CompatibleDistributions}\big[\mathcal{G}\big]\\
    \text{iff}\quad \exists_{P'}\:& :\:\: P'_{\boldsymbol{AB}\vert\boldsymbol{A}^\#} \in \textsf{ValidExtensions}_{\mathcal{G'}}\big[P_{\boldsymbol{AB}}\big],\\
    \text{i.e.}\quad & P_{\boldsymbol{AB}}(\boldsymbol{ab})=P'_{\boldsymbol{AB}\vert\boldsymbol{A}^\#}(\boldsymbol{ab} \;\;\vert\,\boldsymbol{a})\\
    \text{and}\quad & P'_{\boldsymbol{A}\boldsymbol{B}\vert\boldsymbol{A}^\#}\in \textsf{CompatibleDistributions}\big[\mathcal{G'}\big].
  \end{split}\end{align}
\end{samepage}

This result allows us to witness causal incompatibility relative to $\mathcal{G}$ by witnessing causal incompatibility relative to $\mathcal{G}'$. Since $\mathcal{G}'$ is of network form, we can readily apply quantum inflation to it.

\subsection{Latent nodes with parents}\label{sec:nonexog}
The remaining case that needs to be considered is that of latent nonexogenous causal structures, where latent nodes can have parents.
The Evans exogenization procedure \cite{Evans2018NMP} allows \emph{classical} latent nonexogenous structures to be transformed into latent exogenous causal structures with the same predictive power. The procedure consists in replacing all arrows from a parent node to a latent node with arrows from the parent node to the children of the latent node. This operation is repeated for all parents of all latent nodes such that finally all latent variables become parentless.

\begin{figure}[!b]
  \begin{center}
    \hfill
    \subfigure[\label{fig:NonExog}]
    {\centering
      \includegraphics[scale=0.45]{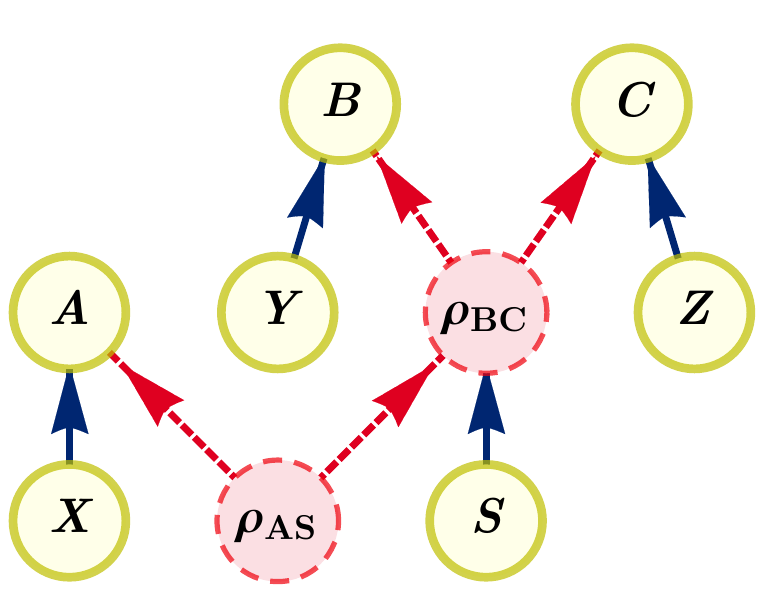}
    }
    \hfill
    \subfigure[\label{fig:ExogAnalog}]
    {\centering
      \includegraphics[scale=0.45]{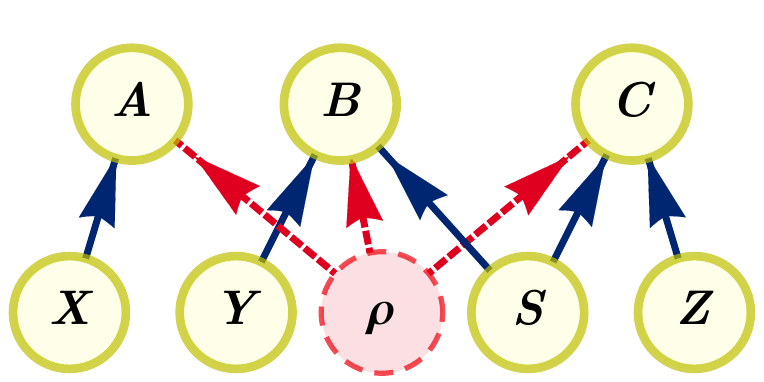}
    }
    \hfill
  \end{center}
  \caption[]{In (a), there is a causal structure with $\rho_{BC}$ being a nonexogenous unobserved quantum node. In (b), there is a different causal structure, corresponding to the classical latent reduction of the former. While these two graphs are equivalent if the unobserved nodes are classical, they are demonstrably inequivalent when the unobserved nodes are quantum.}
  \label{fig:Exog}
\end{figure}

When applied to quantum latent variables, however, exogenization results in a new quantum network that, in general, does not predict the same distributions of observed events as its predecessor.
This result is again related to the impossibility of broadcasting quantum information~\cite{NoCloningGeneral2006}. The example in Fig.~\ref{fig:Exog}, evidencing this compatibility mismatch, is wholly due to Stefano Pironio.

To make the issue explicit, in the original network shown in Fig.~\ref{fig:NonExog} the variable $S$ has the interpretation of a setting, which adjusts the state $\rho_{BC}$ before it is sent to $B$ and $C$.
Thus, it is possible for ${P(A,B|X,Y,S{=}0)}$ to maximally violate a Bell inequality for $A$ and $B$, and ${P(A,C|X,Z,S{=}1)}$ to maximally violate a Bell inequality for $A$ and $C$.
If one applies the Evans exogenization procedure to the original network, one obtains the network in Fig.~\ref{fig:ExogAnalog}. However, no quantum state prepared independently of $S$ can maximally violate a Bell inequality between $A$ and $B$ and at the same time maximally violate another Bell inequality between $A$ and $C$, due to the monogamy of quantum correlations~\cite{Toner2009,Seevinck2010}.
Consequently, it is not possible to reproduce such correlations within the causal network in Fig.~\ref{fig:ExogAnalog}, a product of applying Evans exogenization to Fig.~\ref{fig:NonExog}.

One way to deal with this predicament is to regard observable variables with unobserved children as random variables indicating the classical control of a quantum channel.
If a nonroot latent node has latent parents, then it is understood as a controlled channel on the Hilbert space defined by the latent parents.
If, instead, a nonroot latent node has exclusively observable parents, then the latent node is regarded as a controlled source of quantum state preparation, which we model as a source of quantum states followed by a quantum channel controlled by the observable parents.
In the same way, latent variables with latent children are treated as uncontrolled quantum channels (we defer to Appendix~\ref{app:nonexog} the discussion that illustrates the need to account for this type of channels).
Thus, in Fig.~\ref{fig:NonExog}, we treat the latent variable $\rho_{AS}$ as a joint quantum state and the root variable $S$ as the classical control for a quantum channel acting on the $BC$ subsystems.
That is, one understands ${\rho_{ABC|S=s}=\hat{U}_s \rho_{AS} \hat{U}^{\dagger}_s}$, where, for all values of $s$, $\hat{U}_s$ is a unitary operator that commutes with any operator acting solely on $A$'s subsystem.
As such, the joint distribution of the values of the visible variables $A$, $B$, and $C$ conditioned on the root visible nodes can be understood as generated by
\begin{align*}
  P(A&{=}a,B{=}b,C{=}c|X{=}x,Y{=}y,Z{=}z,S{=}s)\\
  & =\expec*{ \hat{U}^{\dagger}_s \hat{A}^a_x \hat{B}^b_y \hat{C}^c_z \hat{U}_s}_{\rho_{AS}} =\expec*{\hat{A}^a_x \hat{U}^{\dagger}_s \hat{B}^b_y \hat{C}^c_z \hat{U}_s}_{\rho_{AS}}.
\end{align*}
This interpretation can be made without loss of generality, since the subspace $S$ of the complete Hilbert space $AS$ can be understood as containing the subspaces corresponding to $B$ and $C$.

As in the exogenous case, an $n^\text{th}$-order inflation of a causal structure with nonexogenous quantum variables requires taking $n$ copies of the unobserved root nodes and of the unobserved nodes with only visible parents. Each unitary operator $\hat{U}_s$ in the original causal structure gives rise to operators of the form $\hat{O}^{\{S^j\}|U}_{s}$ in the inflated graph, where $j$ denotes the copy of the Hilbert space where $\hat{U}_s$ acts. The unitary or outcome operators associated to the descendants of any such ``unitary node'' (for instance, $B$ and $C$, in Fig.~\ref{fig:Exog}) inherit the copy label $j$ of the Hilbert space $S^j$.

With this last prescription, the symmetry relabeling rule [Eq.~\eqref{eq:symconstraint}] still holds. However, the projection rules [Eqs. \eqref{rule1} and \eqref{rule2}] hold only if the noncommuting variable $k$ in question corresponds to an outcome variable in the original graph. If $k$ corresponds to a unitary variable, then the operator $\hat{O}^{\boldsymbol{s}|k}_{m}$ must be subject to the constraints
\begin{equation}
\hat{O}^{\boldsymbol{s}|k}_{m}(\hat{O}^{\boldsymbol{s}|k}_{m})^\dagger=(\hat{O}^{\boldsymbol{s}|k}_{m})^\dagger\hat{O}^{\boldsymbol{s}|k}_{m}=\id.
\end{equation}
The commutation rule \eqref{commutation:copies} remains valid upon qualifying that the Hilbert spaces listed in $\boldsymbol{s}_1\cup\boldsymbol{s}_2$ must be simultaneously coexisting in the original graph.
For example, in Fig.~\ref{fig:NonExog}, the operators corresponding to the Hilbert spaces associated to the outcome variables $B$ and $C$ coexist after the transformation $U_s$ is applied over system $S$.
It follows that the corresponding measurement operators $O^{\boldsymbol{s}|B}_{b|y}$ and $O^{\boldsymbol{s}'|C}_{c|z}$ commute.
Finally, rule \eqref{consistent} expressing consistency with identifiable monomials must also be amended to take into account that descendants of a unitary variable must be bracketed by the corresponding unitary and its adjoint.

Note that the aforementioned operator and statistical constraints are all polynomial, and, thus, they can all be enforced in the framework of NPO theory.

Interruption, classical exogenization, and the treatment for quantum exogenous variables hereby presented cover all possible nontrivial causal influences in arbitrary quantum causal structures composed of preparations, transformations, and measurements as described in Sec.~\ref{sec:structures}. Quantum inflation is, therefore, a technique of full applicability to bound the quantum correlations achievable in general causal scenarios.

\section{SDP FOR CLASSICAL COMPATIBILITY}\label{sec:classical}
The quantum inflation technique can be easily adapted for solving the problem of causal compatibility with an arbitrary \emph{classical} causal structure.
It is known that any correlation achievable with only classical latent variables can be realized in terms of commuting measurements acting on a quantum state~\cite{baccari2017classical}.
Therefore, in order to detect correlations incompatible with classical structures, one must generalize the commutation relations in the original quantum inflation method to the constraint that any pair of measurement operators commutes. That is,
\begin{equation*}
    \hat{O}^{\boldsymbol{s}_1|k_1}_{i_1|m_1}  \!\!\cdot \hat{O}^{\boldsymbol{s}_2|k_2}_{i_2|m_2} =
\hat{O}^{\boldsymbol{s}_2|k_2}_{i_2|m_2}  \!\!\cdot \hat{O}^{\boldsymbol{s}_1|k_1}_{i_1|m_1},
\end{equation*}
for all $\boldsymbol{s}_1$, $\boldsymbol{s}_2$, $k_1$, $k_2$, $m_1$, $m_2$, $i_1$, and $i_2$.
This generalization defines a hierarchy of constraints that classical correlations compatible with a given causal structure must satisfy.

Contrary to the quantum case, the NPO hierarchy with commuting operators associated to a fixed inflation level is guaranteed to converge at a finite level.
In fact, for NPO hierarchy levels higher than $N\cdot m\cdot(d{-}1)$---where $N$ is the number of parties, $m$ is the number of settings per party, and $d$ is the number of outcomes per measurement---application of the commutation relations allows one to reduce any product of the operators involved into one of shorter length.
For a fixed inflation level, the problem solved at the highest level of the NPO hierarchy is analogous to the linear program solved in classical inflation~\cite{wolfe2016inflation} at the same inflation level.

In contrast with the original classical inflation technique, the classical variant of the quantum inflation technique described in this work uses semidefinite programming and exhibits far more efficient scaling with the inflation hierarchy than the original linear programming approach~\cite{navascues2017inflation}.
One must note that this gain in efficiency comes at the expense of introducing further relaxations in the problem.
Nevertheless, this classical variant of quantum inflation is capable of recovering a variety of seminal results of classical causal compatibility, such as the incompatibility of the so-called W and GHZ distributions (described in the results sections) with classical realizations in the triangle scenario.
It also identifies the distribution described by Fritz~\cite{fritz2012bell}, known to have a quantum realization in the triangle scenario, as incompatible with classical realizations.
For all these results, the relaxed SDP formulation is far less memory demanding than the raw linear programming formulation.
Furthermore, the SDP approach is the only method that can be used when using inflation to assess causal compatibility in the presence of terminal nodes which can take continuous values.

In conclusion, not only can quantum inflation be leveraged to obtain results for networks with classical sources, but we argue that it is the most suitable technique to be used for addressing causal compatibility with classical realizations in large networks.

\section{APPLICATIONS TO QUANTUM CORRELATIONS IN NETWORKS}\label{sec:results}
We illustrate in this section various problems akin to quantum information science where quantum inflation provides a means to its solution or to approximations of it.
We focus here on broad classes of problems (determining whether a distribution can be generated in a quantum network, finding witnesses of network nonlocality, and characterizing the quantum correlations achievable in a particular network) that underlie many practical problems and which we illustrate in concrete examples of relevance.
Except those in Secs.~\ref{sec:SimpleTriangle}~and~\ref{sec:Bilocality}, all the results described are novel, and quantum inflation outperforms alternative methods for characterizing quantum network correlations.\footnote{In the particular case in Sec.~\ref{sec:SimpleTriangle}, we still answer an open problem, although we later find that the same result can also be derived by using nonfanout classical inflations of the triangle scenario. Note that nonfanout classical inflations yield GPT-valid---and, hence, also quantum-valid---constraints~\cite{wolfe2016inflation}.}
Moreover, to date, quantum inflation is the only technique capable of making strong quantitative statements about genuine multipartite quantum nonlocality under local operations and shared randomness (as illustrated in Ref.~\cite{Schmid2020LOSR}), in that it is \emph{the} natural framework for studying the device-independent analog of genuine network multipartite entanglement~\cite{navascues2020gnme}.

\subsection{Quantum causal compatibility}
We begin by showing examples of specific probability distributions that we are able to identify as incompatible with various tripartite quantum causal networks.
While quantum inflation applies to networks with observed variables of arbitrary cardinality, we hereafter consider scenarios involving exclusively two-output observed variables. We adopt the standard notations and label these outputs as $\{0,1\}$, unless otherwise specified.

\subsubsection{The triangle scenario}
\label{sec:SimpleTriangle}
\begin{figure}[t]
  \centering
  \includegraphics[scale=0.45]{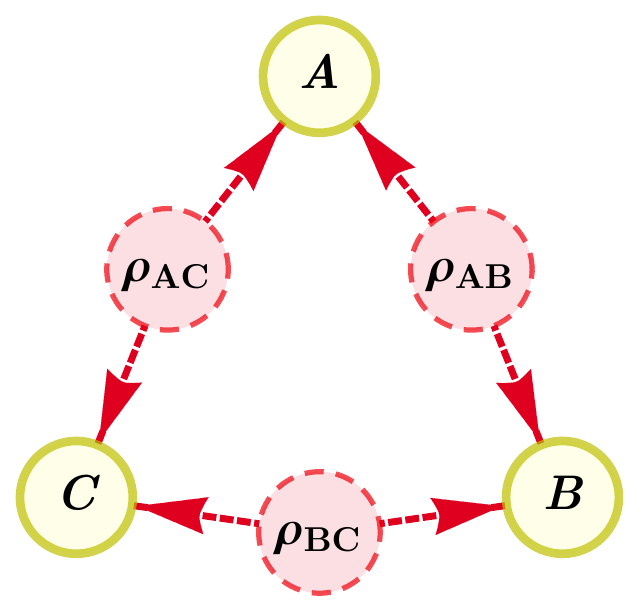}
  \caption{The quantum triangle scenario, without settings. Each of the observable variables $A$, $B$, and $C$ are dependent only on the sources of bipartite entanglement. This figure is essentially a reproduction of Fig.~\ref{fig:TriangleNoSettings}, the only difference being some added specificity regarding the quantum nature of the latent nodes.}
  \label{fig:TriangleNoSettingsQ}
\end{figure}
We start with the simplest version of the triangle scenario consisting of three parties that are influenced in pairs by bipartite latent variables, as depicted in Fig.~\ref{fig:TriangleNoSettingsQ}.
The first example we study in this scenario is the so-called W distribution.
This distribution is defined by the task of all parties outputting the outcome $0$ except one, which should output $1$.
Explicitly, it is
\begin{equation}
    P_\text{W}(a,b,c)\coloneqq\begin{cases}
        \frac{1}{3} & a+b+c=1, \\
        0           & \text{otherwise.}
    \end{cases}
    \label{wdistribution}
\end{equation}
The W distribution is proven in Ref.~\cite{wolfe2016inflation} not to be realizable in the triangle scenario when the latent variables are classical.
Additionally, it is easy to see that it is realizable with tripartite classical randomness.
However, the question of whether the W distribution is realizable in the quantum triangle scenario remained open.
It can be shown that $P_\text{W}$ does not admit a second-order quantum inflation, and a witness that certifies this incompatibility can be found in Eq.~\eqref{eq:Wwitness}.
Therefore, a quantum realization of the W distribution in the triangle scenario is impossible.

Quantum inflation is robust and certifies that, when mixing the W distribution with white noise, the resulting distribution $P_{\text{W},v}(a,b,c)\coloneqq v P_\text{W}(a,b,c)+(1-v)/8$ does not have a quantum realization in the triangle for all visibilities $v$ higher than $3(2-\sqrt{3})\approx0.8039$. This result is obtained by solving the NPO program associated to a second-order inflation and the set of monomials $\mathcal{L}_2$ (see Appendix~\ref{app:sets} for the definition of this set), restricted to operators of length ${\leq}\,3$.

\subsubsection{The triangle scenario with settings}\label{sec:Mermin}
\begin{figure}[t]
  \centering
  \includegraphics[scale=0.45]{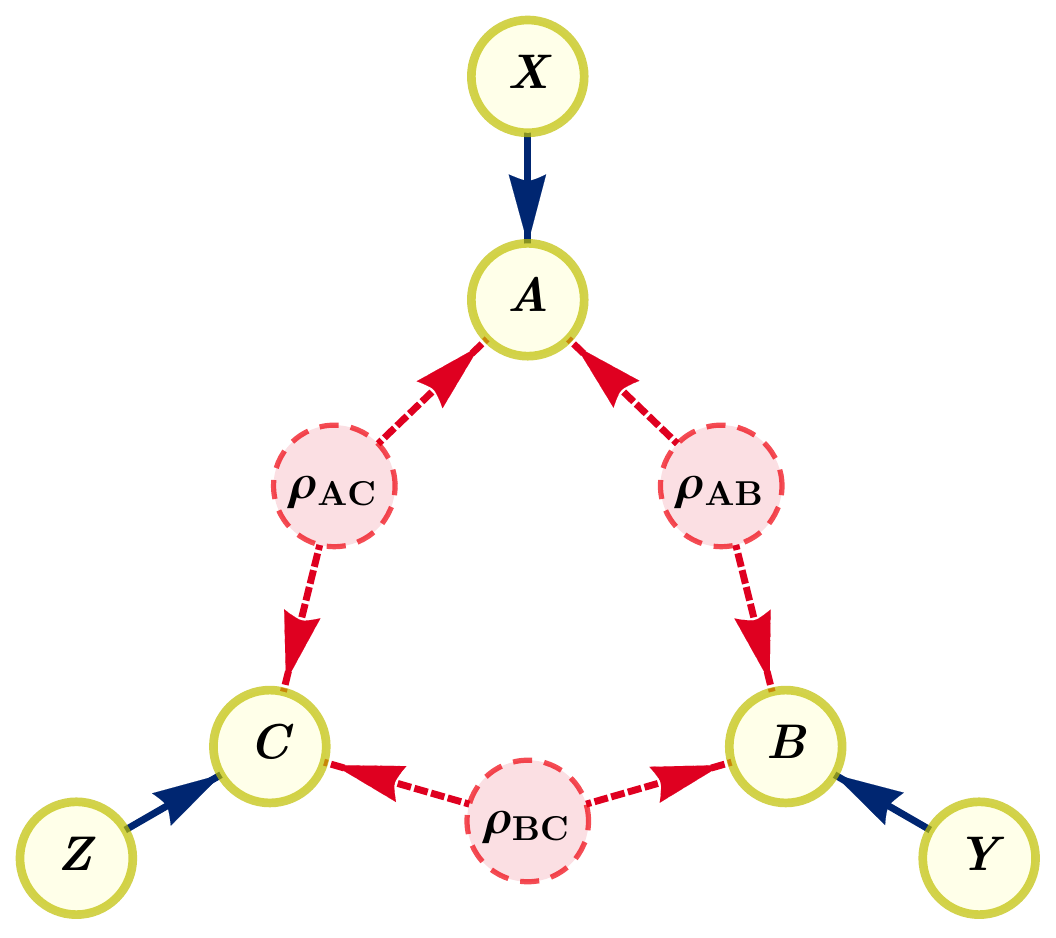}
  \caption{The quantum triangle scenario with settings, sometimes denoted $\mathcal{Q}^{\triangle}$ for brevity.  Each of the observable variables $A$, $B$, and $C$ are now dependent not only on the sources of bipartite entanglement, but on additional observable classical variables $X$, $Y$, and $Z$ representing measurement choices.}
  \label{fig:TriangleWithSettings}
\end{figure}
One can also consider more complicated networks that include additional observable variables to encode for choices of discrete measurement settings.
Figure~\ref{fig:TriangleWithSettings} shows this type of network for the case of the triangle scenario.
In this setup we study the Mermin-GHZ distribution, defined by $P_{\text{Mermin}}(a,b,c|x,y,z)\coloneqq$
\begin{align}\label{eq:MerminBox}
    \begin{cases}
        1/8                            & x+y+z=0 {\mod{}} \;\; 2,\\
        \left(1+(-1)^{a+b+c}\right) /8 & x+y+z=1, \\
        \left(1-(-1)^{a+b+c}\right) /8 & x+y+z=3.
  \end{cases}
\end{align}
Quantum inflation also allows one to prove that the Mermin-GHZ distribution is not compatible with a quantum realization in the triangle scenario with inputs, by showing that $P_\text{Mermin}$ does not admit a second-order quantum inflation; such incompatibility is certified by the violation of Eq.~\eqref{eq:merminwitness}.
Additionally, its noisy version $P_{\text{Mermin},v}\coloneqq v P_\text{Mermin}+(1-v)/8$ can be proven not to have a quantum realization for any visibility $v$ higher than $\sqrt{2/3}\approx 0.8165$.

\subsubsection{The tripartite-line scenario}\label{sec:Bilocality}
\begin{figure}[t]
\centering
\includegraphics[scale=0.45]{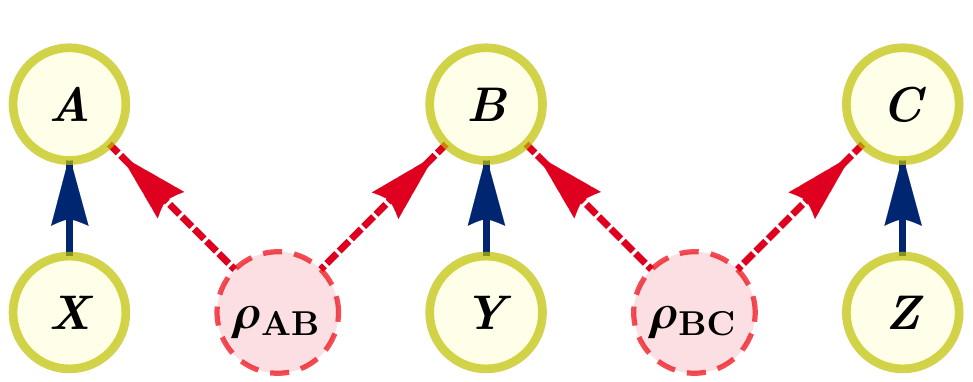}
  \caption[]{\label{fig:TripartiteLine}
  The tripartite-line causal scenario, where two causally independent parties $A$ and $C$ each share some quantum entanglement with a central party $B$, sometimes denoted $\mathcal{Q}^{{\protect\biloc}}$ for brevity. This figure is a reproduction of Fig.~\ref{fig:Bilocality}.}
  \label{fig:BilocalityCopy}
\end{figure}

Quantum inflation is organized as an infinite hierarchy of necessary conditions. There are, however, situations in which it recovers the quantum boundary at a finite step. An example of these situations is provided by the tripartite-line scenario in Fig.~\ref{fig:TripartiteLine}, which underlies phenomena such as entanglement swapping. The tripartite-line scenario constitutes the essential causal unit employed in quantum repeaters and long-distance entanglement distribution networks.
The main characteristic of this structure is that there is no causal connection between the extreme variables $A$ and $C$.
As a consequence of this characteristic, all correlations realizable in the tripartite-line scenario satisfy the following factorization relation
\begin{equation}
    \sum_b P_\text{obs}(a,b,c|x,y,z)=P_\text{obs}(a|x)P_\text{obs}(c|z).
    \label{eq:linecompatible}
\end{equation}

This scenario is thoroughly studied in the literature~\cite{branciard2010bilocality,branciard2012bilocality}.
In fact, it is known that the probability distribution
\begin{equation}
    {P_{\text{2PR}}(a,b,c|x,y,z)\coloneqq\left[1+(-1)^{a+b+c+xy+yz}\right]/8},
\end{equation}
despite satisfying the constraint of Eq.~\eqref{eq:linecompatible}, cannot be realized in the tripartite-line scenario in terms of classical or quantum latent variables.
However, it is known that its mixture with white noise, \mbox{$P_{\text{2PR},v}\coloneqq vP_{\text{2PR}}+(1-v)/8$}, can be realized if the visibility parameter $v$ is sufficiently small~\cite{branciard2012bilocality}.
$P_{\text{2PR},v}$ admits a realization in terms of quantum latent variables for any $0 \,{\leq}\, v \,{\leq}\, 1/2$ and in terms of classical latent variables for any $0 \,{\leq}\, v \,{\leq}\, 1/4$.

Quantum inflation correctly recovers that all $P_{\text{2PR},v}$  with visibility $v\,{>}\,1/2$ are incompatible with the quantum tripartite-line scenario.
It does so by certifying that, for any $v\,{>}\,1/2$, the corresponding $P_{\text{2PR},v}$ does not admit a second-order inflation, and this infeasibility is found already at the NPO hierarchy level corresponding to the set $\mathcal{S}_2$ (see Appendix~\ref{app:sets} for a definition of this monomial set).
That is, in this scenario, a finite order of quantum inflation, namely, the second, already singles out the whole region of quantum violation.

Furthermore, we can also contrast against realizations in terms of classical latent variables by following the predicament in Sec.~\ref{sec:classical}, that is, by imposing that all operators in the problem commute.
This classical version of quantum inflation, when analyzing compatibility with a third-order inflation, solving the NPO problem associated to the corresponding set of monomials $\mathcal{L}_1$ (its definition can be found in Appendix~\ref{app:sets}), witnesses that all $P_{\text{2PR},v}$ with visibility $v\,{>}\,0.328$ cannot be realized in terms of classical hidden variables.
Lower values of this bound can, in principle, be achieved by considering higher inflation orders and larger monomial sets. The required computational capabilities for solving these problems, however, fall outside those provided by standard tabletop computers. We defer the interested reader to Sec.~\ref{sec:conclusions} for a brief discussion on the computational cost of implementing quantum inflation, and to Ref.~\cite[Chap. 5]{alexThesis} for a fuller analysis.
\begin{figure}[t]
  \centering
  \includegraphics[scale=0.3]{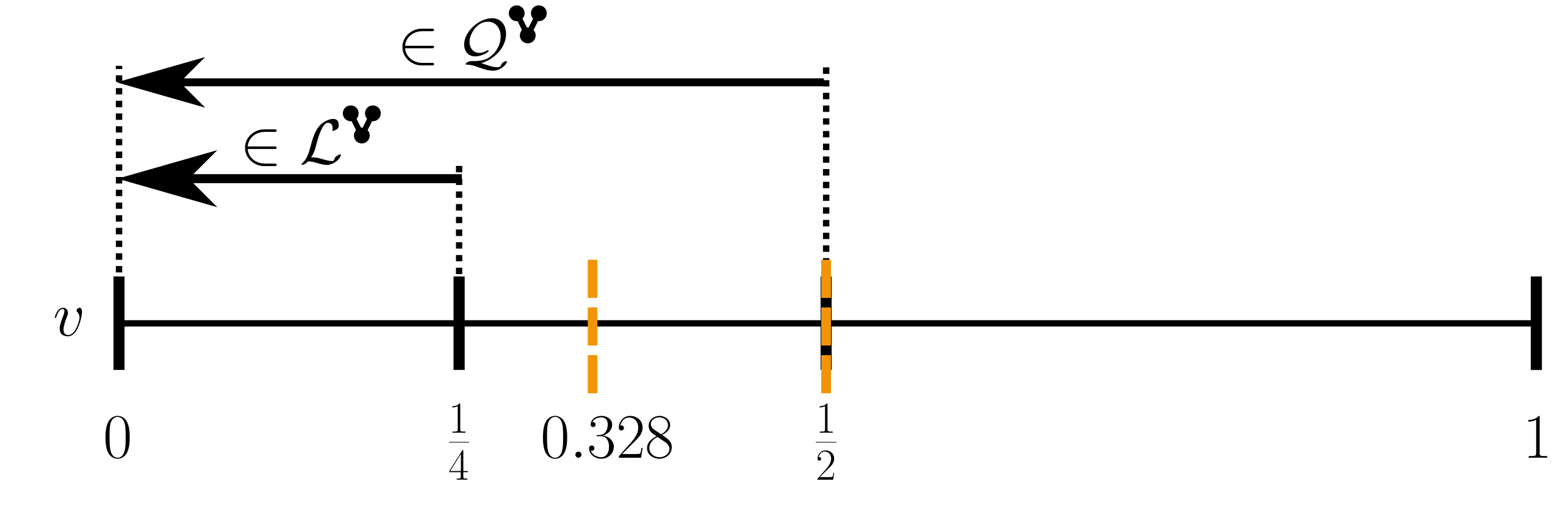}
  \caption{
  Summary of results recoverable with quantum inflation in the tripartite-line scenario in Fig.~\ref{fig:Bilocality} (orange dashed lines).
  $\mathcal{Q}^\protect\biloc$ and $\mathcal{L}^\protect\biloc$ denote, respectively, the sets of quantum and classical correlations that can be generated in the tripartite-line scenario.
  Quantum inflation correctly recovers that all $P_{\text{2PR},v}$ with visibility $v\,{>}\,1/2$ are incompatible with the quantum tripartite line scenario already at the NPO hierarchy finite level $\mathcal{S}_2$ when assessing compatibility with a second-order inflation.
  When imposing that all measurements commute, it witnesses that all $P_{\text{2PR},v}$  with visibility $v\,{>}\,0.328$ cannot be realized in terms of classical hidden variables.}
  \label{fig:line_results}
\end{figure}

\subsection{Witnesses of quantum causal incompatibility}\label{sec:witness}
A feature of semidefinite programming that has been widely employed in quantum information theory is the fact that the solution certificates of SDPs can be interpreted as Bell-like inequalities or witnesses that are capable of identifying correlations not attainable using the considered resources~\cite{npa2,baccari2017classical}.
The certificates obtained when using quantum inflation can have a similar interpretation as witnesses of quantum causal incompatibility.
For instance, the certificate that provides the value of ${v_\text{max}=6-3\sqrt{3}\approx 0.8039}$ for the W distribution in the quantum triangle scenario gives rise to the inequality
\begin{align}
\nonumber&\textbf{W's Certificate}\\
&\begin{psmallmatrix*}[l]
		-2P_\text{inf}(a^{\{A'_1,A''_1\}}{=}0) \\
		 + P_\text{inf}(a^{\{A'_1,A''_1\}}{=}0,a^{\{A'_2,A''_2\}}{=}0) \\
		 - 2 P_\text{inf}(a^{\{A'_1,A''_1\}}{=}0,b^{\{B'_1,B''_1\}}{=}0) \\
		+ 4 P_\text{inf}(a^{\{A'_1,A''_1\}}{=}0,b^{\{B'_2,B''_2\}}{=}0) \\
		- 4 P_\text{inf}(a^{\{A'_1,A''_1\}}{=}0,b^{\{B'_1,B''_1\}}{=}0, \\
		\hskip 1.87cm a^{\{A'_2,A''_2\}}{=}0,b^{\{B'_2,B''_2\}}{=}0) \\
		 + 2 P_\text{inf}(a^{\{A'_1,A''_1\}}{=}0,a^{\{A'_2,A''_2\}}{=}0,b^{\{B'_1,B''_1\}}{=}0) \\
		+ 2 P_\text{inf}(a^{\{A'_1,A''_1\}}{=}0,b^{\{B'_1,B''_1\}}{=}0,b^{\{B'_2,B''_2\}}{=}0) \\
		 - 2 P_\text{inf}(a^{\{A'_1,A''_2\}}{=}0,b^{\{B'_1,B''_2\}}{=}0,c^{\{C'_1,C''_2\}}{=}0) \\
		 + [A\rightarrow B\rightarrow C\rightarrow A] + [A\rightarrow C\rightarrow B\rightarrow A]
		\end{psmallmatrix*} \leq 0,
    \label{eq:Wcert}
\end{align}
where ${+[A\rightarrow B\rightarrow C\rightarrow A]}$ and ${+[A\rightarrow C\rightarrow B\rightarrow A]}$ mean repeating every term in the sum under the cyclic and anticyclic permutations of subsystems, respectively---thereby implicitly tripling the coefficients of any cyclic-invariant terms.
This certificate is defined in terms of a quantum distribution compatible with a second-order inflation of the triangle scenario.
Now, by using the rules of Eq.~\eqref{consistent} for consistency with identifiable monomials, this certificate can be transformed into a witness of tripartite distributions whose violation signals the distribution as being incompatible with a realization in the triangle scenario with quantum latent variables (note that we switch to the expectation-value picture\footnote{This result is obtained by performing the substitutions $p(i=0)=\frac{1}{2}(1+\expec*{I})$, and  $p(i=0,j=0)=\frac{1}{4}(1+\expec*{I}+\expec*{J}+\expec*{IJ})$.}):
\begin{align}
\nonumber&\textbf{W's Witness}\\
&\begin{psmallmatrix*}[l]
		\expec*{A} + \expec*{A}^2 \!-\! \expec*{A}\expec*{AB}\!-\!\expec*{A}\expec*{AC} \\
		-\! 2\expec*{BC} \!+\! \expec*{B}\expec*{C} \!-\! \expec*{BC}^2 \!-\! \expec*{A}\expec*{B}\expec*{C} \\
		 + [A\rightarrow B\rightarrow C\rightarrow A] + [A\rightarrow C\rightarrow B\rightarrow A]
		 	\end{psmallmatrix*}
\leq 3.
	\label{eq:Wwitness}
\end{align}
In fact, the W distribution of Eq.~\eqref{wdistribution} attains a value of $3+\nicefrac{8}{9}$ for this witness.

Remarkably, the witness obtained is polynomial in the elements of the probability distribution.
This result is in stark contrast with witnesses obtained through other techniques, which are either linear in the elements of the probability distribution or in the variables' entropies.

Exploiting SDP duality also enables us to obtain a polynomial witness for the infeasibility of the Mermin-GHZ distribution box discussed in Sec.~\ref{sec:Mermin}, namely,
\begin{samepage}
\begin{align}
\nonumber&\textbf{Mermin Box Polynomial Witness}\\
&\begin{psmallmatrix*}[c]
\begin{psmallmatrix*}[l]
   \expec{A_0B_0C_0}{}^2\\
  +\expec{A_0B_1C_1}{}^2\\
  +\expec{A_1B_0C_1}{}^2\\
  +\expec{A_1B_1C_0}{}^2
 \end{psmallmatrix*}
 +3 \begin{psmallmatrix*}[l]
   \expec{A_1B_1C_1}{}^2 \\
  +\expec{A_0B_0C_1}{}^2\\
  +\expec{A_0B_1C_0}{}^2\\
  +\expec{A_1B_0C_0}{}^2 \end{psmallmatrix*}
  \\
+2\begin{psmallmatrix*}[l]
   \expec{A_0B_1C_1} \left(\expec{A_1B_0C_1}-\expec{A_0B_0C_0}\right)\\
  +\expec{A_1B_0C_1} \left(\expec{A_1B_1C_0}-\expec{A_0B_0C_0}\right)\\
  +\expec{A_1B_1C_0} \left(\expec{A_0B_1C_1}-\expec{A_0B_0C_0}\right)\\
  \end{psmallmatrix*}
  \\
+6 \begin{psmallmatrix*}[l]
   \expec{A_0B_0C_1} \left(\expec{A_1B_0C_0}-\expec{A_1B_1C_1}\right)\\
 + \expec{A_0B_1C_0} \left(\expec{A_0B_0C_1}-\expec{A_1B_1C_1}\right) \\
 + \expec{A_1B_0C_0} \left(\expec{A_0B_1C_0}-\expec{A_1B_1C_1}\right)
 \end{psmallmatrix*}
\end{psmallmatrix*}\leq 32.
\label{eq:merminwitness}
\end{align}\end{samepage}
We apply this witness to the 45 nonlocal extremal points of the tripartite nonsignaling scenario~\cite{tripartiteNS} and notice that it recognizes 15 out of these 45 points as not realizable in the quantum triangle scenario.\footnote{In the ordering of Ref.~\cite{tripartiteNS}, the boxes incompatible with the quantum triangle scenario are numbers 2, 13, 21, 22, 27, 29, 30, 31, 34, 36, 39, 41, 43, 44, and 46.}

\subsection{Optimization in quantum causal scenarios}\label{sec:opt}
As explained in Sec.~\ref{sec:details}, the constraints~\eqref{rules}, \eqref{commutation:copies}, \eqref{eq:symconstraint}, and \eqref{consistent} characterize relaxations of the set of quantum correlations compatible with a given causal structure.
This characterization can be employed to easily bound optimal values of polynomials of the measurement operators in the problem via NPO theory~\cite{npo}.
We provide in what follows several examples of this procedure.

\subsubsection{Optimization of linear functionals}\label{sec:results:optim:linear}
As a first application, we derive upper bounds to the maximum value that certain Bell-like operators can achieve in the quantum triangle scenario.
The results herein are obtained by considering second-order inflations, solving the associated NPO problems with the set of monomials $\mathcal{S}_2\cup\mathcal{L}_1$.
Including the identity operator, we find that each party has nine possible operators at this SDP level, such that the resulting moment matrix involved in the calculations has size $873\,{\times}\,873$.
\begin{align}
\nonumber&\textbf{Mermin's Inequality~\cite{Mermin1990}}\\
	\label{eq:mermin}
	&\begin{psmallmatrix*}[l]
	\expec*{A_1 B_0 C_0}\\+\expec*{A_0 B_1 C_0}\\+\expec*{A_0 B_0 C_1}\\-\expec*{A_1 B_1 C_1}
	\end{psmallmatrix*}
	\leq \begin{cases}
		2 				& \mathcal{L}^\triangle,\;\;\mathcal{L}^\bell \\
		3.085^*		& \mathcal{Q}^\triangle \\
		4 & \mathcal{Q}^\bell,\;\;\mathcal{NS}^\triangle,\;\;\mathcal{NS}^\bell
	\end{cases},
\\\nonumber&\textbf{Svetlichny's Inequality~\cite{Svetlichny1987}}\\
	\label{eq:svet}
	&\begin{psmallmatrix*}[l]
	\expec*{A_1 B_0 C_0}\\+\expec*{A_0 B_1 C_0}\\+\expec*{A_0 B_0 C_1}\\-\expec*{A_1 B_1 C_1}\\
	-\expec*{A_0 B_1 C_1}\\-\expec*{A_1 B_0 C_1}\\-\expec*{A_1 B_1 C_0}\\+\expec*{A_0 B_0 C_0}
	\end{psmallmatrix*}
	\leq \begin{cases}
		4 				& \mathcal{L}^\triangle,\;\;\mathcal{L}^\bell\\
		4.405^* 	& \mathcal{Q}^\triangle \\
		4\sqrt{2} & \mathcal{Q}^\bell\\
		8 				& \mathcal{NS}^\triangle,\;\;\mathcal{NS}^\bell
	\end{cases}.
\end{align}
In Eqs.~\eqref{eq:mermin} and~\eqref{eq:svet}, the triangle $\triangle$ denotes the causal triangle scenario, while $\bell$ refers to the standard tripartite Bell scenario where all parties receive shares of a joint quantum state. Clearly, for any set $\mathcal X=\{\mathcal L, \mathcal Q, \mathcal{NS}\}$, the bound obtained in $\mathcal X^{\triangle}$ is smaller than or equal to the one obtained in $\mathcal X^{\bell}$. The asterisk means that the values given are upper bounds; that is, quantum inflation shows that Mermin's and Svetlichny's inequalities cannot exceed $3.085$ or $4.405$, respectively, in the quantum triangle scenario, but, at the moment, it is unknown whether those values are attainable.

It is known that both the algebraic maximum of $4$ for Mermin's inequality and the algebraic maximum of $8$ for Svetlichny's inequality can be achieved in the triangle structure, if one considers that the latent nodes distribute nonsignaling resources (see, for instance,~\cite[Sec.~III~C]{TripartiteViaPR}). This result means that no difference between the triangle and the Bell scenarios can be identified with these two inequalities when the latent variables represent nonsignaling resources, i.e., $\max_{\mathcal{NS}^\triangle}=\max_{\mathcal{NS}^\bell}$ for these two inequalities. Consequently, our finding here that $\max_{\mathcal{Q}^\triangle}<\max_{\mathcal{Q}^\bell}$ for these two inequalities cannot be recovered by means of the GPT-valid (nonfanout) variant of the inflation method considered in Ref.~\citep[Sec.~V~D]{wolfe2016inflation}.

In a similar manner, the maximum values attainable by any linear function in the classical triangle and Bell scenarios also coincide, because, even though $\mathcal{L}^\triangle\subsetneq \mathcal{L}^\bell$, the \emph{extremal} correlations in $\mathcal{L}^\bell$ are also members of the set $\mathcal{L}^\triangle$, as the local deterministic strategies are product correlations.

Remarkably, we see a divergence in linear-function optimization over $\triangle$ and $\bell$ when considering quantum latent variables.
In the quantum triangle scenario, it is not possible to saturate such inequalities up to the bounds attainable in the quantum Bell scenario.
This constraint has the implication that having access to tripartite quantum resources allows for improved performance in information tasks relative to having access to arbitrary bipartite quantum states and unlimited shared randomness~\cite{navascues2020gnme}.

\begin{figure}[b]
  \begin{center}
    \hfill
    \subfigure[t][\label{fig:TriangleWithSettingsSR}]
    {\centering
      \includegraphics[scale=0.38]{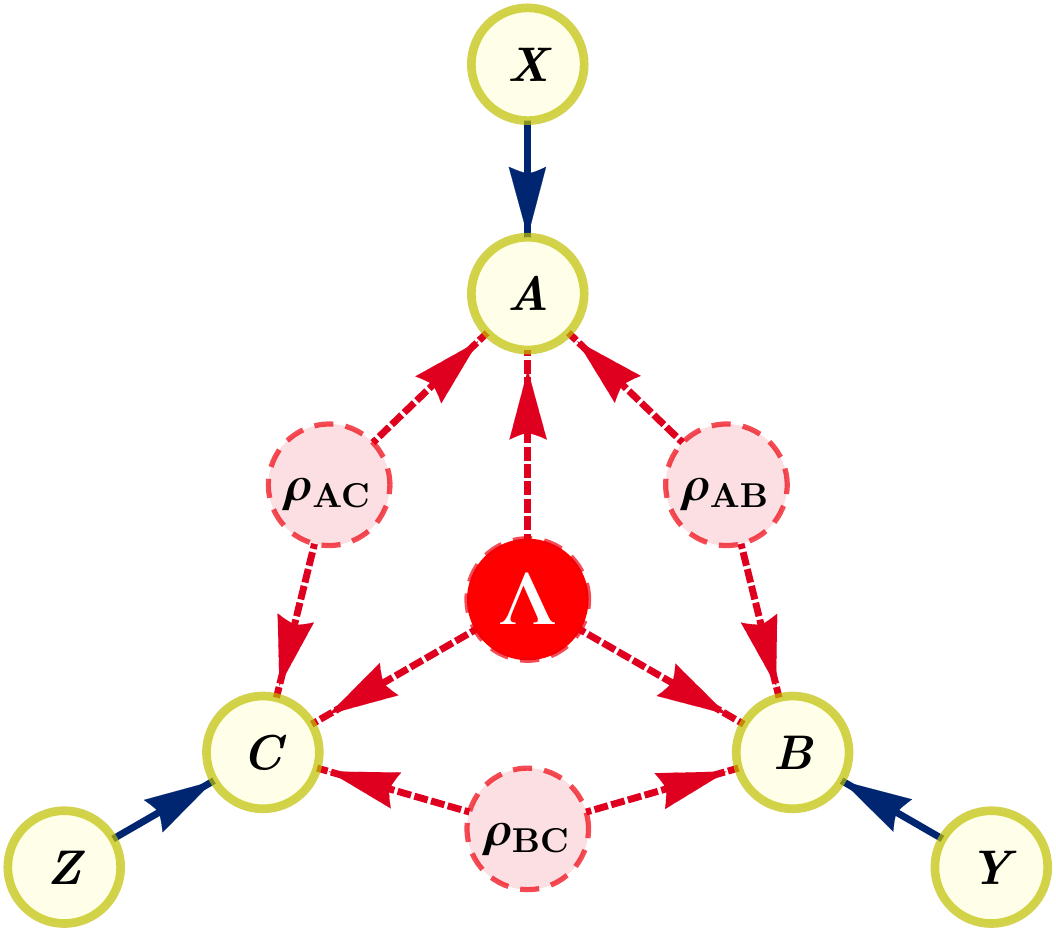}
    }
    \hfill
    \subfigure[t][\label{fig:BilocalityQSR}]
    {\centering
      \includegraphics[scale=0.38]{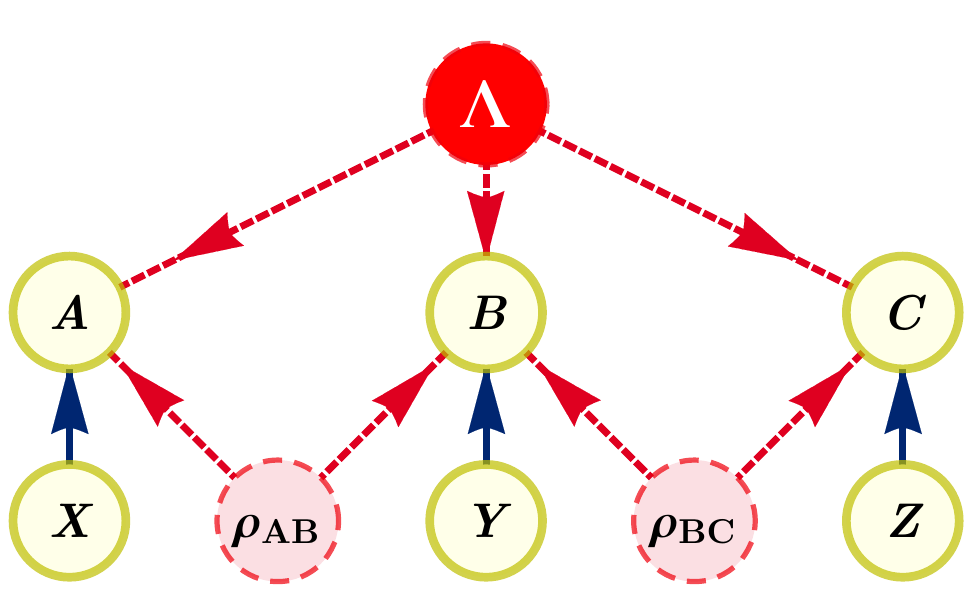}
    }
    \hfill
  \end{center}
  \caption{(a) The quantum triangle scenario with settings and shared randomness, sometimes denoted $\mathcal{Q}^{\triangle+\text{SR}}$ for brevity.
  (b) The quantum tripartite-line scenario with shared randomness, sometimes denoted $\mathcal{Q}^{{\protect\biloc}+\text{SR}}$ for brevity.
  Such ``hybrid'' DAGs denote the presence of quantum sources along with the presence of a classical source of shared randomness. The classical latent node is denoted $\Lambda$, and is depicted in white text on a dark red background to distinguish it from the other quantum latent nodes. Though $\mathcal{Q}^{\triangle+\text{SR}}$ is inequivalent to $\mathcal{Q}^\triangle$ depicted in Fig.~\ref{fig:TriangleWithSettings}, and similarly $\mathcal{Q}^{{\protect\biloc}+\text{SR}}$ is inequivalent to $\mathcal{Q}^{\protect\biloc}$ depicted in Fig.~\ref{fig:BilocalityCopy}, nevertheless both causal structures in each pair are equivalent up to \emph{linear}-function optimization.}
  \label{fig:SRDAGs}
\end{figure}

Recall that, while polynomial functions may differ over the triangle scenarios with and without shared randomness, \emph{linear}-function optimization is independent of the presence or absence of shared randomness.\footnote{Linear-function optimization problems are agnostic to the presence or absence of shared randomness only when there are no additional constraints on $P_\text{obs}$, as is the case with our maximizing the Mermin or Svetlichny functionals here.
However, constrained optimization is \emph{not} agnostic to the presence or absence of shared randomness, even if the objective function is linear in $P_\text{obs}$. In the absence of shared randomness, constraints on $P_\text{obs}$ are incorporated into the quantum inflation semidefinite program via Eq.~\eqref{consistent}, i.e., the rule regarding consistency with identifiable monomials. However, Eq.~\eqref{consistent} does not hold in the presence of shared randomness; it is replaced by the weaker Eq.~\eqref{consistentSR}. The cryptography example in Sec.~\ref{sec:cryptography} explicitly demonstrates the dramatic impact of adding shared randomness to a linear-function optimization problem involving constraints on $P_\text{obs}$.}
That is, even though $\mathcal{Q}^\triangle\subsetneq \mathcal{Q}^{\triangle+\text{SR}}\subsetneq \mathcal{Q}^{\bell}$, it is the case that $\mathcal{Q}^{\triangle+\text{SR}}$ is the convex hull of $\mathcal{Q}^\triangle$. The results above, thus, demonstrate that the Mermin-GHZ box of Eq.~\eqref{eq:MerminBox} is not achievable in $\mathcal{Q}^{\triangle+\text{SR}}$, a result relevant to Ref.~\cite{Schmid2020LOSR}.

The values 3.085 and 4.405 given for $\mathcal{Q}^\triangle$ in~Eqs.~\eqref{eq:mermin} and \eqref{eq:svet} represent upper bounds to the real maximum values achievable. We can readily establish lower bounds of $2\sqrt{2}$ and $4$, respectively, by considering explicit bipartite quantum correlations for Alice and and Bob and taking Charlie to always answer $+1$. It is an open question whether, when increasing the inflation and NPO hierarchies, one will find that the values for $\mathcal{Q}^\triangle$ will collapse to those lower bounds.

It should be noted that the quantum inflation technique can readily be adapted to witness the incompatibility of a given distribution with a causal structure involving \emph{classical shared randomness} in addition to quantum latent nodes, such as the quantum triangle scenario with settings and shared randomness or the quantum tripartite-line scenario with shared randomness depicted in Fig.~\ref{fig:SRDAGs}. In particular, Eqs.~\eqref{rules}--\eqref{eq:symconstraint} hold regardless of the presence or absence of shared randomness. However, Eq.~\eqref{consistent} is valid only in the absence of shared randomness. The analog of Eq.~\eqref{consistent} for hybrid causal structures with shared randomness simply replaces higher-degree monomials with degree-1 monomials, i.e.,
\begin{align}
&\expec*{\prod_{k}\hat{O}^{\vec{\boldsymbol{1}}_{k}|k}_{i_{k}|m_{k}}}_\rho={P_\text{obs}}{\left(\bigcap_k (i_{k}|m_{k},k)\right)}.
\label{consistentSR}
\end{align}

\subsubsection{Optimization of nonlinear functionals}\label{sec:results:nonlinear}
Quantum inflation can also be used to optimize nonlinear witnesses $f$. The essential idea can be seen already in Sec.~\ref{sec:witness}, where Eq.~\eqref{consistent} relates linear functions in an inflation to polynomials in the corresponding original scenario. Note that, for this optimization to be possible, in general one must consider at least an order-$q$ inflation when optimizing a polynomial of degree $q$.

Consider, for instance, minimizing the $2$-norm between a distribution ${P_\text{obs}(a,b,c)}$ achievable in the quantum triangle scenario and that of the W distribution, that is:
\begin{align}\begin{split}
f(P_\text{obs})\equiv \sum_{a,b,c=0,1} & \abs*{P_\text{obs}(a,b,c) -P_\text{W}(a,b,c)}^2\\
=\sum_{a,b,c=0,1} & \begin{psmallmatrix*}[l]
P_\text{obs}(a,b,c)^2+P_\text{W}(a,b,c)^2\\
 \quad -2P_\text{obs}(a,b,c)P_\text{W}(a,b,c)
\end{psmallmatrix*},
\end{split}\end{align}
where one can readily verify that ${\sum_{a,b,c} P_\text{W}(a,b,c)^2 = 1/3}$.

One can estimate the minimum value of this function using second-order inflation or higher, via Eq.~\eqref{consistent}. Explicitly, the constraints from Eq.~\eqref{consistent} that apply to this problem are
\begin{align*}
&P_\text{obs}(a,b,c)^2 = \expec*{E^{1,1}_aE^{2,2}_aF^{1,1}_bF^{2,2}_bG^{1,1}_cG^{2,2}_c},
\\\text{and }  &P_\text{obs}(a,b,c)P_\text{W}(a,b,c) = P_\text{W}(a,b,c)\expec*{E^{1,1}_aF^{1,1}_bG^{1,1}_c}.
\end{align*}
While obtaining nontrivial bounds on  $f(P_\text{obs})$ is, in principle, possible, in practice, it appears to require levels of the NPO hierarchy too computationally expensive.

\section{Applications to quantum cryptography in networks}\label{sec:cryptography}
In this section, we showcase a concrete application of quantum inflation: the security analysis of multipartite quantum key distribution protocols where the parties not all share a joint quantum state but are instead arranged in a network.
Because entangled systems can be transmitted only over relatively short distances and only a small number of parties can practically share an entangled state distributed by a single source, multisource networks are a promising solution for long-distance quantum key distribution.
However, the fact that the underlying causal structure is no longer the Bell scenario has a profound impact in the associated security proofs~\cite{lee2018crypto}.

\begin{figure}[!b]
  \begin{center}
    \hfill
    \subfigure[t][\label{fig:eavesdrop}]
    {\centering
      \includegraphics[scale=0.38]{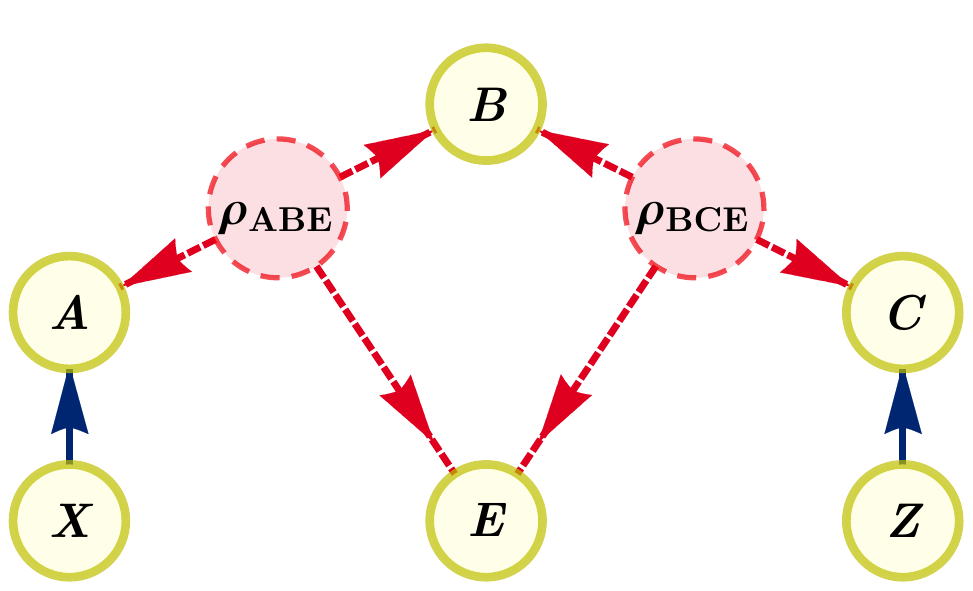}
    }
    \hfill
    \subfigure[t][\label{fig:eavesdropSR}]
    {\centering
      \includegraphics[scale=0.38]{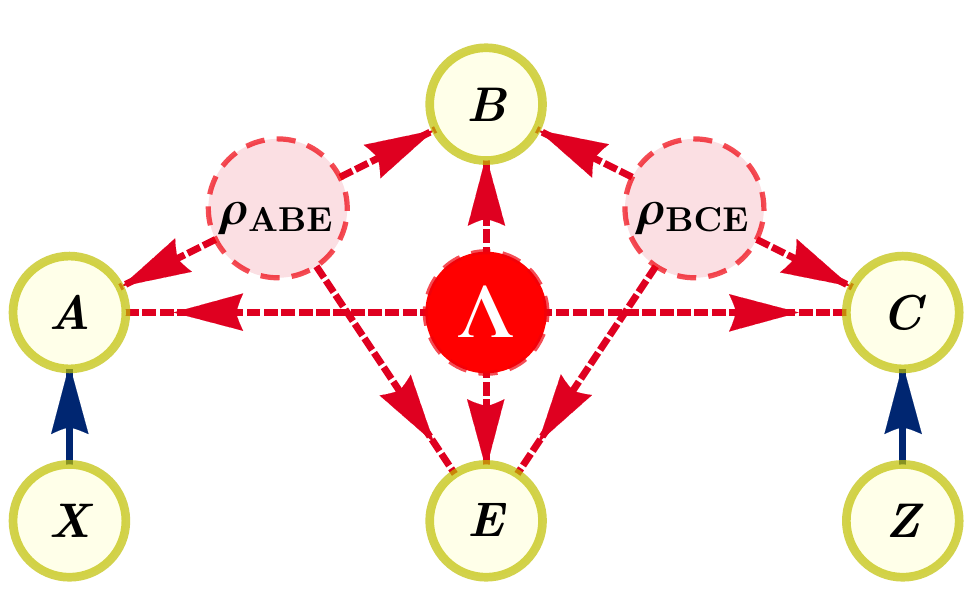}
    }
    \hfill
  \end{center}
  \caption[]{(a) $A$ and $C$ employ the repeater $B$ in order to generate a secret shared key. While $A$ and $C$ have a choice of which measurement to perform in the states they receive---denoted by $X$ and $Z$---$B$ performs a four-outcome Bell state measurement. Unbeknowns to $A$, $B$, and $C$, a fourth party $E$ may be eavesdropping the sources in an attempt to extract information about the secret key that $A$ and $C$ are establishing.
  (b) The eavesdropped network scenario with shared randomness. Some tripartite distributions which are cryptographically secure under the assumption of no shared randomness become insecure if hidden dependence on shared randomness cannot be excluded on physical grounds.}
  \label{fig:crypto}
\end{figure}

Imagine that two parties, $A$ and $C$, attempt to establish a secret key.
They are too far apart, so they employ a quantum repeater $B$ in order to establish correlations.
However, a malicious party $E$ may be eavesdropping what in reality are tripartite sources that leak one of the subsystems generated,\footnote{These ``third subsystems'' need not strictly be additional particles generated at each source but can, for instance, be the fields surrounding them~\cite{thinh2016vacuumleakage}. Eve is allowed to perform a joint measurement on her two shares, one received from each source (call them $E'$ and $E''$, using the notation in Sec.~\ref{sec:simple}), producing the value of the random variable $E$.} as in Fig.~\ref{fig:eavesdrop}.
The $ABC|XZ$ marginal within the eavesdropped quantum repeater scenario is essentially a special case of the tripartite-line scenario in Fig.~\ref{fig:Bilocality}.
An important question to answer is how reliably can $E$ guess the key generated by $A$ and $C$ by manipulating the leaked subsystems.

Reference~\cite{lee2018crypto} tackles this problem, finding a trade-off between the violation of previously known network inequalities and the amount of information that $E$ can infer about the outcomes of $A$ and $C$ upon performing a measurement, under the assumption that the eavesdropper is bounded only by nonsignaling constraints.
Since no quantum constraints are considered, the obtained bounds on the information achievable by the eavesdropper are very conservative. In fact, if $A$, $B$, and $C$ observe statistics compatible with a quantum entanglement swapping experiment, the information bounds derived in Ref.~\cite{lee2018crypto} turn out to be trivial. To derive tighter bounds for the more realistic case of quantum eavesdroppers, one needs to use tools bounding the set of allowed quantum correlations in the considered network, something that quantum inflation provides.
In the following, we analyze the same scenario using the semidefinite relaxations of the set of quantum correlations achievable in the scenario in Fig.~\ref{fig:eavesdrop} provided by quantum inflation,\footnote{This scenario can, in principle, also be numerically characterized via the scalar extension method in Ref.~\cite{Pozas2019}.} finding tighter bounds on the eavesdropper predictability on the measurement outcomes by the honest parties $A$ and $C$.

We are interested in bounding how much information $E$ can infer about $A$'s and $C$'s outcomes in the network in Fig.~\ref{fig:eavesdrop} from their choice of measurements.
We follow Ref.~\cite{lee2018crypto} and estimate this information via the total variation distance $\Delta_{\text{E},\text{E{\textbar}AC}}$ between the distribution of $E$'s outcome with and without conditioning on the inputs and outputs of $A$ and $C$, that is,
\begin{align}
\Delta_{\text{E},\text{E{\textbar}AC}}^{a,c,x,z}\coloneqq \frac{1}{2}\sum_{e=1}^{\abs{E}}\left|P_{\text{E{\textbar}ACXZ}}(e|a,c,x,z)-P_{\text{E}}(e)\right|,
\end{align}
when parties $A$, $B$, and $C$ observe some specific probability distribution $P_{\text{ABC{\textbar}XZ}}(a,b,c|x,z)$. For the sake of example, we focus on bounding the total variation distance conditioned upon ${a{=}c{=}x{=}z{=}0}$, i.e., $\Delta_{\text{E},\text{E{\textbar}AC}}^{0,0,0,0}$.
That is, we want to maximize $\Delta_{\text{E},\text{E{\textbar}AC}}^{0,0,0,0}$ over all the probability distributions $\left\{P_\text{ABCE{\textbar}XZ}\right\}$ that can be quantumly realized in the network in Fig.~\ref{fig:eavesdrop}, and whose marginal on $A$, $B$, and $C$ matches the given $P_{\text{ABC{\textbar}XZ}}$.

Superficially, it might appear that the unspecified cardinality of $|E|$ constitutes an obstacle to exploring the space of all $\left\{P_\text{ABCE{\textbar}XZ}\right\}$ compatible with the network in Fig.~\ref{fig:eavesdrop}. We bypass this concern, however, by exploiting an alternative formulation of total variation distance~\cite{Levin2017}, namely,
\begin{align}\begin{split}
\Delta_{\text{E},\text{E{\textbar}AC}}^{a,c,x,z}&= P_{\text{E{\textbar}ACXZ}}({\boldsymbol{0}}^\star|a,c,x,z)-P_{\text{E}}({\boldsymbol{0}}^\star)\,,\\
\end{split}\end{align}
where ${\boldsymbol{0}}^\star$ is the \emph{bin} (coarse graining over a subset of possible values of $E$) which maximizes \mbox{$P_{\text{E{\textbar}ACXZ}}({\boldsymbol{0}}^\star|a,c,x,z)-P_{\text{E}}({\boldsymbol{0}}^\star)$}, i.e.,
\begin{align}\begin{split}
{\boldsymbol{0}}^\star &\coloneqq \bigvee e'\: : \:P_{\text{E{\textbar}ACXZ}}(e'|a,c,x,z)\geq P_{\text{E}}(e')\,.
\end{split}\end{align}
Note that ${\boldsymbol{0}}^\star$ is an implicit function of the values $\{a,c,x,z\}$ which identify the particular total variation distance $\Delta_{\text{E},\text{E{\textbar}AC}}^{a,c,x,z}$.
Since we are concerned only about binning $E$ into $e\in {\boldsymbol{0}}^\star$ versus $e\notin {\boldsymbol{0}}^\star$, we can effectively model $E$ as dichotomic for the purposes of optimization, allowing us to set up the following SDP:
\begin{align}\begin{split}\label{eq:cryptoSDP}
    \max \quad & \frac{P_{\text{ACE{\textbar}XZ}}({\boldsymbol{0}}^\star,0,0|0,0)}{P_{\text{AC{\textbar}XZ}}(0,0|0,0)}-P_{\text{E}}({\boldsymbol{0}}^\star)
    \\&=\frac{\expec*{A^{1}_{0|0}C^{1}_{0|0}E^{1,1}_{{\boldsymbol{0}}^\star}}}{P_{\text{AC{\textbar}XZ}}(0,0|0,0)}-\expec*{E^{1,1}_{{\boldsymbol{0}}^\star}} \\
    \text{s.t.} \quad & \Gamma^{(n,I)}\succeq 0, \text{ and Eqs.~\eqref{rules}, \eqref{commutation:copies}, \eqref{eq:symconstraint}.}\\
    &\text{Eq.~\eqref{consistent} is also imposed, but only on}
    \\&\text{the marginal }P_{\text{ABC{\textbar}XZ}},\text{ i.e. excluding}
    \\&\text{identifiable monomials involving \(E\).}
\end{split}\end{align}
Note that we distinguish between \emph{identifiable monomials}---which are defined by the order of inflation via Eq.~\eqref{consistent}---versus \emph{known probabilities}---which may be defined by a \emph{partial} specification of the global probability distribution over the observable nodes in a given causal structure. That is, one may find that a \emph{strict subset} of all identifiable monomials are preassigned numerical values when tackling constrained optimization problems, such as is the case with Eq.~\eqref{eq:cryptoSDP}.
In Eq.~\eqref{eq:cryptoSDP}, $\Gamma^{(n,I)}$ is the moment matrix associated to the set of operators $\left\{A^{i}_{a|x},B^{j,k}_b,C^{l}_{c|z},E^{m,o}_e\right\}$ with $1\leq i,j,k,l,m,o\leq I$, for some $I^\text{th}$-order inflation\footnote{Recall that, in the eavesdropped repeater of Fig.~\ref{fig:eavesdrop}, the indices $i$, $j$, and $m$ correspond to copies of the state $\rho_{ABE}$ and the indices $k$, $l$, and $o$ correspond to copies of the state $\rho_{BCE}$.} and some NPO hierarchy level $n$.

\begin{figure}[t]
    \includegraphics[width=\columnwidth]{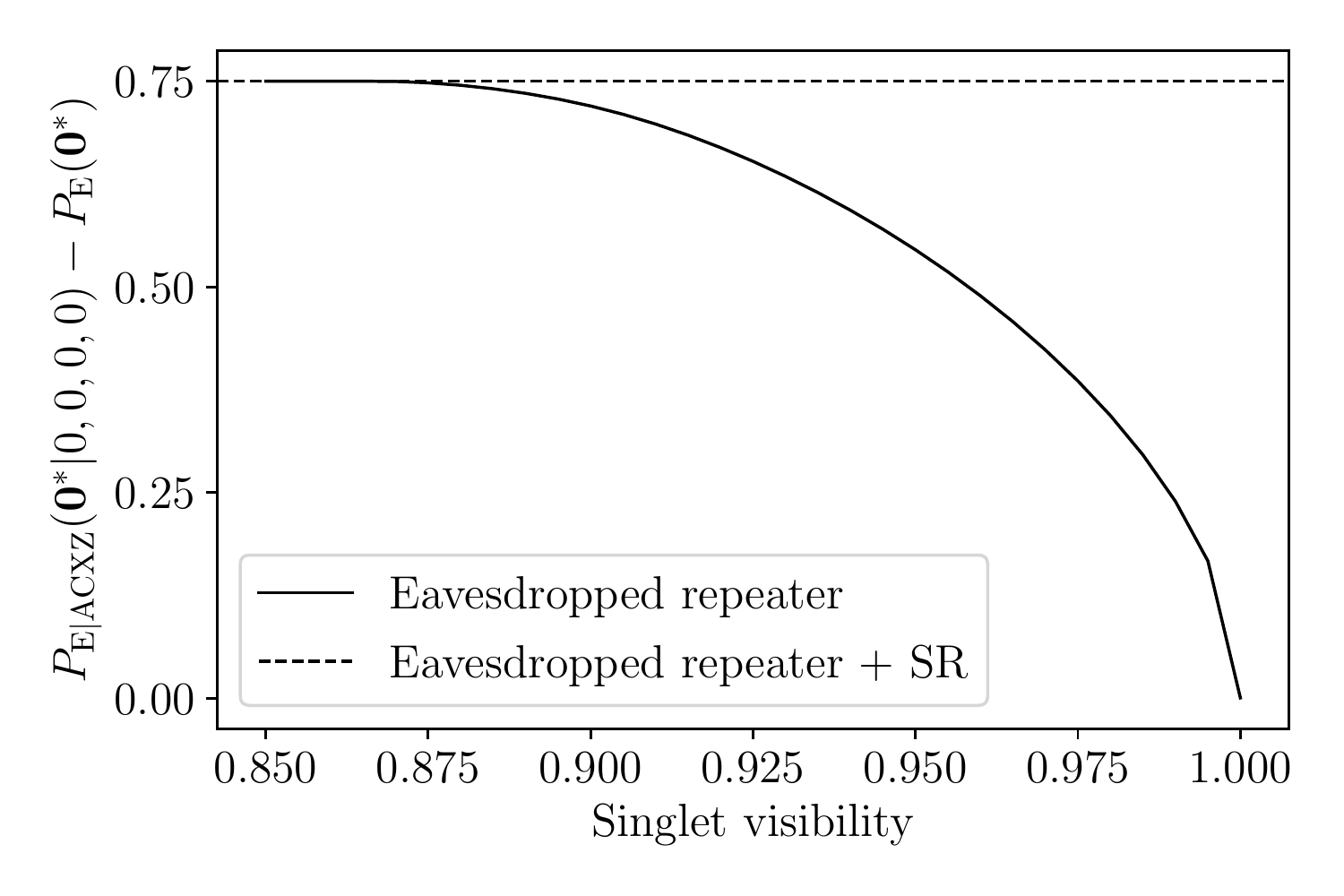}
    \caption{Upper bounds to the total variation distance ${\Delta_{\text{E},\text{E{\textbar}AC}}^{0,0,0,0}\coloneqq {P_{\text{E{\textbar}ACXZ}}({\boldsymbol{0}}^\star|0,0,0,0)-P_{\text{E}}({\boldsymbol{0}}^\star)}}$ when the parties $A$, $B$, and $C$ observe the distribution depicted in Eq.~\eqref{eq:cryptodistribution} with singlet visibility $v$.
    The solid curve tracks the upper bound in the eavesdropped repeater network depicted in Fig.~\ref{fig:eavesdrop}, whereas the dashed line tracks the worst case bound, when all the parties are coordinated through classical shared randomness, as in Fig.~\ref{fig:eavesdropSR}.
    The computations for the eavesdropped repeater network are carried out by considering its second-order quantum inflation and the moment matrix generated by operator sequences associated to the corresponding local level $\mathcal{L}_1$ not exceeding length two. This computation results in a moment matrix of size $857\times857$.
    On the other hand, the bound ${\Delta_{\text{E},\text{E{\textbar}AC}}^{a,c,x,z}\leq \min \{1,\; 2(2-\mathcal{R})\}}$ offered by Ref.~\cite{lee2018crypto} merely implies the trivial ${\Delta_{\text{E},\text{E{\textbar}AC}}^{0,0,0,0}\leq 1}$ for this example, as $2(2-\mathcal{R})$ evaluates to $2(2-v\sqrt{2})$, and $1\leq 2(2-v\sqrt{2})$ for all $v\in[0,1]$.}
    \label{fig:cryptoresults}
\end{figure}

In Fig.~\ref{fig:cryptoresults}, we depict upper bounds for $\Delta_{\text{E},\text{E{\textbar}AC}}^{0,0,0,0}$ when the parties $A$, $B$, and $C$ measure the standard distribution in repeater networks (see Appendix~\ref{app:crypto} for its definition) with some visibility $v$.
Importantly, when $A$, $B$, and $C$ observe the distribution corresponding to the sources sending noiseless singlets (i.e., ${v{=}1}$), we certify that the distance between the distributions cannot exceed \emph{zero}, and, thus, there is no better strategy for $E$ to guess $A$'s and $C$'s outcomes than making a uniformly random guess.
Thus, for ${v{=}1}$, we prove that the key established between $A$ and $C$ is completely secure, as expected.
Note that, in contrast with Ref.~\cite{lee2018crypto}, we allow $E$ to be a quantum eavesdropper, meaning that Eve is allowed to perform a joint entangled measurement on her subsystems of both $\rho_{ABE}$ and $\rho_{BCE}$.

For comparison, imagine if it were impossible to ensure the absence of any classical shared randomness between the parties. Dropping the no-shared-randomness assumptions means that we now consider the network underlying the process to be that of Fig.~\ref{fig:eavesdropSR}. In that causal scenario, we can readily establish that ${\max_P\: \Delta^{0,0,0,0}_{\text{E},\text{E{\textbar}AC}}=3/4}$ when $P_{\text{ABC{\textbar}XZ}}$ is the standard distribution in repeater networks~\eqref{eq:cryptodistribution}, for any visibility $v$. This result follows from the fact that such $P_{\text{ABC{\textbar}XZ}}$ can be simulated classically with tripartite shared randomness; i.e., $P_{\text{ABC{\textbar}XZ}}$ does not violate any tripartite Bell inequality. Since $P_{\text{AC{\textbar}XZ}}$ \emph{could} be a deterministic function of $\Lambda$, it follows that there exists a causal model for Fig.~\ref{fig:eavesdropSR} in which the eavesdropper $E$ can perfectly predict the outcomes of $A$, $B$, and $C$ leveraging her access to $\Lambda$. Perfect predictability (i.e., \emph{zero} security) leads to $P_{\text{E{\textbar}ACXZ}}({\boldsymbol{0}}^\star|0,0,0,0)=1$ and $P_\text{E}({\boldsymbol{0}}^\star)=P_\text{AC}(0,0)$ and, more generally, to the absolute maximal possible total variation distance $\Delta_{\text{E},\text{E{\textbar}AC}}^{a,c,x,z}=1-P_{\text{AC{\textbar}XZ}}(0,0|0,0)=3/4$.

In summary, our analysis demonstrates the practical value for quantum cryptography applications of being able to bound the quantum correlations achievable in networks by means of quantum inflation. It also highlights the critical role that network structure assumptions play in establishing security proofs, as just the addition of shared randomness transforms the distribution from being perfectly secure to completely insecure.

\section{Bounding Causal Effects under Quantum Confounding}\label{sec:mediationanalysis}
We now turn to explore an application of quantum inflation unrelated to quantum information theory.
It is the quantum version of \emph{mediation analysis}, which uses observational data and a promise of causal structure to estimate causal relationships and predict counterfactual intervention. Informally, mediation analysis quantifies the relative strengths of different causal pathways which could play a role in establishing some observed correlations.
While this aspect has not been explored in the context of quantum mechanics (with the very recent exception of Ref.~\cite{Quantifying2020}), mediation analysis is a widespread tool in medical and biochemical studies~\cite{huang2016mediation,hutton2018mediation,sohn2019microbiome}.
To quantify causal strengths in disciplines concerned with microscopic phenomena, it is crucial to account for causes mediated by quantum carriers, something possible using quantum inflation.

Reichenbach's common causal principle states that all nonspurious statistical correlations must admit some causal explanation, and this principle can be generalized to include quantum causal explanations as well~\cite{Wood2015,Cavalcanti2014RCCP,Allen2017RCCP,WolfeBellQuantified}. Often, however, one may wish to distinguish between various causal possibilities for establishing statistical correlation. Moreover, even knowledge of the true causal structure is still inferior to a complete understanding of functional relationships between observable variables.

\begin{figure}[b]
  \begin{center}
     \centering\includegraphics[scale=0.43]{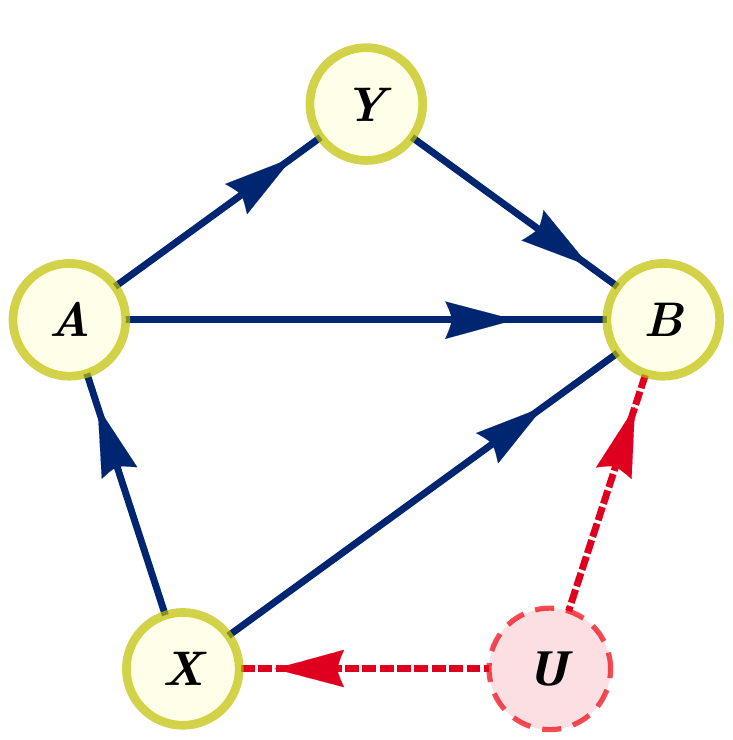}
  \end{center}
  \caption[]{A graph admitting multiple causal explanations for correlation between $A$ and $B$.
  \label{fig:ExampleForCausalPathways}
  }
\end{figure}

Consider the multiple possible causal explanations for correlation between variables $A$ and $B$ in Fig.~\ref{fig:ExampleForCausalPathways}. We have
{\setlength{\parindent}{0em}
\begin{compactitem}[\textbullet\hspace{1em}]
    \item the \emph{direct effect} of $A$ on $B$, associated with the \emph{edge} ${A\to B}$,
    \item the \emph{indirect effect} of $A$ on $B$, associated with the \emph{direct path} ${A\to Y\to B}$,
    \item the \emph{common dependence} of $A$ and $B$ on $X$, associated with the \emph{forking path} ${A\leftarrow X\to B}$, and
    \item the common dependence of $A$ and $B$ on $U$, associated with the forking path ${A\leftarrow X\leftarrow U\to B}$.
\end{compactitem}
}

Of course, in reality, the true causal explanation of a correlation typically involves \emph{all} these routes.  The subfield of causal inference known as \term{mediation analysis} provides definitions of effect strength parameters for quantifying the relative significance of different causal pathways~\cite{janzing2013,miles2015partial,malinsky2019potential,bhattacharya2020semiparametric}. Such measures are generally intended to distinguish between
\begin{compactitem}[\textbullet\hspace{1em}]
    \item the \emph{total causal effect} of $A$ on $B$, associated with \emph{all} paths originating at $A$ and terminating at $B$, and
    \item everything else, namely the \emph{total common cause dependence} shared between $A$ and $B$, associated with \emph{all} their common ancestors.
\end{compactitem}
Latent \term{confounding} is the possibility of partially explaining a pair of variables' observed correlation in terms of the functional dependence on their unobserved common causes. Estimating (or identifying) the strength of causal effects in the presence of latent confounding is the subject of extensive research~\cite{shpitser2016,shpitser2018identification,stensrud2019separable,malinsky2019potential,bhattacharya2020semiparametric,Cai2007,kang2012inequality,miles2015partial}; our contribution in this section is to unlock the possibility of effect estimation in the presence of quantum latent confounding by using quantum inflation.

The most well-studied measure of effect strength is the \term{average causal effect} (ACE) which quantifies how much one variable $B$ functionally depends on another $A$. It is defined as the variation in the expectation value of $B$ under intervention of different values for $A$. Formally,
\begin{align}\begin{split}
&{\operatorname{ACE}}[a_1,a_2,B]\coloneqq \expec{B\textbf{ do } a^\#_1{\shortto} B}-\expec{B\textbf{ do } a^\#_2{\shortto} B}
\\&\text{with}\qquad{\operatorname{ACE}}[A,B]\coloneqq\displaystyle\max\limits_{a_1, a_2} \;\;{\operatorname{ACE}[a_1,a_2,B]}.
\end{split}\end{align}
Note that the distribution of $B$ under \emph{intervention} on $A$ has a meaning distinct from conditioning on $A$. When $A$ and $B$ share some common causal ancestry, typically
\begin{align}
P(B\vert\braces{A{=}a})\neq P(B\textbf{ do } \braces{A^\#{=}a}{\shortto} B),
\end{align}
where the notation $\braces{A^\#{=}a}{\shortto} B$ indicates that the value of $A$ \emph{as seen by $B$} is been artificially set to $a$, independent of the actual value of $A$ that may be observed. For a review of do-conditionals and distinction between passive (observational) and active (interventional) conditioning, we refer the reader to Refs.~\cite{shpitser2016,shpitser2018identification,stensrud2019separable,malinsky2019potential}.
The fundamental lemmas of mediation analysis constrain the possible values of interventional do-conditionals from knowledge of both the underlying causal structure and the distribution under passive observation. Every do-conditional can be expressed in terms of extending the original distribution to a particular interruption of the original graph.

When estimating the do-conditional ${\braces{\boldsymbol{A}^\#{=}\boldsymbol{a}^\#}{\shortto}\boldsymbol{B}}$, the interrupted graph $\mathcal{G'}$ is formed by replacing the  ${\boldsymbol{A}{\to} \boldsymbol{B}}$ edge set in $\mathcal{G}$ with ${\boldsymbol{A}^\#{\to} \boldsymbol{B}}$, such that $\mathcal{G'}$ contains the additional variables $\boldsymbol{A}^\#$. Whenever $P_{\boldsymbol{ABC}}$ is defined over the observed variables ${\{\boldsymbol{A},\boldsymbol{B},\boldsymbol{C}\}}$, then the extended distribution $P'_{\boldsymbol{ABC}\vert\boldsymbol{A}^\#}$ further pertains to the variables $\boldsymbol{A}^\#$ as well.

\begin{figure*}[t]
  \begin{center}
    \subfigure[\label{fig:ExampleForIdentifiability}]
    {\centering
      \begin{minipage}[t]{0.21\textwidth}
      \centering\includegraphics[scale=0.45]{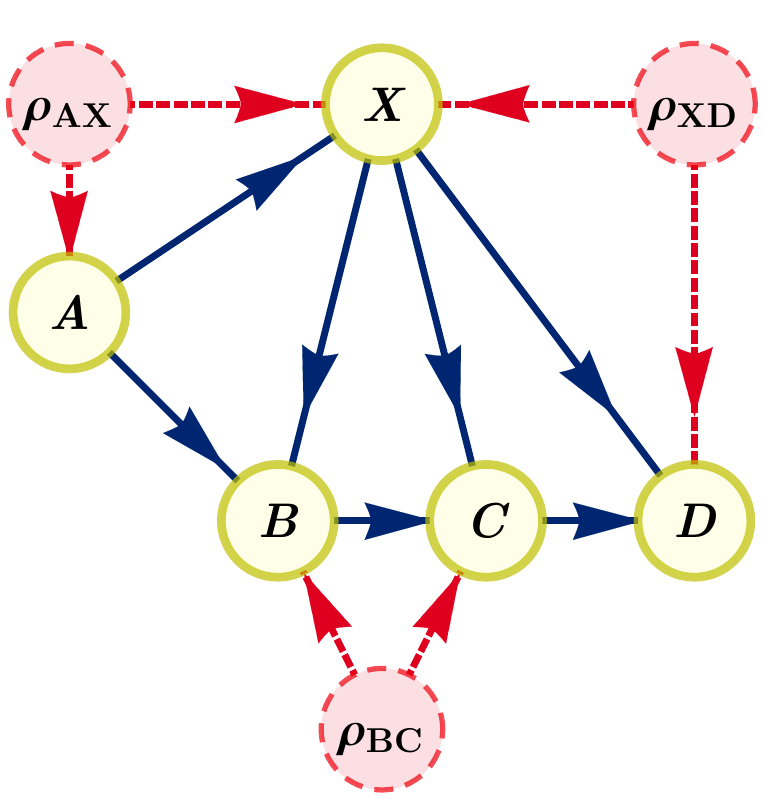}\end{minipage}
    }
    \subfigure[\label{fig:TheIdentifiableCase}]
    {\centering
      \begin{minipage}[t]{0.23\textwidth}
      \centering\includegraphics[scale=0.45]{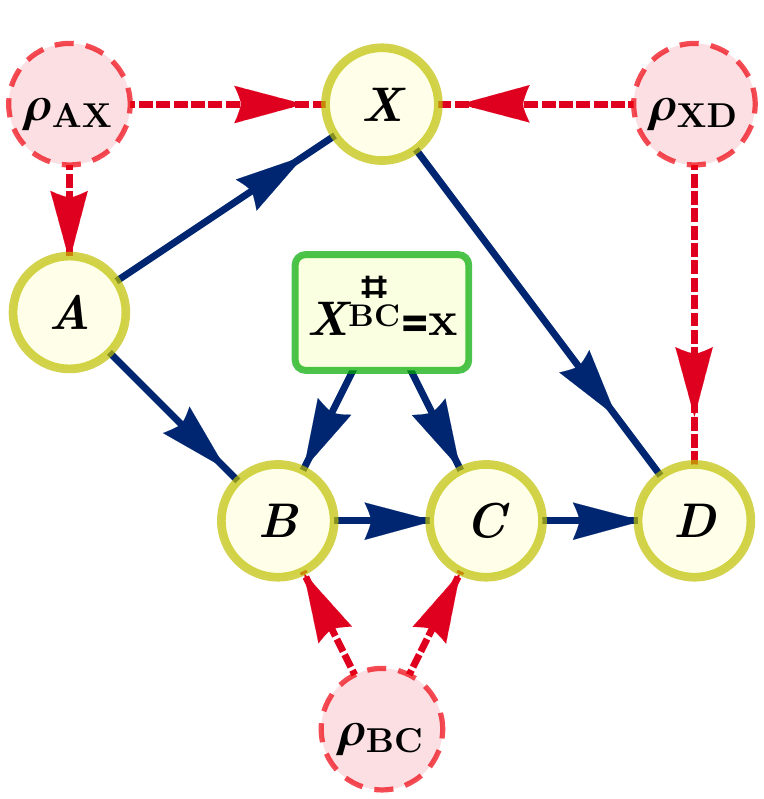}\end{minipage}
    }
    \subfigure[\label{fig:NoGoSameDistrict}]
      {\centering
        \begin{minipage}[t]{0.26\textwidth}
        \includegraphics[scale=0.45]{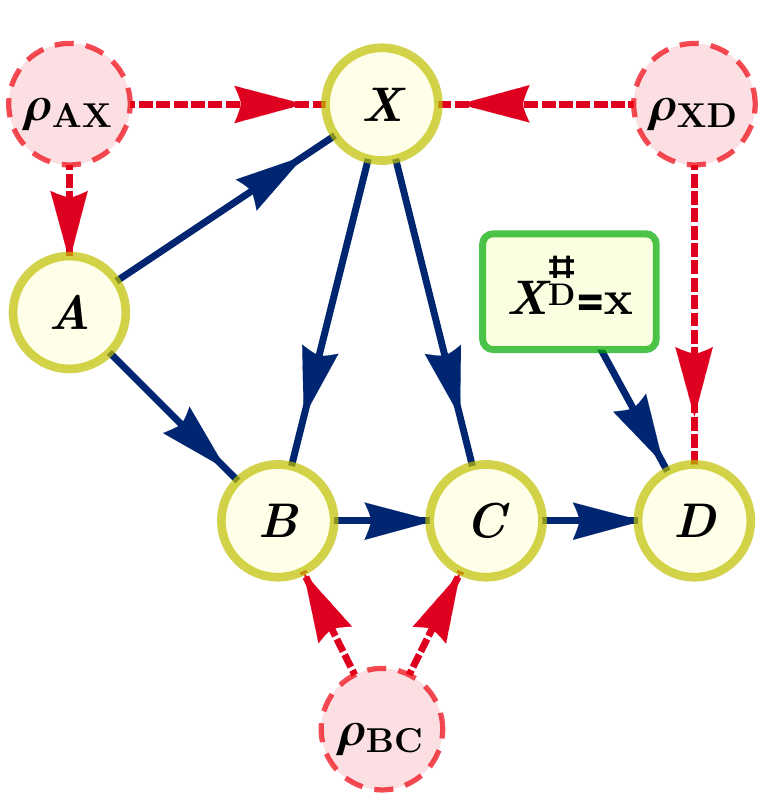}
        \end{minipage}
    }
    \hskip 0.1in
    \subfigure[\label{fig:NoGoKiteCase}]
    {\centering
      \begin{minipage}[t]{0.23\textwidth}
      \centering\includegraphics[scale=0.45]{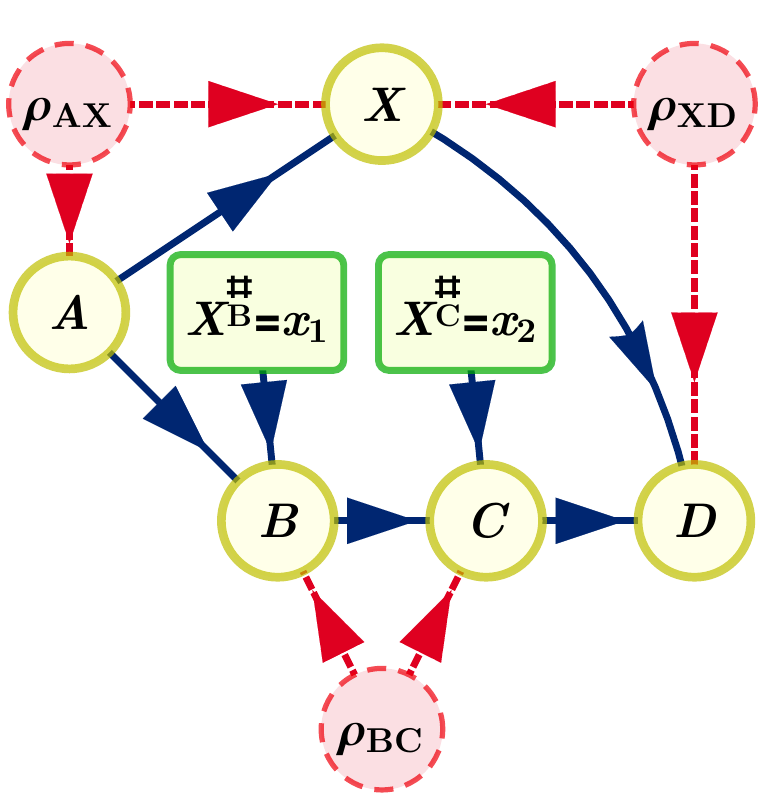}\end{minipage}
    }
  \end{center}
  \caption[]{
A survey of hypothetical interventions on edges emanating from the node $X$ in a pedagogically motivated  example causal structure.  %
  \\(a)~The example causal structure without any interventions.%
  \\(b)\phantom{~}The do-conditional ${P}{\left(\textit{ABCDX} \textbf{ do } \braces{X^\#{=}x^\#}{\shortto}\{BC\}\right)}$ where both edges ${X\to B}$ and ${X\to C}$ are intervened upon with the same value of $X$ being sent down both intervened edges. This do-conditional is identifiable and has a unique value computable from observational statistics alone, namely
 ${{P}{(\textit{ABCD}X{=}x \textbf{ do } \braces{X^\#{=}x^\#}{\shortto}\{BC\})}\,=\,{P(\textit{ABCD}X{=}x)P(BC \vert A,X{=}x^\#)}/{P(BC \vert A,X{=}x)}}.$
  \\(c)~The do-conditional ${P}{\left(\textit{ABCDX} \textbf{ do } \braces{X^\#{=}x^\#}{\shortto}D\right)}$ where the edge ${X\to D}$  is intervened upon. This do-conditional is nonidentifiable (as $X$ and $D$ share a latent parent) and, hence, is interesting to estimate using quantum inflation.
  \\(d)~The do-conditional ${P}{(\textit{ABCDX} \textbf{ do } \braces{X^\#{=}x^\#_1}{\shortto}B,\braces{X^\#{=}x^\#_2}{\shortto}C)}$ where the pair of edges ${X\to B}$ and ${X\to C}$ are intervened upon with a different value of $X$ being sent down the two intervened edges. This do-conditional is nonidentifiable (as $B$ and $C$ share a latent parent) and, hence, is interesting to estimate using quantum inflation. %
  \label{fig:interventionexamples}
  }
\end{figure*}

The fundamental principle of mediation analysis is that every viable do-conditional constitutes some extended distribution or suitable marginal thereof. Accordingly, the potential range of a do-conditional is defined by variation over all \emph{valid extensions} of the original distributions.

Note that the interrupted graph $\mathcal{G}'$ representing the intervention on $\mathcal{G}$ has a vertex set consisting of all nodes of $\mathcal{G}$---say, $\boldsymbol{ABC}$---along with the auxiliary intervention variables $\boldsymbol{A}^\#$. When the do-conditional pertains only to a \emph{marginal} of the interrupted graph $\mathcal{G}'$, the do-conditional $P'$ refers to a strict subset of the observed variables in $\mathcal{G}'$. Without loss of generality, therefore, we consider characterizing the set of do-conditionals of the form $P_{\boldsymbol{AB}\textbf{ do } \boldsymbol{A}^\#\shortto \boldsymbol{B}}$ which are compatible with some observed distribution $P_{\boldsymbol{ABC}}$ relative to the intervention $\boldsymbol{A}^\#\shortto \boldsymbol{B}$ depicted in $\mathcal{G'}$.\par

\medskip
\begin{samepage}
  \noindent\textbf{Fundamental Lemma of Mediation Analysis:}
  \begin{align}\nonumber
    &{P_{\boldsymbol{AB}\textbf{ do } \boldsymbol{A}^\#\shortto \boldsymbol{B}} \in \textsf{CompatibleDoConditionals}_{\mathcal{G'}}\big[P_{\boldsymbol{ABC}}\big]}\\\nonumber
    &\text{iff}\quad\exists_{P'}\::\:\:{P'_{\boldsymbol{ABC}\vert\boldsymbol{A}^\#} \in \textsf{ValidExtensions}_{\mathcal{G'}}\big[P_{\boldsymbol{ABC}}\big]}\\
    &\quad\text{such that}
    \\\nonumber &P_{\boldsymbol{AB}\textbf{ do } \boldsymbol{A}^\#\shortto \boldsymbol{B}}(\boldsymbol{ab}\textbf{ do }\boldsymbol{a}^\#)
    =\sum_{\boldsymbol{c}}P'_{\boldsymbol{ABC}\vert\boldsymbol{A}^\#}({\boldsymbol{abc}\vert\boldsymbol{a}^\#}).
  \end{align}
\end{samepage}

Recall that $P'_{\boldsymbol{ABC}\vert\boldsymbol{A}^\#}$ is said to be valid extension of the original distribution $P_{\boldsymbol{ABC}}$ to $\mathcal{G}'$ only if it is compatible with the interrupted graph and recovers the original distribution under postselection. That is,
\begin{align}\begin{split}
&{P'_{\boldsymbol{ABC}\vert\boldsymbol{A}^\#} \in \textsf{ValidExtensions}_{\mathcal{G'}}\big[P_{\boldsymbol{ABC}}\big]}\\
&\text{when}\quad{P_{\boldsymbol{ABC}}(\boldsymbol{abc})=P'_{\boldsymbol{ABC}\vert\boldsymbol{A}^\#}(\boldsymbol{abc} \;\;\vert\,\boldsymbol{a})}\\
&\text{ and}\quad{P'_{\boldsymbol{A}\boldsymbol{B}\boldsymbol{C}\vert\boldsymbol{A}^\#}\in \textsf{CompatibleDistributions}\big[\mathcal{G'}\big]}.
\end{split}\label{eq:mediation}
\end{align}
which pretty much recapitulates Eq.~\eqref{eq:interruption}, except here we distinguish between the nodes pertinent to the conditional ($\boldsymbol{AB}$) and all other observed nodes ($\boldsymbol{C}$).

For the purposes of this work, we emphasize that mediation analysis explicitly maps the problem of bounding causal effects to the problem of causal compatibility relative to an interrupted graph. In particular, mediation analysis implies that the quantum inflation technique can provide upper and lower bounds for the \emph{quantum} ACE, i.e., can constrain causal effects in the presence of quantum confounding. For instance, quantum ACE can be lower (respectively, upper) bounded by minimizing (respectively, maximizing) ${{\operatorname{ACE}}[a_1,a_2,B]}$ over extended distributions which are quantum compatible with the graph interruption corresponding to replacing the edge $A\to B$ with the edge ${A^\#{\to} B}$.

An important result in mediation analysis is that some do-conditionals are constrained to a single point value (called \emph{identifiable}) regardless of the particular observed statistics. The criteria for determining identifiable do-conditionals (and efficient algorithms for computing their values) is the subject of Refs.~\cite{shpitser2016,shpitser2018identification,stensrud2019separable,malinsky2019potential,bhattacharya2020semiparametric}.
Nonidentifiable do-conditionals can typically be constrained only to lie within certain numerical spans~\cite{Cai2007,kang2012inequality,miles2015partial}.\footnote{Fine-tuned instances of the observable distribution can constrain a nonidentifiable do-conditional to the point where it can take only some extremal fixed value.} Visual examples of some identifiable and nonidentifiable do-conditionals are provided in Fig.~\ref{fig:interventionexamples}.

We note that the criteria for determining identifiability are valid independent of the (non)classicality of the latent variables.
Consequently, quantum and classical mediation analysis can differ only in terms of the inequalities they imply for ranges of nonidentifiable do-conditionals.
Reference~\cite{Chaves2017} highlights such a quantum-classical gap regarding the nonidentifiable do-conditional of the instrumental scenario. \citet{Chaves2017} explicitly construct a quantum causal model for the instrumental scenario where the quantum average causal effect is zero even though the resulting observational distribution would necessarily imply a strictly positive classical ACE.

\begin{figure}[t]
  \begin{center}
    \subfigure[\label{fig:TriangleNoSettingsPlus}]
    {\centering
      \begin{minipage}[t]{0.48\linewidth}
      \centering\includegraphics[scale=0.5]{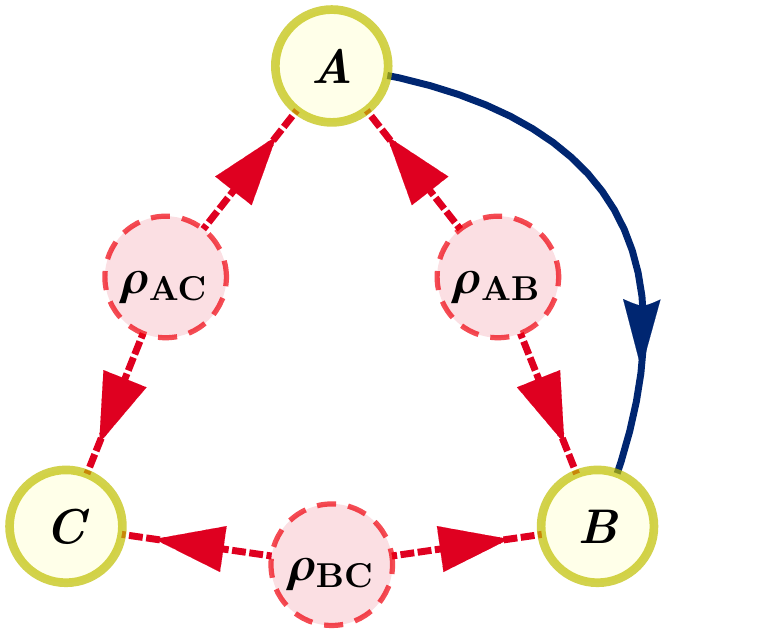}\end{minipage}
    }\hfill
    \subfigure[\label{fig:TriangleNoSettingsPlusInterrupted}]
    {\centering
      \begin{minipage}[t]{0.49\linewidth}
      \centering\includegraphics[scale=0.5]{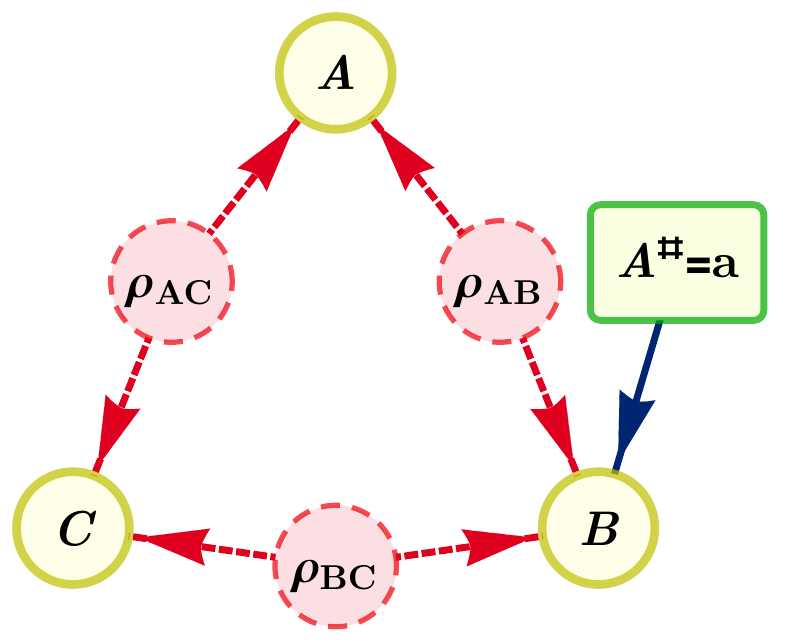}\end{minipage}
    }
  \end{center}
  \caption[]{ %
  (a)~The triangle scenario supplemented with a directed edge from $A\to B$.
  (b)~The relevant interrupted graph used to both define and compute the ACE of $A$ on $B$. %
  \label{fig:triangleintervention}
  }
\end{figure}

To illustrate the utility of quantum inflation for mediation analysis, consider the quantum triangle scenario supplemented by a directed edge ${A\to B}$, illustrated in  Fig.~\ref{fig:TriangleNoSettingsPlus}. The GHZ distribution
\begin{equation}
    P_\text{GHZ}(a,b,c)\coloneqq\begin{cases}
        \frac{1}{2} & a{=}b{=}c \\
        0           & \text{otherwise}
    \end{cases}
    \label{GHZdistribution}
\end{equation}
is compatible with this graph. To achieve this correlation, however, we anticipate that $B$ must \emph{functionally} depend on $A$. However, ${P_{\boldsymbol{B}\textbf{ do } \boldsymbol{A}^\#\shortto \boldsymbol{B}}}$ is not identifiable, as $A$ and $B$ share an unobserved common parent. We employ the quantum inflation technique to verify that the ${\operatorname{ACE}[A,B]=1}$ for the GHZ distribution without assuming classicality, thus teaching us that the quantum ACE is \emph{not} less than the classical ACE in this example.
This result means that, when reproducing the GHZ distribution in the quantum triangle scenario with signaling per Fig.~\ref{fig:TriangleNoSettingsPlus}, the functional dependence of $B$ on the signal from $A$ cannot be reduced relative to the strong dependence of $B$ on $A$ required to reproduce the GHZ distribution in the classical triangle scenario with signaling. This result is in stark contrast to the example in Ref.~\cite{Chaves2017}, where an instrumental scenario correlation requires reduced functional dependence between the observable variables when the latent node is quantum versus when it is classical.

Authors' note: Following the initial publication of the manuscript on the arXiv, the authors became aware of the related work in Ref.~\cite{Quantifying2020} which explores quantifying direct causal influence in the quantum Instrumental scenario.

\section{CONCLUSIONS}\label{sec:conclusions}
We introduced the quantum inflation technique, a systematic method to discern whether an observable distribution is compatible with a causal explanation involving quantum degrees of freedom.
The technique is of general applicability, in that it can be employed to analyze correlations achievable by any quantum causal structure with, potentially, visible-to-visible, latent-to-visible, visible-to-latent, or latent-to-latent connections.
Furthermore, we discussed how a slight modification allows also to study causal realizations in terms of classical latent variables.

We used quantum inflation to study correlations achievable in different quantum causal structures.
First, we proved that the W and Mermin-GHZ distributions cannot be generated in the triangle scenario with quantum latent variables, bounded their noise resistance, and showed how quantum inflation is capable of recovering known results in the entanglement swapping scenario.
Moreover, exploiting the dual formulation of semidefinite programs, we derived polynomial quantum causal witnesses for the triangle scenario with and without inputs.
Second, we showed how quantum inflation can be employed as a tool for constrained polynomial optimization.
We illustrated this fact in a variety of ways.
We bounded the maximum values achievable by Mermin's and Svetlichny's inequalities in the quantum triangle scenario, finding significant gaps relative to the values achievable when one has access to arbitrary three-way entanglement.
We also bounded the amount of information that an eavesdropper can obtain when two distant parties generate secret key using a quantum repeater.
Finally, we illustrated the applicability of quantum inflation for mediation analysis in the presence of quantum common causes.

The implementation of quantum inflation synthesizes two different hierarchies: the one of inflations and, for each inflation, the NPO hierarchy used to determine whether a distribution admits such an inflation.
While asymptotic convergence has been proven for the latter; that of the former is an open question.
Nevertheless, we have identified situations in which tight results can be obtained at finite steps of the hierarchies.
Moreover, recent results~\cite{bowles2020bounding} show that a convergent NPO hierarchy can be defined for the analysis of quantum correlations in sequential Bell scenarios. Given that many nonexogenous causal structures can be mapped to networks linking such sequential scenarios (though not all, as shown in Appendix~\ref{app:nonexog}), Ref.~\cite{bowles2020bounding}'s proof becomes an interesting alternative to quantum inflation for analyzing such special cases.

The number of copies of a given operator increases polynomially with the quantum inflation hierarchy level, where the exponent is given by the number of sources that feed the original operator.
Moreover it is known~\cite{npa2} that, for a fixed inflation level, the complexity of each level in the corresponding NPO hierarchy, $\mathcal{S}_n$, is exponential in the level index $n$.
For these reasons, in terms of practical applicability, standard tabletop computers are typically incapable of handling inflation and NPO hierarchy levels beyond 3.
The optimal choice of which hierarchy to prioritize under scarce computational resources appears to vary depending on the particular problem at hand.
We direct the reader to Ref.~\cite[Chap. 5]{alexThesis} for a fuller discussion on the impacts of varying the different hierarchy level choices and an account of some intuitions in this matter.
Despite this computational load, the SDPs associated to quantum inflation present a high degree of symmetry, and it is expected that exploiting these symmetries explicitly will reduce the computational barriers currently associated with high levels of the inflation hierarchy.

Quantum inflation can find an application in many fields.
Clear first applications are generalizations of entanglement theory and quantum information protocols to networks~\cite{navascues2020gnme,Schmid2020LOSR}.
From a more general perspective, and due to the central role that causality has in science, we expect quantum inflation to become a fundamental tool for analyzing causality in any situation where a quantum behavior is presumed.

\vspace{-10pt}
\begin{acknowledgments}
We acknowledge useful discussions with Stefano Pironio, Rob Spekkens, David Schmid, and T.C. Fraser. This work is supported by Fundaci\'o Obra Social ``la Caixa'' (LCF/BQ/ES15/10360001), the ERC through the AdG CERQUTE and the CoG QITBOX, the European Union's Horizon 2020 research and innovation programm---Grant Agreement No. 648913---the AXA Chair in Quantum Information Science, the Government of Spain (FIS2020-TRANQI and Severo Ochoa CEX2019-000910-S), Fundacio Cellex, Fundacio Mir-Puig, Generalitat de Catalunya (CERCA, AGAUR SGR 1381), and the Austrian Science fund (FWF) stand-alone project P 30947.
This research was supported by Perimeter Institute for Theoretical Physics. Research at Perimeter Institute is supported in part by the Government of Canada through the Department of Innovation, Science and Economic Development Canada and by the Province of Ontario through the Ministry of Colleges and Universities.
This publication was made possible through the support of a grant from the John Templeton Foundation. The opinions expressed in this publication are those of the authors and do not necessarily reflect the views of the John Templeton Foundation.
\end{acknowledgments}

\bibliographystyle{apsrev4-2-wolfe}
\nocite{apsrev41Control}
\setlength{\bibsep}{2pt plus 2pt minus 1pt}
\bibliography{bibliography}

\begin{thebibliography}{67}%
\makeatletter
\providecommand \@ifxundefined [1]{%
 \@ifx{#1\undefined}
}%
\providecommand \@ifnum [1]{%
 \ifnum #1\expandafter \@firstoftwo
 \else \expandafter \@secondoftwo
 \fi
}%
\providecommand \@ifx [1]{%
 \ifx #1\expandafter \@firstoftwo
 \else \expandafter \@secondoftwo
 \fi
}%
\providecommand \natexlab [1]{#1}%
\providecommand \enquote  [1]{``#1''}%
\providecommand \bibnamefont  [1]{#1}%
\providecommand \bibfnamefont [1]{#1}%
\providecommand \citenamefont [1]{#1}%
\providecommand \href@noop [0]{\@secondoftwo}%
\providecommand \href [0]{\begingroup \@sanitize@url \@href}%
\providecommand \@href[1]{\@@startlink{#1}\@@href}%
\providecommand \@@href[1]{\endgroup#1\@@endlink}%
\providecommand \@sanitize@url [0]{\catcode `\\12\catcode `\$12\catcode
  `\&12\catcode `\#12\catcode `\^12\catcode `\_12\catcode `\%12\relax}%
\providecommand \@@startlink[1]{}%
\providecommand \@@endlink[0]{}%
\providecommand \url  [0]{\begingroup\@sanitize@url \@url }%
\providecommand \@url [1]{\endgroup\@href {#1}{\urlprefix }}%
\providecommand \urlprefix  [0]{URL }%
\providecommand \Eprint [0]{\href }%
\providecommand \doibase [0]{https://doi.org/}%
\providecommand \selectlanguage [0]{\@gobble}%
\providecommand \bibinfo  [0]{\@secondoftwo}%
\providecommand \bibfield  [0]{\@secondoftwo}%
\providecommand \translation [1]{[#1]}%
\providecommand \BibitemOpen [0]{}%
\providecommand \bibitemStop [0]{}%
\providecommand \bibitemNoStop [0]{.\EOS\space}%
\providecommand \EOS [0]{\spacefactor3000\relax}%
\providecommand \BibitemShut  [1]{\csname bibitem#1\endcsname}%
\let\auto@bib@innerbib\@empty
\bibitem [{\citenamefont {Pearl}(2009)}]{pearl}%
  \BibitemOpen
  \bibfield  {author} {\bibinfo {author} {\bibfnamefont {J.}~\bibnamefont
  {Pearl}},\ }\href {https://doi.org/10.1017/CBO9780511803161} {\emph {\bibinfo
  {title} {{Causality: Models, Reasoning, and Inference}}}}\ (\bibinfo
  {publisher} {Cambridge University Press},\ \bibinfo {year}
  {2009})\BibitemShut {NoStop}%
\bibitem [{\citenamefont {Morgan}\ and\ \citenamefont
  {Winship}(2007)}]{morgan2007counterfactuals}%
  \BibitemOpen
  \bibfield  {author} {\bibinfo {author} {\bibfnamefont {S.}~\bibnamefont
  {Morgan}}\ and\ \bibinfo {author} {\bibfnamefont {C.}~\bibnamefont
  {Winship}},\ }\href
  {https://www.powells.com/book/counterfactuals-and-causal-inference-9780521671934}
  {\emph {\bibinfo {title} {{Counterfactuals and Causal Inference: Methods and
  Principles for Social Research}}}}\ (\bibinfo  {publisher} {Cambridge
  University Press},\ \bibinfo {year} {2007})\BibitemShut {NoStop}%
\bibitem [{\citenamefont {Shpitser}\ and\ \citenamefont
  {Pearl}(2008)}]{ShpitserPearlIdentification}%
  \BibitemOpen
  \bibfield  {author} {\bibinfo {author} {\bibfnamefont {I.}~\bibnamefont
  {Shpitser}}\ and\ \bibinfo {author} {\bibfnamefont {J.}~\bibnamefont
  {Pearl}},\ }\bibfield  {title} {\enquote {\bibinfo {title} {Complete
  identification methods for the causal hierarchy},}\ }\href
  {http://www.jmlr.org/papers/volume9/shpitser08a/shpitser08a.pdf} {\bibfield
  {journal} {\bibinfo  {journal} {J. Mach. Learn. Res.}\ }\textbf {\bibinfo
  {volume} {9}},\ \bibinfo {pages} {1941} (\bibinfo {year} {2008})}\BibitemShut
  {NoStop}%
\bibitem [{\citenamefont {Bell}(1966)}]{Bell66}%
  \BibitemOpen
  \bibfield  {author} {\bibinfo {author} {\bibfnamefont {J.~S.}\ \bibnamefont
  {Bell}},\ }\bibfield  {title} {\enquote {\bibinfo {title} {On the problem of
  hidden variables in quantum mechanics},}\ }\href
  {https://link.aps.org/doi/10.1103/RevModPhys.38.447} {\bibfield  {journal}
  {\bibinfo  {journal} {Rev. Mod. Phys.}\ }\textbf {\bibinfo {volume} {38}},\
  \bibinfo {pages} {447} (\bibinfo {year} {1966})}\BibitemShut {NoStop}%
\bibitem [{\citenamefont {Brunner}\ \emph {et~al.}(2014)\citenamefont
  {Brunner}, \citenamefont {Cavalcanti}, \citenamefont {Pironio}, \citenamefont
  {Scarani},\ and\ \citenamefont {Wehner}}]{BellReview}%
  \BibitemOpen
  \bibfield  {author} {\bibinfo {author} {\bibfnamefont {N.}~\bibnamefont
  {Brunner}}, \bibinfo {author} {\bibfnamefont {D.}~\bibnamefont {Cavalcanti}},
  \bibinfo {author} {\bibfnamefont {S.}~\bibnamefont {Pironio}}, \bibinfo
  {author} {\bibfnamefont {V.}~\bibnamefont {Scarani}},\ and\ \bibinfo {author}
  {\bibfnamefont {S.}~\bibnamefont {Wehner}},\ }\bibfield  {title} {\enquote
  {\bibinfo {title} {Bell nonlocality},}\ }\href
  {https://link.aps.org/doi/10.1103/RevModPhys.86.419} {\bibfield  {journal}
  {\bibinfo  {journal} {Rev. Mod. Phys.}\ }\textbf {\bibinfo {volume} {86}},\
  \bibinfo {pages} {419} (\bibinfo {year} {2014})}\BibitemShut {NoStop}%
\bibitem [{\citenamefont {Fritz}(2012)}]{fritz2012bell}%
  \BibitemOpen
  \bibfield  {author} {\bibinfo {author} {\bibfnamefont {T.}~\bibnamefont
  {Fritz}},\ }\bibfield  {title} {\enquote {\bibinfo {title} {Beyond {B}ell's
  theorem: correlation scenarios},}\ }\href
  {http://stacks.iop.org/1367-2630/14/i=10/a=103001} {\bibfield  {journal}
  {\bibinfo  {journal} {New J. Phys.}\ }\textbf {\bibinfo {volume} {14}},\
  \bibinfo {pages} {103001} (\bibinfo {year} {2012})}\BibitemShut {NoStop}%
\bibitem [{\citenamefont {Henson}\ \emph {et~al.}(2014)\citenamefont {Henson},
  \citenamefont {Lal},\ and\ \citenamefont {Pusey}}]{HLP}%
  \BibitemOpen
  \bibfield  {author} {\bibinfo {author} {\bibfnamefont {J.}~\bibnamefont
  {Henson}}, \bibinfo {author} {\bibfnamefont {R.}~\bibnamefont {Lal}},\ and\
  \bibinfo {author} {\bibfnamefont {M.~F.}\ \bibnamefont {Pusey}},\ }\bibfield
  {title} {\enquote {\bibinfo {title} {{Theory-independent limits on
  correlations from generalized Bayesian networks}},}\ }\href
  {https://doi.org/10.1088%2F1367-2630%2F16%2F11%2F113043} {\bibfield
  {journal} {\bibinfo  {journal} {New J. Phys.}\ }\textbf {\bibinfo {volume}
  {16}},\ \bibinfo {pages} {113043} (\bibinfo {year} {2014})}\BibitemShut
  {NoStop}%
\bibitem [{\citenamefont {Wood}\ and\ \citenamefont
  {Spekkens}(2015)}]{Wood2015}%
  \BibitemOpen
  \bibfield  {author} {\bibinfo {author} {\bibfnamefont {C.~J.}\ \bibnamefont
  {Wood}}\ and\ \bibinfo {author} {\bibfnamefont {R.~W.}\ \bibnamefont
  {Spekkens}},\ }\bibfield  {title} {\enquote {\bibinfo {title} {{The lesson of
  causal discovery algorithms for quantum correlations: causal explanations of
  Bell-inequality violations require fine-tuning}},}\ }\href
  {https://doi.org/10.1088/1367-2630/17/3/033002} {\bibfield  {journal}
  {\bibinfo  {journal} {New J. Phys.}\ }\textbf {\bibinfo {volume} {17}},\
  \bibinfo {pages} {033002} (\bibinfo {year} {2015})}\BibitemShut {NoStop}%
\bibitem [{\citenamefont {Chaves}\ \emph
  {et~al.}(2015{\natexlab{a}})\citenamefont {Chaves}, \citenamefont {Kueng},
  \citenamefont {{Bohr Brask}},\ and\ \citenamefont
  {Gross}}]{Chaves2015relaxing}%
  \BibitemOpen
  \bibfield  {author} {\bibinfo {author} {\bibfnamefont {R.}~\bibnamefont
  {Chaves}}, \bibinfo {author} {\bibfnamefont {R.}~\bibnamefont {Kueng}},
  \bibinfo {author} {\bibfnamefont {J.}~\bibnamefont {{Bohr Brask}}},\ and\
  \bibinfo {author} {\bibfnamefont {D.}~\bibnamefont {Gross}},\ }\bibfield
  {title} {\enquote {\bibinfo {title} {Unifying framework for relaxations of
  the causal assumptions in {Bell}'s theorem},}\ }\href
  {https://link.aps.org/doi/10.1103/PhysRevLett.114.140403} {\bibfield
  {journal} {\bibinfo  {journal} {Phys. Rev. Lett.}\ }\textbf {\bibinfo
  {volume} {114}},\ \bibinfo {pages} {140403} (\bibinfo {year}
  {2015}{\natexlab{a}})}\BibitemShut {NoStop}%
\bibitem [{\citenamefont {Wolfe}\ \emph {et~al.}(2020)\citenamefont {Wolfe},
  \citenamefont {Schmid}, \citenamefont {Sainz}, \citenamefont {Kunjwal},\ and\
  \citenamefont {Spekkens}}]{WolfeBellQuantified}%
  \BibitemOpen
  \bibfield  {author} {\bibinfo {author} {\bibfnamefont {E.}~\bibnamefont
  {Wolfe}}, \bibinfo {author} {\bibfnamefont {D.}~\bibnamefont {Schmid}},
  \bibinfo {author} {\bibfnamefont {A.~B.}\ \bibnamefont {Sainz}}, \bibinfo
  {author} {\bibfnamefont {R.}~\bibnamefont {Kunjwal}},\ and\ \bibinfo {author}
  {\bibfnamefont {R.~W.}\ \bibnamefont {Spekkens}},\ }\bibfield  {title}
  {\enquote {\bibinfo {title} {Quantifying {B}ell: the {R}esource {T}heory of
  {N}onclassicality of {C}ommon-{C}ause {B}oxes},}\ }\href
  {https://doi.org/10.22331/q-2020-06-08-280} {\bibfield  {journal} {\bibinfo
  {journal} {{Quantum}}\ }\textbf {\bibinfo {volume} {4}},\ \bibinfo {pages}
  {280} (\bibinfo {year} {2020})}\BibitemShut {NoStop}%
\bibitem [{\citenamefont {{Wolfe}}\ \emph {et~al.}(2019)\citenamefont
  {{Wolfe}}, \citenamefont {{Spekkens}},\ and\ \citenamefont
  {{Fritz}}}]{wolfe2016inflation}%
  \BibitemOpen
  \bibfield  {author} {\bibinfo {author} {\bibfnamefont {E.}~\bibnamefont
  {{Wolfe}}}, \bibinfo {author} {\bibfnamefont {R.~W.}\ \bibnamefont
  {{Spekkens}}},\ and\ \bibinfo {author} {\bibfnamefont {T.}~\bibnamefont
  {{Fritz}}},\ }\bibfield  {title} {\enquote {\bibinfo {title} {The {Inflation
  Technique} for causal inference with latent variables},}\ }\href
  {http://dx.doi.org/10.1515/jci-2017-0020} {\bibfield  {journal} {\bibinfo
  {journal} {J. Causal Inference}\ }\textbf {\bibinfo {volume} {7}},\ \bibinfo
  {pages} {20170020} (\bibinfo {year} {2019})}\BibitemShut {NoStop}%
\bibitem [{\citenamefont {Navascués}\ and\ \citenamefont
  {Wolfe}(2020)}]{navascues2017inflation}%
  \BibitemOpen
  \bibfield  {author} {\bibinfo {author} {\bibfnamefont {M.}~\bibnamefont
  {Navascués}}\ and\ \bibinfo {author} {\bibfnamefont {E.}~\bibnamefont
  {Wolfe}},\ }\bibfield  {title} {\enquote {\bibinfo {title} {The inflation
  technique completely solves the causal compatibility problem},}\ }\href
  {https://doi.org/https://doi.org/10.1515/jci-2018-0008} {\bibfield  {journal}
  {\bibinfo  {journal} {J. Causal Inference}\ }\textbf {\bibinfo {volume}
  {8}},\ \bibinfo {pages} {70 } (\bibinfo {year} {2020})}\BibitemShut {NoStop}%
\bibitem [{\citenamefont {Rosset}\ \emph {et~al.}(2018)\citenamefont {Rosset},
  \citenamefont {Gisin},\ and\ \citenamefont {Wolfe}}]{rosset2016finite}%
  \BibitemOpen
  \bibfield  {author} {\bibinfo {author} {\bibfnamefont {D.}~\bibnamefont
  {Rosset}}, \bibinfo {author} {\bibfnamefont {N.}~\bibnamefont {Gisin}},\ and\
  \bibinfo {author} {\bibfnamefont {E.}~\bibnamefont {Wolfe}},\ }\bibfield
  {title} {\enquote {\bibinfo {title} {Universal bound on the cardinality of
  local hidden variables in networks},}\ }\href
  {https://arxiv.org/abs/1709.00707} {\bibfield  {journal} {\bibinfo  {journal}
  {Quant. Info. {\&} Comp.}\ }\textbf {\bibinfo {volume} {18}},\ \bibinfo
  {pages} {910} (\bibinfo {year} {2018})}\BibitemShut {NoStop}%
\bibitem [{\citenamefont {Slofstra}(2019)}]{Slofstra}%
  \BibitemOpen
  \bibfield  {author} {\bibinfo {author} {\bibfnamefont {W.}~\bibnamefont
  {Slofstra}},\ }\bibfield  {title} {\enquote {\bibinfo {title} {The set of
  quantum correlations is not closed},}\ }\href
  {https://doi.org/10.1017/fmp.2018.3} {\bibfield  {journal} {\bibinfo
  {journal} {For. Math., Pi}\ }\textbf {\bibinfo {volume} {7}},\ \bibinfo
  {pages} {e1} (\bibinfo {year} {2019})}\BibitemShut {NoStop}%
\bibitem [{\citenamefont {{Ji}}\ \emph {et~al.}(2020)\citenamefont {{Ji}},
  \citenamefont {{Natarajan}}, \citenamefont {{Vidick}}, \citenamefont
  {{Wright}},\ and\ \citenamefont {{Yuen}}}]{Ji2020Connes}%
  \BibitemOpen
  \bibfield  {author} {\bibinfo {author} {\bibfnamefont {Z.}~\bibnamefont
  {{Ji}}}, \bibinfo {author} {\bibfnamefont {A.}~\bibnamefont {{Natarajan}}},
  \bibinfo {author} {\bibfnamefont {T.}~\bibnamefont {{Vidick}}}, \bibinfo
  {author} {\bibfnamefont {J.}~\bibnamefont {{Wright}}},\ and\ \bibinfo
  {author} {\bibfnamefont {H.}~\bibnamefont {{Yuen}}},\ }\bibfield  {title}
  {\enquote {\bibinfo {title} {{MIP*=RE}},}\ }\href
  {https://arxiv.org/abs/2001.04383} {\bibfield  {journal} {\bibinfo  {journal}
  {arXiv:2001.04383}\ } (\bibinfo {year} {2020})}\BibitemShut {NoStop}%
\bibitem [{\citenamefont {Barnum}\ \emph {et~al.}(1996)\citenamefont {Barnum},
  \citenamefont {Caves}, \citenamefont {Fuchs}, \citenamefont {Jozsa},\ and\
  \citenamefont {Schumacher}}]{BroadcastingMixed}%
  \BibitemOpen
  \bibfield  {author} {\bibinfo {author} {\bibfnamefont {H.}~\bibnamefont
  {Barnum}}, \bibinfo {author} {\bibfnamefont {C.~M.}\ \bibnamefont {Caves}},
  \bibinfo {author} {\bibfnamefont {C.~A.}\ \bibnamefont {Fuchs}}, \bibinfo
  {author} {\bibfnamefont {R.}~\bibnamefont {Jozsa}},\ and\ \bibinfo {author}
  {\bibfnamefont {B.}~\bibnamefont {Schumacher}},\ }\bibfield  {title}
  {\enquote {\bibinfo {title} {Noncommuting mixed states cannot be
  broadcast},}\ }\href {https://link.aps.org/doi/10.1103/PhysRevLett.76.2818}
  {\bibfield  {journal} {\bibinfo  {journal} {Phys. Rev. Lett.}\ }\textbf
  {\bibinfo {volume} {76}},\ \bibinfo {pages} {2818} (\bibinfo {year}
  {1996})}\BibitemShut {NoStop}%
\bibitem [{\citenamefont {{Barnum}}\ \emph {et~al.}(2006)\citenamefont
  {{Barnum}}, \citenamefont {{Barrett}}, \citenamefont {{Leifer}},\ and\
  \citenamefont {{Wilce}}}]{NoCloningGeneral2006}%
  \BibitemOpen
  \bibfield  {author} {\bibinfo {author} {\bibfnamefont {H.}~\bibnamefont
  {{Barnum}}}, \bibinfo {author} {\bibfnamefont {J.}~\bibnamefont {{Barrett}}},
  \bibinfo {author} {\bibfnamefont {M.}~\bibnamefont {{Leifer}}},\ and\
  \bibinfo {author} {\bibfnamefont {A.}~\bibnamefont {{Wilce}}},\ }\bibfield
  {title} {\enquote {\bibinfo {title} {{Cloning and broadcasting in generic
  probabilistic theories}},}\ }\href {http://arxiv.org/abs/quant-ph/0611295}
  {\bibfield  {journal} {\bibinfo  {journal} {quant-ph/0611295}\ } (\bibinfo
  {year} {2006})}\BibitemShut {NoStop}%
\bibitem [{\citenamefont {Chaves}\ \emph
  {et~al.}(2015{\natexlab{b}})\citenamefont {Chaves}, \citenamefont {Majenz},\
  and\ \citenamefont {Gross}}]{Chaves2015}%
  \BibitemOpen
  \bibfield  {author} {\bibinfo {author} {\bibfnamefont {R.}~\bibnamefont
  {Chaves}}, \bibinfo {author} {\bibfnamefont {C.}~\bibnamefont {Majenz}},\
  and\ \bibinfo {author} {\bibfnamefont {D.}~\bibnamefont {Gross}},\ }\bibfield
   {title} {\enquote {\bibinfo {title} {Information{\textendash}theoretic
  implications of quantum causal structures},}\ }\href
  {https://doi.org/10.1038/ncomms6766} {\bibfield  {journal} {\bibinfo
  {journal} {Nat. Commun.}\ }\textbf {\bibinfo {volume} {6}} (\bibinfo {year}
  {2015}{\natexlab{b}})}\BibitemShut {NoStop}%
\bibitem [{\citenamefont {Pozas-Kerstjens}\ \emph {et~al.}(2019)\citenamefont
  {Pozas-Kerstjens}, \citenamefont {Rabelo}, \citenamefont {Rudnicki},
  \citenamefont {Chaves}, \citenamefont {Cavalcanti}, \citenamefont
  {Navascu\'es},\ and\ \citenamefont {Ac\'{\i}n}}]{Pozas2019}%
  \BibitemOpen
  \bibfield  {author} {\bibinfo {author} {\bibfnamefont {A.}~\bibnamefont
  {Pozas-Kerstjens}}, \bibinfo {author} {\bibfnamefont {R.}~\bibnamefont
  {Rabelo}}, \bibinfo {author} {\bibfnamefont {L.}~\bibnamefont {Rudnicki}},
  \bibinfo {author} {\bibfnamefont {R.}~\bibnamefont {Chaves}}, \bibinfo
  {author} {\bibfnamefont {D.}~\bibnamefont {Cavalcanti}}, \bibinfo {author}
  {\bibfnamefont {M.}~\bibnamefont {Navascu\'es}},\ and\ \bibinfo {author}
  {\bibfnamefont {A.}~\bibnamefont {Ac\'{\i}n}},\ }\bibfield  {title} {\enquote
  {\bibinfo {title} {Bounding the sets of classical and quantum correlations in
  networks},}\ }\href {https://doi.org/10.1103/PhysRevLett.123.140503}
  {\bibfield  {journal} {\bibinfo  {journal} {Phys. Rev. Lett.}\ }\textbf
  {\bibinfo {volume} {123}},\ \bibinfo {pages} {140503} (\bibinfo {year}
  {2019})}\BibitemShut {NoStop}%
\bibitem [{\citenamefont {\AA{}berg}\ \emph {et~al.}(2020)\citenamefont
  {\AA{}berg}, \citenamefont {Nery}, \citenamefont {Duarte},\ and\
  \citenamefont {Chaves}}]{aberg2020covariance}%
  \BibitemOpen
  \bibfield  {author} {\bibinfo {author} {\bibfnamefont {J.}~\bibnamefont
  {\AA{}berg}}, \bibinfo {author} {\bibfnamefont {R.}~\bibnamefont {Nery}},
  \bibinfo {author} {\bibfnamefont {C.}~\bibnamefont {Duarte}},\ and\ \bibinfo
  {author} {\bibfnamefont {R.}~\bibnamefont {Chaves}},\ }\bibfield  {title}
  {\enquote {\bibinfo {title} {Semidefinite tests for quantum network
  topologies},}\ }\href {https://doi.org/10.1103/PhysRevLett.125.110505}
  {\bibfield  {journal} {\bibinfo  {journal} {Phys. Rev. Lett.}\ }\textbf
  {\bibinfo {volume} {125}},\ \bibinfo {pages} {110505} (\bibinfo {year}
  {2020})}\BibitemShut {NoStop}%
\bibitem [{\citenamefont {Ness}\ \emph {et~al.}(2016)\citenamefont {Ness},
  \citenamefont {Sachs},\ and\ \citenamefont {Vitek}}]{biomolecular}%
  \BibitemOpen
  \bibfield  {author} {\bibinfo {author} {\bibfnamefont {R.~O.}\ \bibnamefont
  {Ness}}, \bibinfo {author} {\bibfnamefont {K.}~\bibnamefont {Sachs}},\ and\
  \bibinfo {author} {\bibfnamefont {O.}~\bibnamefont {Vitek}},\ }\bibfield
  {title} {\enquote {\bibinfo {title} {From correlation to causality:
  Statistical approaches to learning regulatory relationships in large-scale
  biomolecular investigations},}\ }\href
  {https://doi.org/10.1021/acs.jproteome.5b00911} {\bibfield  {journal}
  {\bibinfo  {journal} {J. Proteome Res.}\ }\textbf {\bibinfo {volume} {15}},\
  \bibinfo {pages} {683} (\bibinfo {year} {2016})}\BibitemShut {NoStop}%
\bibitem [{\citenamefont {Hartemink}\ \emph {et~al.}(2001)\citenamefont
  {Hartemink}, \citenamefont {Gifford}, \citenamefont {Jaakkola},\ and\
  \citenamefont {Young}}]{genomics}%
  \BibitemOpen
  \bibfield  {author} {\bibinfo {author} {\bibfnamefont {A.~J.}\ \bibnamefont
  {Hartemink}}, \bibinfo {author} {\bibfnamefont {D.~K.}\ \bibnamefont
  {Gifford}}, \bibinfo {author} {\bibfnamefont {T.~S.}\ \bibnamefont
  {Jaakkola}},\ and\ \bibinfo {author} {\bibfnamefont {R.~A.}\ \bibnamefont
  {Young}},\ }\enquote {\bibinfo {title} {Using graphical models and genomic
  expression data to statistically validate models of genetic regulatory
  networks},}\ in\ \href {https://doi.org/10.1142/9789814447362_0042} {\emph
  {\bibinfo {booktitle} {Biocomputing}}}\ (\bibinfo  {publisher} {World
  Scientific},\ \bibinfo {year} {2001})\ pp.\ \bibinfo {pages}
  {422--433}\BibitemShut {NoStop}%
\bibitem [{\citenamefont {Lambert}\ \emph {et~al.}(2013)\citenamefont
  {Lambert}, \citenamefont {Chen}, \citenamefont {Cheng}, \citenamefont {Li},
  \citenamefont {Chen},\ and\ \citenamefont {Nori}}]{qbio1}%
  \BibitemOpen
  \bibfield  {author} {\bibinfo {author} {\bibfnamefont {N.}~\bibnamefont
  {Lambert}}, \bibinfo {author} {\bibfnamefont {Y.-N.}\ \bibnamefont {Chen}},
  \bibinfo {author} {\bibfnamefont {Y.-C.}\ \bibnamefont {Cheng}}, \bibinfo
  {author} {\bibfnamefont {C.-M.}\ \bibnamefont {Li}}, \bibinfo {author}
  {\bibfnamefont {G.-Y.}\ \bibnamefont {Chen}},\ and\ \bibinfo {author}
  {\bibfnamefont {F.}~\bibnamefont {Nori}},\ }\bibfield  {title} {\enquote
  {\bibinfo {title} {Quantum biology},}\ }\href
  {https://doi.org/10.1038/nphys2474} {\bibfield  {journal} {\bibinfo
  {journal} {Nat. Phys.}\ }\textbf {\bibinfo {volume} {9}},\ \bibinfo {pages}
  {10} (\bibinfo {year} {2013})}\BibitemShut {NoStop}%
\bibitem [{\citenamefont {Cao}\ \emph {et~al.}(2020)\citenamefont {Cao},
  \citenamefont {Cogdell}, \citenamefont {Coker}, \citenamefont {Duan},
  \citenamefont {Hauer}, \citenamefont {Kleinekath{\"o}fer}, \citenamefont
  {Jansen}, \citenamefont {Man{\v c}al}, \citenamefont {Miller}, \citenamefont
  {Ogilvie}, \citenamefont {Prokhorenko}, \citenamefont {Renger}, \citenamefont
  {Tan}, \citenamefont {Tempelaar}, \citenamefont {Thorwart}, \citenamefont
  {Thyrhaug}, \citenamefont {Westenhoff},\ and\ \citenamefont
  {Zigmantas}}]{qbio2}%
  \BibitemOpen
  \bibfield  {author} {\bibinfo {author} {\bibfnamefont {J.}~\bibnamefont
  {Cao}}, \bibinfo {author} {\bibfnamefont {R.~J.}\ \bibnamefont {Cogdell}},
  \bibinfo {author} {\bibfnamefont {D.~F.}\ \bibnamefont {Coker}}, \bibinfo
  {author} {\bibfnamefont {H.-G.}\ \bibnamefont {Duan}}, \bibinfo {author}
  {\bibfnamefont {J.}~\bibnamefont {Hauer}}, \bibinfo {author} {\bibfnamefont
  {U.}~\bibnamefont {Kleinekath{\"o}fer}}, \bibinfo {author} {\bibfnamefont
  {T.~L.~C.}\ \bibnamefont {Jansen}}, \bibinfo {author} {\bibfnamefont
  {T.}~\bibnamefont {Man{\v c}al}}, \bibinfo {author} {\bibfnamefont
  {R.~J.~D.}\ \bibnamefont {Miller}}, \bibinfo {author} {\bibfnamefont {J.~P.}\
  \bibnamefont {Ogilvie}}, \bibinfo {author} {\bibfnamefont {V.~I.}\
  \bibnamefont {Prokhorenko}}, \bibinfo {author} {\bibfnamefont
  {T.}~\bibnamefont {Renger}}, \bibinfo {author} {\bibfnamefont {H.-S.}\
  \bibnamefont {Tan}}, \bibinfo {author} {\bibfnamefont {R.}~\bibnamefont
  {Tempelaar}}, \bibinfo {author} {\bibfnamefont {M.}~\bibnamefont {Thorwart}},
  \bibinfo {author} {\bibfnamefont {E.}~\bibnamefont {Thyrhaug}}, \bibinfo
  {author} {\bibfnamefont {S.}~\bibnamefont {Westenhoff}},\ and\ \bibinfo
  {author} {\bibfnamefont {D.}~\bibnamefont {Zigmantas}},\ }\bibfield  {title}
  {\enquote {\bibinfo {title} {Quantum biology revisited},}\ }\href
  {https://advances.sciencemag.org/content/6/14/eaaz4888} {\bibfield  {journal}
  {\bibinfo  {journal} {Sci. Adv.}\ }\textbf {\bibinfo {volume} {6}} (\bibinfo
  {year} {2020})}\BibitemShut {NoStop}%
\bibitem [{\citenamefont {Branciard}\ \emph {et~al.}(2010)\citenamefont
  {Branciard}, \citenamefont {Gisin},\ and\ \citenamefont
  {Pironio}}]{branciard2010bilocality}%
  \BibitemOpen
  \bibfield  {author} {\bibinfo {author} {\bibfnamefont {C.}~\bibnamefont
  {Branciard}}, \bibinfo {author} {\bibfnamefont {N.}~\bibnamefont {Gisin}},\
  and\ \bibinfo {author} {\bibfnamefont {S.}~\bibnamefont {Pironio}},\
  }\bibfield  {title} {\enquote {\bibinfo {title} {Characterizing the nonlocal
  correlations created via entanglement swapping},}\ }\href
  {https://link.aps.org/doi/10.1103/PhysRevLett.104.170401} {\bibfield
  {journal} {\bibinfo  {journal} {Phys. Rev. Lett.}\ }\textbf {\bibinfo
  {volume} {104}},\ \bibinfo {pages} {170401} (\bibinfo {year}
  {2010})}\BibitemShut {NoStop}%
\bibitem [{\citenamefont {Branciard}\ \emph {et~al.}(2012)\citenamefont
  {Branciard}, \citenamefont {Rosset}, \citenamefont {Gisin},\ and\
  \citenamefont {Pironio}}]{branciard2012bilocality}%
  \BibitemOpen
  \bibfield  {author} {\bibinfo {author} {\bibfnamefont {C.}~\bibnamefont
  {Branciard}}, \bibinfo {author} {\bibfnamefont {D.}~\bibnamefont {Rosset}},
  \bibinfo {author} {\bibfnamefont {N.}~\bibnamefont {Gisin}},\ and\ \bibinfo
  {author} {\bibfnamefont {S.}~\bibnamefont {Pironio}},\ }\bibfield  {title}
  {\enquote {\bibinfo {title} {{Bilocal versus nonbilocal correlations in
  entanglement-swapping experiments}},}\ }\href
  {https://link.aps.org/doi/10.1103/PhysRevA.85.032119} {\bibfield  {journal}
  {\bibinfo  {journal} {Phys. Rev. A}\ }\textbf {\bibinfo {volume} {85}},\
  \bibinfo {pages} {032119} (\bibinfo {year} {2012})}\BibitemShut {NoStop}%
\bibitem [{\citenamefont {Pironio}\ \emph {et~al.}(2010)\citenamefont
  {Pironio}, \citenamefont {Navascu\'es},\ and\ \citenamefont
  {Ac{\'i}n}}]{npo}%
  \BibitemOpen
  \bibfield  {author} {\bibinfo {author} {\bibfnamefont {S.}~\bibnamefont
  {Pironio}}, \bibinfo {author} {\bibfnamefont {M.}~\bibnamefont
  {Navascu\'es}},\ and\ \bibinfo {author} {\bibfnamefont {A.}~\bibnamefont
  {Ac{\'i}n}},\ }\bibfield  {title} {\enquote {\bibinfo {title} {Convergent
  relaxations of polynomial optimization problems with non-commuting
  variables},}\ }\href {https://doi.org/10.1137/090760155} {\bibfield
  {journal} {\bibinfo  {journal} {SIAM J. Optim.}\ }\textbf {\bibinfo {volume}
  {20}},\ \bibinfo {pages} {2157} (\bibinfo {year} {2010})}\BibitemShut
  {NoStop}%
\bibitem [{\citenamefont {Navascu\'es}\ \emph {et~al.}(2007)\citenamefont
  {Navascu\'es}, \citenamefont {Pironio},\ and\ \citenamefont
  {Ac\'{i}n}}]{npa}%
  \BibitemOpen
  \bibfield  {author} {\bibinfo {author} {\bibfnamefont {M.}~\bibnamefont
  {Navascu\'es}}, \bibinfo {author} {\bibfnamefont {S.}~\bibnamefont
  {Pironio}},\ and\ \bibinfo {author} {\bibfnamefont {A.}~\bibnamefont
  {Ac\'{i}n}},\ }\bibfield  {title} {\enquote {\bibinfo {title} {Bounding the
  set of quantum correlations},}\ }\href
  {https://link.aps.org/doi/10.1103/PhysRevLett.98.010401} {\bibfield
  {journal} {\bibinfo  {journal} {Phys. Rev. Lett.}\ }\textbf {\bibinfo
  {volume} {98}},\ \bibinfo {pages} {010401} (\bibinfo {year}
  {2007})}\BibitemShut {NoStop}%
\bibitem [{\citenamefont {Navascu\'es}\ \emph {et~al.}(2008)\citenamefont
  {Navascu\'es}, \citenamefont {Pironio},\ and\ \citenamefont
  {Ac{\'i}n}}]{npa2}%
  \BibitemOpen
  \bibfield  {author} {\bibinfo {author} {\bibfnamefont {M.}~\bibnamefont
  {Navascu\'es}}, \bibinfo {author} {\bibfnamefont {S.}~\bibnamefont
  {Pironio}},\ and\ \bibinfo {author} {\bibfnamefont {A.}~\bibnamefont
  {Ac{\'i}n}},\ }\bibfield  {title} {\enquote {\bibinfo {title} {A convergent
  hierarchy of semidefinite programs characterizing the set of quantum
  correlations},}\ }\href {https://doi.org/10.1088/1367-2630/10/7/073013}
  {\bibfield  {journal} {\bibinfo  {journal} {New J. Phys.}\ }\textbf {\bibinfo
  {volume} {10}},\ \bibinfo {pages} {073013} (\bibinfo {year}
  {2008})}\BibitemShut {NoStop}%
\bibitem [{\citenamefont {Vandenberghe}\ and\ \citenamefont
  {Boyd}(1996)}]{sdp}%
  \BibitemOpen
  \bibfield  {author} {\bibinfo {author} {\bibfnamefont {L.}~\bibnamefont
  {Vandenberghe}}\ and\ \bibinfo {author} {\bibfnamefont {S.}~\bibnamefont
  {Boyd}},\ }\bibfield  {title} {\enquote {\bibinfo {title} {Semidefinite
  programming},}\ }\href {https://doi.org/10.1137/1038003} {\bibfield
  {journal} {\bibinfo  {journal} {SIAM Rev.}\ }\textbf {\bibinfo {volume}
  {38}},\ \bibinfo {pages} {49} (\bibinfo {year} {1996})}\BibitemShut {NoStop}%
\bibitem [{\citenamefont {Van~Himbeeck}\ \emph {et~al.}(2019)\citenamefont
  {Van~Himbeeck}, \citenamefont {Bohr~Brask}, \citenamefont {Pironio},
  \citenamefont {Ramanathan}, \citenamefont {Sainz},\ and\ \citenamefont
  {Wolfe}}]{Himbeeck2018instrumental}%
  \BibitemOpen
  \bibfield  {author} {\bibinfo {author} {\bibfnamefont {T.}~\bibnamefont
  {Van~Himbeeck}}, \bibinfo {author} {\bibfnamefont {J.}~\bibnamefont
  {Bohr~Brask}}, \bibinfo {author} {\bibfnamefont {S.}~\bibnamefont {Pironio}},
  \bibinfo {author} {\bibfnamefont {R.}~\bibnamefont {Ramanathan}}, \bibinfo
  {author} {\bibfnamefont {A.~B.}\ \bibnamefont {Sainz}},\ and\ \bibinfo
  {author} {\bibfnamefont {E.}~\bibnamefont {Wolfe}},\ }\bibfield  {title}
  {\enquote {\bibinfo {title} {Quantum violations in the {I}nstrumental
  scenario and their relations to the {B}ell scenario},}\ }\href
  {https://doi.org/10.22331/q-2019-09-16-186} {\bibfield  {journal} {\bibinfo
  {journal} {{Quantum}}\ }\textbf {\bibinfo {volume} {3}},\ \bibinfo {pages}
  {186} (\bibinfo {year} {2019})}\BibitemShut {NoStop}%
\bibitem [{\citenamefont {Agresti}\ \emph {et~al.}(2019)\citenamefont
  {Agresti}, \citenamefont {Carvacho}, \citenamefont {Poderini}, \citenamefont
  {Aolita}, \citenamefont {Chaves},\ and\ \citenamefont
  {Sciarrino}}]{Agresti2019}%
  \BibitemOpen
  \bibfield  {author} {\bibinfo {author} {\bibfnamefont {I.}~\bibnamefont
  {Agresti}}, \bibinfo {author} {\bibfnamefont {G.}~\bibnamefont {Carvacho}},
  \bibinfo {author} {\bibfnamefont {D.}~\bibnamefont {Poderini}}, \bibinfo
  {author} {\bibfnamefont {L.}~\bibnamefont {Aolita}}, \bibinfo {author}
  {\bibfnamefont {R.}~\bibnamefont {Chaves}},\ and\ \bibinfo {author}
  {\bibfnamefont {F.}~\bibnamefont {Sciarrino}},\ }\bibfield  {title} {\enquote
  {\bibinfo {title} {Experimental connection between the {Instrumental} and
  {Bell} inequalities},}\ }\href
  {https://doi.org/10.3390/proceedings2019012027} {\bibfield  {journal}
  {\bibinfo  {journal} {Proc. MDPI AG}\ }\textbf {\bibinfo {volume} {12}},\
  \bibinfo {pages} {27} (\bibinfo {year} {2019})}\BibitemShut {NoStop}%
\bibitem [{\citenamefont {Gachechiladze}\ \emph {et~al.}(2020)\citenamefont
  {Gachechiladze}, \citenamefont {Miklin},\ and\ \citenamefont
  {Chaves}}]{Quantifying2020}%
  \BibitemOpen
  \bibfield  {author} {\bibinfo {author} {\bibfnamefont {M.}~\bibnamefont
  {Gachechiladze}}, \bibinfo {author} {\bibfnamefont {N.}~\bibnamefont
  {Miklin}},\ and\ \bibinfo {author} {\bibfnamefont {R.}~\bibnamefont
  {Chaves}},\ }\bibfield  {title} {\enquote {\bibinfo {title} {Quantifying
  causal influences in the presence of a quantum common cause},}\ }\href
  {https://doi.org/10.1103/PhysRevLett.125.230401} {\bibfield  {journal}
  {\bibinfo  {journal} {Phys. Rev. Lett.}\ }\textbf {\bibinfo {volume} {125}},\
  \bibinfo {pages} {230401} (\bibinfo {year} {2020})}\BibitemShut {NoStop}%
\bibitem [{\citenamefont {Richardson}\ and\ \citenamefont
  {Robins}(2013)}]{Richardson2013SingleWI}%
  \BibitemOpen
  \bibfield  {author} {\bibinfo {author} {\bibfnamefont {T.~S.}\ \bibnamefont
  {Richardson}}\ and\ \bibinfo {author} {\bibfnamefont {J.~M.}\ \bibnamefont
  {Robins}},\ }\href {https://www.csss.washington.edu/Papers/wp128.pdf} {\emph
  {\bibinfo {title} {{Single World Intervention Graphs (SWIGs) : A Unification
  of the Counterfactual and Graphical Approaches to Causality}}}}\ (\bibinfo
  {publisher} {Now Publishers Inc},\ \bibinfo {year} {2013})\ \bibinfo {note}
  {{Working Paper \#128, Center for Stat. \& Soc. Sci., U.
  Washington}}\BibitemShut {NoStop}%
\bibitem [{\citenamefont {Barrett}\ \emph {et~al.}(2019)\citenamefont
  {Barrett}, \citenamefont {Lorenz},\ and\ \citenamefont
  {Oreshkov}}]{BarrettQCM}%
  \BibitemOpen
  \bibfield  {author} {\bibinfo {author} {\bibfnamefont {J.}~\bibnamefont
  {Barrett}}, \bibinfo {author} {\bibfnamefont {R.}~\bibnamefont {Lorenz}},\
  and\ \bibinfo {author} {\bibfnamefont {O.}~\bibnamefont {Oreshkov}},\
  }\bibfield  {title} {\enquote {\bibinfo {title} {{Quantum causal models}},}\
  }\href {https://arxiv.org/abs/1906.10726} {\bibfield  {journal} {\bibinfo
  {journal} {arXiv:1906.10726}\ } (\bibinfo {year} {2019})}\BibitemShut
  {NoStop}%
\bibitem [{\citenamefont {Evans}(2012)}]{evans2012graphicalmethods}%
  \BibitemOpen
  \bibfield  {author} {\bibinfo {author} {\bibfnamefont {R.~J.}\ \bibnamefont
  {Evans}},\ }\bibfield  {title} {\enquote {\bibinfo {title} {Graphical methods
  for inequality constraints in marginalized {DAGs}},}\ }in\ \href
  {https://doi.org/10.1109/mlsp.2012.6349796} {\emph {\bibinfo {booktitle}
  {{{IEEE} International Workshop on Machine Learning for Signal
  Processing}}}}\ (\bibinfo {year} {2012})\BibitemShut {NoStop}%
\bibitem [{\citenamefont {Evans}(2018)}]{Evans2018NMP}%
  \BibitemOpen
  \bibfield  {author} {\bibinfo {author} {\bibfnamefont {R.~J.}\ \bibnamefont
  {Evans}},\ }\bibfield  {title} {\enquote {\bibinfo {title} {{Margins of
  discrete Bayesian networks}},}\ }\href {https://doi.org/10.1214/17-aos1631}
  {\bibfield  {journal} {\bibinfo  {journal} {Annals Stat.}\ }\textbf {\bibinfo
  {volume} {46}},\ \bibinfo {pages} {2623} (\bibinfo {year}
  {2018})}\BibitemShut {NoStop}%
\bibitem [{\citenamefont {Toner}(2009)}]{Toner2009}%
  \BibitemOpen
  \bibfield  {author} {\bibinfo {author} {\bibfnamefont {B.}~\bibnamefont
  {Toner}},\ }\bibfield  {title} {\enquote {\bibinfo {title} {Monogamy of
  non-local quantum correlations},}\ }\href
  {https://doi.org/10.1098/rspa.2008.0149} {\bibfield  {journal} {\bibinfo
  {journal} {Proc. Roy. Soc. A}\ }\textbf {\bibinfo {volume} {465}},\ \bibinfo
  {pages} {59} (\bibinfo {year} {2009})}\BibitemShut {NoStop}%
\bibitem [{\citenamefont {Seevinck}(2010)}]{Seevinck2010}%
  \BibitemOpen
  \bibfield  {author} {\bibinfo {author} {\bibfnamefont {M.~P.}\ \bibnamefont
  {Seevinck}},\ }\bibfield  {title} {\enquote {\bibinfo {title} {Monogamy of
  correlations versus monogamy of entanglement},}\ }\href
  {https://doi.org/10.1007/s11128-009-0161-6} {\bibfield  {journal} {\bibinfo
  {journal} {Quant. Info. Proc.}\ }\textbf {\bibinfo {volume} {9}},\ \bibinfo
  {pages} {273} (\bibinfo {year} {2010})}\BibitemShut {NoStop}%
\bibitem [{\citenamefont {Baccari}\ \emph {et~al.}(2017)\citenamefont
  {Baccari}, \citenamefont {Cavalcanti}, \citenamefont {Wittek},\ and\
  \citenamefont {Ac\'in}}]{baccari2017classical}%
  \BibitemOpen
  \bibfield  {author} {\bibinfo {author} {\bibfnamefont {F.}~\bibnamefont
  {Baccari}}, \bibinfo {author} {\bibfnamefont {D.}~\bibnamefont {Cavalcanti}},
  \bibinfo {author} {\bibfnamefont {P.}~\bibnamefont {Wittek}},\ and\ \bibinfo
  {author} {\bibfnamefont {A.}~\bibnamefont {Ac\'in}},\ }\bibfield  {title}
  {\enquote {\bibinfo {title} {Efficient device-independent entanglement
  detection for multipartite systems},}\ }\href
  {https://link.aps.org/doi/10.1103/PhysRevX.7.021042} {\bibfield  {journal}
  {\bibinfo  {journal} {Phys. Rev. X}\ }\textbf {\bibinfo {volume} {7}},\
  \bibinfo {pages} {021042} (\bibinfo {year} {2017})}\BibitemShut {NoStop}%
\bibitem [{\citenamefont {Schmid}\ \emph {et~al.}(2020)\citenamefont {Schmid},
  \citenamefont {Fraser}, \citenamefont {Kunjwal}, \citenamefont {Sainz},
  \citenamefont {Wolfe},\ and\ \citenamefont {Spekkens}}]{Schmid2020LOSR}%
  \BibitemOpen
  \bibfield  {author} {\bibinfo {author} {\bibfnamefont {D.}~\bibnamefont
  {Schmid}}, \bibinfo {author} {\bibfnamefont {T.~C.}\ \bibnamefont {Fraser}},
  \bibinfo {author} {\bibfnamefont {R.}~\bibnamefont {Kunjwal}}, \bibinfo
  {author} {\bibfnamefont {A.~B.}\ \bibnamefont {Sainz}}, \bibinfo {author}
  {\bibfnamefont {E.}~\bibnamefont {Wolfe}},\ and\ \bibinfo {author}
  {\bibfnamefont {R.~W.}\ \bibnamefont {Spekkens}},\ }\bibfield  {title}
  {\enquote {\bibinfo {title} {{Understanding the interplay of entanglement and
  nonlocality: motivating and developing a new branch of entanglement
  theory}},}\ }\href {https://arxiv.org/abs/2004.09194} {\bibfield  {journal}
  {\bibinfo  {journal} {arXiv:2004.09194}\ } (\bibinfo {year}
  {2020})}\BibitemShut {NoStop}%
\bibitem [{\citenamefont {Navascu\'es}\ \emph {et~al.}(2020)\citenamefont
  {Navascu\'es}, \citenamefont {Wolfe}, \citenamefont {Rosset},\ and\
  \citenamefont {Pozas-Kerstjens}}]{navascues2020gnme}%
  \BibitemOpen
  \bibfield  {author} {\bibinfo {author} {\bibfnamefont {M.}~\bibnamefont
  {Navascu\'es}}, \bibinfo {author} {\bibfnamefont {E.}~\bibnamefont {Wolfe}},
  \bibinfo {author} {\bibfnamefont {D.}~\bibnamefont {Rosset}},\ and\ \bibinfo
  {author} {\bibfnamefont {A.}~\bibnamefont {Pozas-Kerstjens}},\ }\bibfield
  {title} {\enquote {\bibinfo {title} {{Genuine Network Multipartite
  Entanglement}},}\ }\href {https://doi.org/10.1103/PhysRevLett.125.240505}
  {\bibfield  {journal} {\bibinfo  {journal} {Phys. Rev. Lett.}\ }\textbf
  {\bibinfo {volume} {125}},\ \bibinfo {pages} {240505} (\bibinfo {year}
  {2020})}\BibitemShut {NoStop}%
\bibitem [{\citenamefont {Pozas-Kerstjens}(2019)}]{alexThesis}%
  \BibitemOpen
  \bibfield  {author} {\bibinfo {author} {\bibfnamefont {A.}~\bibnamefont
  {Pozas-Kerstjens}},\ }\emph {\bibinfo {title} {Quantum information outside
  quantum information}},\ \href {https://www.tdx.cat/handle/10803/667696}
  {Ph.D. thesis},\ \bibinfo  {school} {Universitat Polit\`ecnica de Catalunya}
  (\bibinfo {year} {2019})\BibitemShut {NoStop}%
\bibitem [{\citenamefont {Pironio}\ \emph {et~al.}(2011)\citenamefont
  {Pironio}, \citenamefont {Bancal},\ and\ \citenamefont
  {Scarani}}]{tripartiteNS}%
  \BibitemOpen
  \bibfield  {author} {\bibinfo {author} {\bibfnamefont {S.}~\bibnamefont
  {Pironio}}, \bibinfo {author} {\bibfnamefont {J.-D.}\ \bibnamefont
  {Bancal}},\ and\ \bibinfo {author} {\bibfnamefont {V.}~\bibnamefont
  {Scarani}},\ }\bibfield  {title} {\enquote {\bibinfo {title} {Extremal
  correlations of the tripartite no-signaling polytope},}\ }\href
  {https://doi.org/10.1088%2F1751-8113%2F44%2F6%2F065303} {\bibfield  {journal}
  {\bibinfo  {journal} {J. Phys. A}\ }\textbf {\bibinfo {volume} {44}},\
  \bibinfo {pages} {065303} (\bibinfo {year} {2011})}\BibitemShut {NoStop}%
\bibitem [{\citenamefont {Mermin}(1990)}]{Mermin1990}%
  \BibitemOpen
  \bibfield  {author} {\bibinfo {author} {\bibfnamefont {N.~D.}\ \bibnamefont
  {Mermin}},\ }\bibfield  {title} {\enquote {\bibinfo {title} {Quantum
  mysteries revisited},}\ }\href {https://doi.org/10.1119/1.16503} {\bibfield
  {journal} {\bibinfo  {journal} {Amer. J. Phys.}\ }\textbf {\bibinfo {volume}
  {58}},\ \bibinfo {pages} {731} (\bibinfo {year} {1990})}\BibitemShut
  {NoStop}%
\bibitem [{\citenamefont {Svetlichny}(1987)}]{Svetlichny1987}%
  \BibitemOpen
  \bibfield  {author} {\bibinfo {author} {\bibfnamefont {G.}~\bibnamefont
  {Svetlichny}},\ }\bibfield  {title} {\enquote {\bibinfo {title}
  {{Distinguishing three-body from two-body nonseparability by a Bell-type
  inequality}},}\ }\href {https://doi.org/10.1103/PhysRevD.35.3066} {\bibfield
  {journal} {\bibinfo  {journal} {Phys. Rev. D}\ }\textbf {\bibinfo {volume}
  {35}},\ \bibinfo {pages} {3066} (\bibinfo {year} {1987})}\BibitemShut
  {NoStop}%
\bibitem [{\citenamefont {Barrett}\ \emph {et~al.}(2005)\citenamefont
  {Barrett}, \citenamefont {Linden}, \citenamefont {Massar}, \citenamefont
  {Pironio}, \citenamefont {Popescu},\ and\ \citenamefont
  {Roberts}}]{TripartiteViaPR}%
  \BibitemOpen
  \bibfield  {author} {\bibinfo {author} {\bibfnamefont {J.}~\bibnamefont
  {Barrett}}, \bibinfo {author} {\bibfnamefont {N.}~\bibnamefont {Linden}},
  \bibinfo {author} {\bibfnamefont {S.}~\bibnamefont {Massar}}, \bibinfo
  {author} {\bibfnamefont {S.}~\bibnamefont {Pironio}}, \bibinfo {author}
  {\bibfnamefont {S.}~\bibnamefont {Popescu}},\ and\ \bibinfo {author}
  {\bibfnamefont {D.}~\bibnamefont {Roberts}},\ }\bibfield  {title} {\enquote
  {\bibinfo {title} {Nonlocal correlations as an information-theoretic
  resource},}\ }\href {https://link.aps.org/doi/10.1103/PhysRevA.71.022101}
  {\bibfield  {journal} {\bibinfo  {journal} {Phys. Rev. A}\ }\textbf {\bibinfo
  {volume} {71}},\ \bibinfo {pages} {022101} (\bibinfo {year}
  {2005})}\BibitemShut {NoStop}%
\bibitem [{\citenamefont {Lee}\ and\ \citenamefont
  {Hoban}(2018)}]{lee2018crypto}%
  \BibitemOpen
  \bibfield  {author} {\bibinfo {author} {\bibfnamefont {C.~M.}\ \bibnamefont
  {Lee}}\ and\ \bibinfo {author} {\bibfnamefont {M.~J.}\ \bibnamefont
  {Hoban}},\ }\bibfield  {title} {\enquote {\bibinfo {title} {Towards
  device-independent information processing on general quantum networks},}\
  }\href {https://doi.org/10.1103/PhysRevLett.120.020504} {\bibfield  {journal}
  {\bibinfo  {journal} {Phys. Rev. Lett.}\ }\textbf {\bibinfo {volume} {120}},\
  \bibinfo {pages} {020504} (\bibinfo {year} {2018})}\BibitemShut {NoStop}%
\bibitem [{\citenamefont {Thinh}\ \emph {et~al.}(2016)\citenamefont {Thinh},
  \citenamefont {Bancal},\ and\ \citenamefont
  {Mart\'{\i}n-Mart\'{\i}nez}}]{thinh2016vacuumleakage}%
  \BibitemOpen
  \bibfield  {author} {\bibinfo {author} {\bibfnamefont {L.~P.}\ \bibnamefont
  {Thinh}}, \bibinfo {author} {\bibfnamefont {J.-D.}\ \bibnamefont {Bancal}},\
  and\ \bibinfo {author} {\bibfnamefont {E.}~\bibnamefont
  {Mart\'{\i}n-Mart\'{\i}nez}},\ }\bibfield  {title} {\enquote {\bibinfo
  {title} {Certified randomness from a two-level system in a relativistic
  quantum field},}\ }\href {https://doi.org/10.1103/PhysRevA.94.022321}
  {\bibfield  {journal} {\bibinfo  {journal} {Phys. Rev. A}\ }\textbf {\bibinfo
  {volume} {94}},\ \bibinfo {pages} {022321} (\bibinfo {year}
  {2016})}\BibitemShut {NoStop}%
\bibitem [{\citenamefont {Levin}\ and\ \citenamefont
  {Peres}(2017)}]{Levin2017}%
  \BibitemOpen
  \bibfield  {author} {\bibinfo {author} {\bibfnamefont {D.~A.}\ \bibnamefont
  {Levin}}\ and\ \bibinfo {author} {\bibfnamefont {Y.}~\bibnamefont {Peres}},\
  }\href {https://darkwing.uoregon.edu/~dlevin/MARKOV/} {\emph {\bibinfo
  {title} {{Markov Chains and Mixing Times}}}}\ (\bibinfo  {publisher}
  {American Mathematical Society},\ \bibinfo {year} {2017})\BibitemShut
  {NoStop}%
\bibitem [{\citenamefont {Huang}\ and\ \citenamefont
  {Pan}(2016)}]{huang2016mediation}%
  \BibitemOpen
  \bibfield  {author} {\bibinfo {author} {\bibfnamefont {Y.-T.}\ \bibnamefont
  {Huang}}\ and\ \bibinfo {author} {\bibfnamefont {W.-C.}\ \bibnamefont
  {Pan}},\ }\bibfield  {title} {\enquote {\bibinfo {title} {Hypothesis test of
  mediation effect in causal mediation model with high-dimensional continuous
  mediators},}\ }\href {https://doi.org/10.1111/biom.12421} {\bibfield
  {journal} {\bibinfo  {journal} {Biometrics}\ }\textbf {\bibinfo {volume}
  {72}},\ \bibinfo {pages} {402} (\bibinfo {year} {2016})}\BibitemShut
  {NoStop}%
\bibitem [{\citenamefont {Hutton}\ \emph {et~al.}(2018)\citenamefont {Hutton},
  \citenamefont {Fatima}, \citenamefont {Major}, \citenamefont {Topless},
  \citenamefont {Stamp}, \citenamefont {Merriman},\ and\ \citenamefont
  {Dalbeth}}]{hutton2018mediation}%
  \BibitemOpen
  \bibfield  {author} {\bibinfo {author} {\bibfnamefont {J.}~\bibnamefont
  {Hutton}}, \bibinfo {author} {\bibfnamefont {T.}~\bibnamefont {Fatima}},
  \bibinfo {author} {\bibfnamefont {T.~J.}\ \bibnamefont {Major}}, \bibinfo
  {author} {\bibfnamefont {R.}~\bibnamefont {Topless}}, \bibinfo {author}
  {\bibfnamefont {L.~K.}\ \bibnamefont {Stamp}}, \bibinfo {author}
  {\bibfnamefont {T.~R.}\ \bibnamefont {Merriman}},\ and\ \bibinfo {author}
  {\bibfnamefont {N.}~\bibnamefont {Dalbeth}},\ }\bibfield  {title} {\enquote
  {\bibinfo {title} {Mediation analysis to understand genetic relationships
  between habitual coffee intake and gout},}\ }\href
  {https://doi.org/10.1186/s13075-018-1629-5} {\bibfield  {journal} {\bibinfo
  {journal} {Arthritis Res. Ther.}\ }\textbf {\bibinfo {volume} {20}} (\bibinfo
  {year} {2018})}\BibitemShut {NoStop}%
\bibitem [{\citenamefont {Sohn}\ and\ \citenamefont
  {Li}(2019)}]{sohn2019microbiome}%
  \BibitemOpen
  \bibfield  {author} {\bibinfo {author} {\bibfnamefont {M.~B.}\ \bibnamefont
  {Sohn}}\ and\ \bibinfo {author} {\bibfnamefont {H.}~\bibnamefont {Li}},\
  }\bibfield  {title} {\enquote {\bibinfo {title} {Compositional mediation
  analysis for microbiome studies},}\ }\href
  {https://doi.org/10.1214/18-AOAS1210} {\bibfield  {journal} {\bibinfo
  {journal} {Ann. Appl. Stat.}\ }\textbf {\bibinfo {volume} {13}},\ \bibinfo
  {pages} {661} (\bibinfo {year} {2019})}\BibitemShut {NoStop}%
\bibitem [{\citenamefont {Cavalcanti}\ and\ \citenamefont
  {Lal}(2014)}]{Cavalcanti2014RCCP}%
  \BibitemOpen
  \bibfield  {author} {\bibinfo {author} {\bibfnamefont {E.~G.}\ \bibnamefont
  {Cavalcanti}}\ and\ \bibinfo {author} {\bibfnamefont {R.}~\bibnamefont
  {Lal}},\ }\bibfield  {title} {\enquote {\bibinfo {title} {{On modifications
  of Reichenbach's principle of common cause in light of Bell's theorem}},}\
  }\href {https://doi.org/10.1088/1751-8113/47/42/424018} {\bibfield  {journal}
  {\bibinfo  {journal} {J. Phys. A}\ }\textbf {\bibinfo {volume} {47}},\
  \bibinfo {pages} {424018} (\bibinfo {year} {2014})}\BibitemShut {NoStop}%
\bibitem [{\citenamefont {Allen}\ \emph {et~al.}(2017)\citenamefont {Allen},
  \citenamefont {Barrett}, \citenamefont {Horsman}, \citenamefont {Lee},\ and\
  \citenamefont {Spekkens}}]{Allen2017RCCP}%
  \BibitemOpen
  \bibfield  {author} {\bibinfo {author} {\bibfnamefont {J.-M.~A.}\
  \bibnamefont {Allen}}, \bibinfo {author} {\bibfnamefont {J.}~\bibnamefont
  {Barrett}}, \bibinfo {author} {\bibfnamefont {D.~C.}\ \bibnamefont
  {Horsman}}, \bibinfo {author} {\bibfnamefont {C.~M.}\ \bibnamefont {Lee}},\
  and\ \bibinfo {author} {\bibfnamefont {R.~W.}\ \bibnamefont {Spekkens}},\
  }\bibfield  {title} {\enquote {\bibinfo {title} {Quantum common causes and
  quantum causal models},}\ }\href {https://doi.org/10.1103/PhysRevX.7.031021}
  {\bibfield  {journal} {\bibinfo  {journal} {Phys. Rev. X}\ }\textbf {\bibinfo
  {volume} {7}},\ \bibinfo {pages} {031021} (\bibinfo {year}
  {2017})}\BibitemShut {NoStop}%
\bibitem [{\citenamefont {Janzing}\ \emph {et~al.}(2013)\citenamefont
  {Janzing}, \citenamefont {Balduzzi}, \citenamefont {Grosse-Wentrup},\ and\
  \citenamefont {Sch\"olkopf}}]{janzing2013}%
  \BibitemOpen
  \bibfield  {author} {\bibinfo {author} {\bibfnamefont {D.}~\bibnamefont
  {Janzing}}, \bibinfo {author} {\bibfnamefont {D.}~\bibnamefont {Balduzzi}},
  \bibinfo {author} {\bibfnamefont {M.}~\bibnamefont {Grosse-Wentrup}},\ and\
  \bibinfo {author} {\bibfnamefont {B.}~\bibnamefont {Sch\"olkopf}},\
  }\bibfield  {title} {\enquote {\bibinfo {title} {Quantifying causal
  influences},}\ }\href {https://doi.org/10.1214/13-AOS1145} {\bibfield
  {journal} {\bibinfo  {journal} {Ann. Statist.}\ }\textbf {\bibinfo {volume}
  {41}},\ \bibinfo {pages} {2324} (\bibinfo {year} {2013})}\BibitemShut
  {NoStop}%
\bibitem [{\citenamefont {Miles}\ \emph {et~al.}(2015)\citenamefont {Miles},
  \citenamefont {Kanki}, \citenamefont {Meloni},\ and\ \citenamefont
  {Tchetgen~Tchetgen}}]{miles2015partial}%
  \BibitemOpen
  \bibfield  {author} {\bibinfo {author} {\bibfnamefont {C.~H.}\ \bibnamefont
  {Miles}}, \bibinfo {author} {\bibfnamefont {P.}~\bibnamefont {Kanki}},
  \bibinfo {author} {\bibfnamefont {S.}~\bibnamefont {Meloni}},\ and\ \bibinfo
  {author} {\bibfnamefont {E.~J.}\ \bibnamefont {Tchetgen~Tchetgen}},\
  }\bibfield  {title} {\enquote {\bibinfo {title} {On partial identification of
  the pure direct effect},}\ }\href {https://arxiv.org/abs/1509.01652}
  {\bibfield  {journal} {\bibinfo  {journal} {arXiv:1509.01652}\ } (\bibinfo
  {year} {2015})}\BibitemShut {NoStop}%
\bibitem [{\citenamefont {Malinsky}\ \emph {et~al.}(2019)\citenamefont
  {Malinsky}, \citenamefont {Shpitser},\ and\ \citenamefont
  {Richardson}}]{malinsky2019potential}%
  \BibitemOpen
  \bibfield  {author} {\bibinfo {author} {\bibfnamefont {D.}~\bibnamefont
  {Malinsky}}, \bibinfo {author} {\bibfnamefont {I.}~\bibnamefont {Shpitser}},\
  and\ \bibinfo {author} {\bibfnamefont {T.}~\bibnamefont {Richardson}},\
  }\bibfield  {title} {\enquote {\bibinfo {title} {A potential outcomes
  calculus for identifying conditional path-specific effects},}\ }\href
  {https://arxiv.org/abs/1903.03662} {\bibfield  {journal} {\bibinfo  {journal}
  {arXiv:1903.03662}\ } (\bibinfo {year} {2019})}\BibitemShut {NoStop}%
\bibitem [{\citenamefont {Bhattacharya}\ \emph {et~al.}(2020)\citenamefont
  {Bhattacharya}, \citenamefont {Nabi},\ and\ \citenamefont
  {Shpitser}}]{bhattacharya2020semiparametric}%
  \BibitemOpen
  \bibfield  {author} {\bibinfo {author} {\bibfnamefont {R.}~\bibnamefont
  {Bhattacharya}}, \bibinfo {author} {\bibfnamefont {R.}~\bibnamefont {Nabi}},\
  and\ \bibinfo {author} {\bibfnamefont {I.}~\bibnamefont {Shpitser}},\
  }\bibfield  {title} {\enquote {\bibinfo {title} {Semiparametric inference for
  causal effects in graphical models with hidden variables},}\ }\href
  {https://arxiv.org/abs/2003.12659} {\bibfield  {journal} {\bibinfo  {journal}
  {arXiv:2003.12659}\ } (\bibinfo {year} {2020})}\BibitemShut {NoStop}%
\bibitem [{\citenamefont {Shpitser}\ and\ \citenamefont
  {Tchetgen~Tchetgen}(2016)}]{shpitser2016}%
  \BibitemOpen
  \bibfield  {author} {\bibinfo {author} {\bibfnamefont {I.}~\bibnamefont
  {Shpitser}}\ and\ \bibinfo {author} {\bibfnamefont {E.~J.}\ \bibnamefont
  {Tchetgen~Tchetgen}},\ }\bibfield  {title} {\enquote {\bibinfo {title}
  {Causal inference with a graphical hierarchy of interventions},}\ }\href
  {https://doi.org/10.1214/15-AOS1411} {\bibfield  {journal} {\bibinfo
  {journal} {Ann. Statist.}\ }\textbf {\bibinfo {volume} {44}},\ \bibinfo
  {pages} {2433} (\bibinfo {year} {2016})}\BibitemShut {NoStop}%
\bibitem [{\citenamefont {Shpitser}\ and\ \citenamefont
  {Sherman}(2018)}]{shpitser2018identification}%
  \BibitemOpen
  \bibfield  {author} {\bibinfo {author} {\bibfnamefont {I.}~\bibnamefont
  {Shpitser}}\ and\ \bibinfo {author} {\bibfnamefont {E.}~\bibnamefont
  {Sherman}},\ }\bibfield  {title} {\enquote {\bibinfo {title} {Identification
  of personalized effects associated with causal pathways},}\ }in\ \href
  {http://auai.org/uai2018/proceedings/papers/198.pdf} {\emph {\bibinfo
  {booktitle} {Uncertainty in Artificial Intelligence}}},\ Vol.\ \bibinfo
  {volume} {2018}\ (\bibinfo {year} {2018})\BibitemShut {NoStop}%
\bibitem [{\citenamefont {Stensrud}\ \emph {et~al.}(2019)\citenamefont
  {Stensrud}, \citenamefont {Young}, \citenamefont {Didelez}, \citenamefont
  {Robins},\ and\ \citenamefont {Hern\'an}}]{stensrud2019separable}%
  \BibitemOpen
  \bibfield  {author} {\bibinfo {author} {\bibfnamefont {M.~J.}\ \bibnamefont
  {Stensrud}}, \bibinfo {author} {\bibfnamefont {J.~G.}\ \bibnamefont {Young}},
  \bibinfo {author} {\bibfnamefont {V.}~\bibnamefont {Didelez}}, \bibinfo
  {author} {\bibfnamefont {J.~M.}\ \bibnamefont {Robins}},\ and\ \bibinfo
  {author} {\bibfnamefont {M.~A.}\ \bibnamefont {Hern\'an}},\ }\bibfield
  {title} {\enquote {\bibinfo {title} {Separable effects for causal inference
  in the presence of competing events},}\ }\href
  {https://arxiv.org/abs/1901.09472} {\bibfield  {journal} {\bibinfo  {journal}
  {arXiv:1901.09472}\ } (\bibinfo {year} {2019})}\BibitemShut {NoStop}%
\bibitem [{\citenamefont {Cai}\ \emph {et~al.}(2008)\citenamefont {Cai},
  \citenamefont {Kuroki}, \citenamefont {Pearl},\ and\ \citenamefont
  {Tian}}]{Cai2007}%
  \BibitemOpen
  \bibfield  {author} {\bibinfo {author} {\bibfnamefont {Z.}~\bibnamefont
  {Cai}}, \bibinfo {author} {\bibfnamefont {M.}~\bibnamefont {Kuroki}},
  \bibinfo {author} {\bibfnamefont {J.}~\bibnamefont {Pearl}},\ and\ \bibinfo
  {author} {\bibfnamefont {J.}~\bibnamefont {Tian}},\ }\bibfield  {title}
  {\enquote {\bibinfo {title} {Bounds on direct effects in the presence of
  confounded intermediate variables},}\ }\href
  {https://doi.org/10.1111/j.1541-0420.2007.00949.x} {\bibfield  {journal}
  {\bibinfo  {journal} {Biometrics}\ }\textbf {\bibinfo {volume} {64}},\
  \bibinfo {pages} {695} (\bibinfo {year} {2008})}\BibitemShut {NoStop}%
\bibitem [{\citenamefont {Kang}\ and\ \citenamefont
  {Tian}(2006)}]{kang2012inequality}%
  \BibitemOpen
  \bibfield  {author} {\bibinfo {author} {\bibfnamefont {C.}~\bibnamefont
  {Kang}}\ and\ \bibinfo {author} {\bibfnamefont {J.}~\bibnamefont {Tian}},\
  }\bibfield  {title} {\enquote {\bibinfo {title} {Inequality constraints in
  causal models with hidden variables},}\ }in\ \href
  {https://dl.acm.org/doi/10.5555/3020419.3020448} {\emph {\bibinfo {booktitle}
  {Uncertainty in Artificial Intelligence}}}\ (\bibinfo {year}
  {2006})\BibitemShut {NoStop}%
\bibitem [{\citenamefont {Chaves}\ \emph {et~al.}(2018)\citenamefont {Chaves},
  \citenamefont {Carvacho}, \citenamefont {Agresti}, \citenamefont {Di~Giulio},
  \citenamefont {Aolita}, \citenamefont {Giacomini},\ and\ \citenamefont
  {Sciarrino}}]{Chaves2017}%
  \BibitemOpen
  \bibfield  {author} {\bibinfo {author} {\bibfnamefont {R.}~\bibnamefont
  {Chaves}}, \bibinfo {author} {\bibfnamefont {G.}~\bibnamefont {Carvacho}},
  \bibinfo {author} {\bibfnamefont {I.}~\bibnamefont {Agresti}}, \bibinfo
  {author} {\bibfnamefont {V.}~\bibnamefont {Di~Giulio}}, \bibinfo {author}
  {\bibfnamefont {L.}~\bibnamefont {Aolita}}, \bibinfo {author} {\bibfnamefont
  {S.}~\bibnamefont {Giacomini}},\ and\ \bibinfo {author} {\bibfnamefont
  {F.}~\bibnamefont {Sciarrino}},\ }\bibfield  {title} {\enquote {\bibinfo
  {title} {Quantum violation of an instrumental test},}\ }\href
  {https://doi.org/10.1038/s41567-017-0008-5} {\bibfield  {journal} {\bibinfo
  {journal} {Nat. Phys.}\ }\textbf {\bibinfo {volume} {14}},\ \bibinfo {pages}
  {291} (\bibinfo {year} {2018})}\BibitemShut {NoStop}%
\bibitem [{\citenamefont {Bowles}\ \emph {et~al.}(2020)\citenamefont {Bowles},
  \citenamefont {Baccari},\ and\ \citenamefont
  {Salavrakos}}]{bowles2020bounding}%
  \BibitemOpen
  \bibfield  {author} {\bibinfo {author} {\bibfnamefont {J.}~\bibnamefont
  {Bowles}}, \bibinfo {author} {\bibfnamefont {F.}~\bibnamefont {Baccari}},\
  and\ \bibinfo {author} {\bibfnamefont {A.}~\bibnamefont {Salavrakos}},\
  }\bibfield  {title} {\enquote {\bibinfo {title} {Bounding sets of sequential
  quantum correlations and device-independent randomness certification},}\
  }\href {https://doi.org/10.22331/q-2020-10-19-344} {\bibfield  {journal}
  {\bibinfo  {journal} {{Quantum}}\ }\textbf {\bibinfo {volume} {4}},\ \bibinfo
  {pages} {344} (\bibinfo {year} {2020})}\BibitemShut {NoStop}%
\bibitem [{\citenamefont {Moroder}\ \emph {et~al.}(2013)\citenamefont
  {Moroder}, \citenamefont {Bancal}, \citenamefont {Liang}, \citenamefont
  {Hofmann},\ and\ \citenamefont {G\"uhne}}]{LocalLevel}%
  \BibitemOpen
  \bibfield  {author} {\bibinfo {author} {\bibfnamefont {T.}~\bibnamefont
  {Moroder}}, \bibinfo {author} {\bibfnamefont {J.-D.}\ \bibnamefont {Bancal}},
  \bibinfo {author} {\bibfnamefont {Y.-C.}\ \bibnamefont {Liang}}, \bibinfo
  {author} {\bibfnamefont {M.}~\bibnamefont {Hofmann}},\ and\ \bibinfo {author}
  {\bibfnamefont {O.}~\bibnamefont {G\"uhne}},\ }\bibfield  {title} {\enquote
  {\bibinfo {title} {Device-independent entanglement quantification and related
  applications},}\ }\href
  {https://link.aps.org/doi/10.1103/PhysRevLett.111.030501} {\bibfield
  {journal} {\bibinfo  {journal} {Phys. Rev. Lett.}\ }\textbf {\bibinfo
  {volume} {111}},\ \bibinfo {pages} {030501} (\bibinfo {year}
  {2013})}\BibitemShut {NoStop}%
\end{thebibliography}%

\appendix
\section{Non-commutative polynomial optimization}
\label{app_npo}
A generic NPO problem can be cast as
\begin{align*}
    p^* =& \underset{(\mathcal{H},X,\rho)}{\min}  \expec*{p(X)}_\rho \\
    \text{such that} \qquad& q_i(X)\succeq 0\quad\forall\,i=1\dots m_q,
\end{align*}
that is, finding a Hilbert space $\mathcal{H}$, a positive-semidefinite operator $\rho:\mathcal{H}\rightarrow\mathcal{H}$ with trace one, and a list of bounded operators $X=(X_1,\dots, X_n)$ in $\mathcal{H}$ (where \mbox{$X_iX_j\neq X_jX_i$} in general) that minimize the expectation value $\expec*{p(X)}_\rho=\Tr\left[\rho\cdot p(X)\right]$ of the polynomial operator $p(X)$ given some polynomial constraints $q_i(X)\succeq 0$, where $q_i(X)\succeq 0$ means that the operator $q_i(X)$ should be positive semidefinite.
One can also add to the optimization statistical constraints of the form $\expec*{r_j(X)}_\rho \geq 0$, for $j=1,\ldots, m_r$.
Note that the NPO formalism can also accommodate equality constraints of the form $q(X)= 0$ or $\expec*{r_j(X)}_\rho = 0$, since they are equivalent to the constraints $q(X),-q(X)\succeq 0$ and $\expec*{r_j(X)}_\rho,\expec*{-r_j(X)}_\rho\geq 0$, respectively.

The procedure for solving these problems is described in Ref.~\cite{npo}, and uses a hierarchy of relaxations where each of the hierarchy's steps is an SDP instance.
The solutions to these problems form a monotonically increasing sequence of lower bounds on the global minimum $p^*$:
\begin{equation*}
    p_1\leq p_2\leq\dots\leq p_\infty\leq p^*.
\end{equation*}
If the constraints $\{q_i(X)\succeq 0\}_i$ imply (explicitly or implicitly; see Ref.~\cite{npo}) that all noncommuting variables $X_1,\dots,X_n$ are bounded (and they do so in all the NPO problems considered in this work), then the sequence of lower bounds is asymptotically convergent. That is, $p_\infty=p^*$.

In our case, given some observed correlations, we deal with a feasibility problem about the existence of a quantum state and measurements subject to polynomial operator and statistical constraints arising from the causal networks and the observed correlations. This feasibility problem can be mapped into an optimization problem in different ways. For instance, while not being the most practical procedure, the easiest way of doing it is by considering a constant polynomial $p(X)=1$ as the function to be optimized. This problem has a solution equal to $1$ provided that the polynomial and statistical constraints are simultaneously satisfiable. Any step in the hierarchy is therefore a necessary SDP test to be satisfied for the causal model to be compatible with the observed correlations. Note that the same formalism can be used to optimize polynomials of the operators, such as, for instance, Bell-like inequalities, over quantum correlations compatible with a given causal structure~\cite{Schmid2020LOSR}.

\section{The treatment of latent variables with latent parents}\label{app:nonexog}
\begin{figure}[b]
  \begin{center}
    \hfill
    \subfigure[\label{fig:RestrictedState}]
    {\centering
      \includegraphics[scale=0.45]{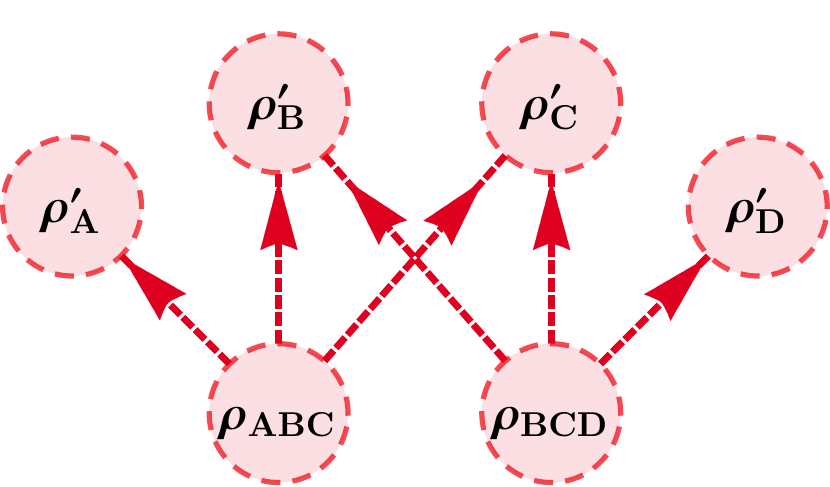}
    }
    \hfill
    \subfigure[\label{fig:EnablingState}]
    {\centering
      \includegraphics[scale=0.45]{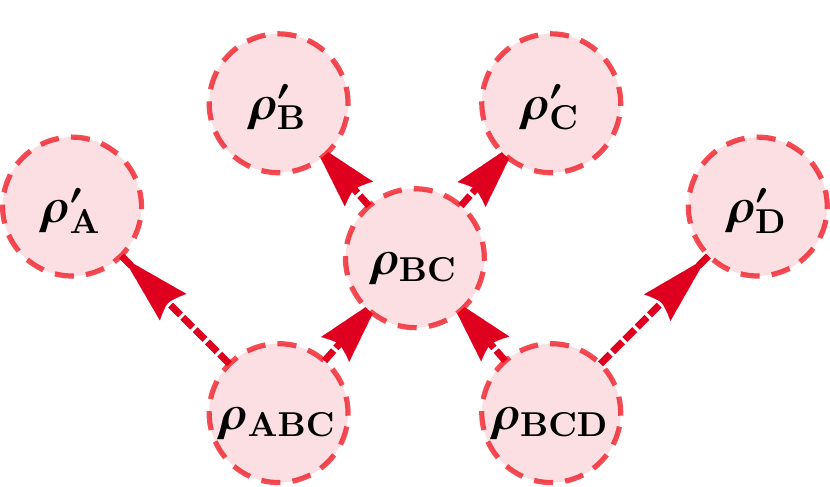}
    }
    \hfill
  \end{center}
  \caption[]{Inequivalent state generating procedures. (a) depicts a two-layer state generating procedure, whereas (b) allows for the application of an arbitrary Bob-Charlie channel, generating the intermediate state $\rho_{BC}$, prior to distributing Hilbert spaces to the four parties. One can confirm that (a) generates a strict subset of the states realizable in (b), and quantum inflation can distinguish these two scenarios in a device-independent manner.}
  \label{fig:ExogMoreSubtle}
\end{figure}

The study of network form quantum causal structures with quantum inflation is, essentially, solving the problem of \emph{classifying entanglement networks} in a device-independent manner~\cite{navascues2020gnme,Schmid2020LOSR}. Generalizing to latent nonexogenous scenarios can, thus, be seen as a further generalization of the problem of classification of entanglement generation in general causal structures. Consider, for instance, four parties and two three-way sources of entanglement. One can imagine two distinct ways to generate the final four-partite quantum state from the initial pair of sources. In the first case, depicted in Fig.~\ref{fig:RestrictedState}, each of the four parties applies arbitrary (local) quantum channels to the states they receive from the sources. In the second case, depicted in Fig.~\ref{fig:EnablingState}, a nonlocal quantum channel spanning Bob and Charlie's systems is applied prior to the final assignment of Hilbert spaces to parties.

One might naively think that this further generality does not add much---after all, Bob and Charlie can share arbitrary entanglement in either scenario---but that is not the case. Only the causal structure depicted in Fig.~\ref{fig:EnablingState} can give rise to the state
\begin{align}\label{eq:latententangledstate}
  \rho_\text{\ref{fig:EnablingState}}&\coloneqq \frac{\ket{AB00}\bra{AB00}+\ket{A0C1}\bra{A0C1}}{2},\\\nonumber
  \text{where}\quad&\ket{AB00} = \frac{\ket{0000}+\ket{1100}}{\sqrt{2}}\\\nonumber
  \text{and}\quad&\ket{A0C1} = \frac{\ket{0001}+\ket{1011}}{\sqrt{2}}.
\end{align}
The state $\rho_\text{\ref{fig:EnablingState}}$ describes maximal ${A{-}B}$ entanglement and ${C{=}\ket{0}}$ when ${D{=}\ket{0}}$ mixed with maximal ${A{-}C}$ entanglement and  ${B{=}\ket{0}}$ when ${D{=}\ket{1}}$. Here, measuring $\rho^\prime_D$ in the computational basis will steer the  $\rho^\prime_{ABC}$ toward either maximal entanglement between $A$ and $B$ or maximal entanglement between $A$ and $C$.
This steering is possible only if the ${B{-}C}$ entanglement structure is determined by a node which shares causal ancestry with $A$ as well with $D$. There is no node in Fig.~\ref{fig:RestrictedState} that satisfies these requirements, which implies that the state~$\rho_\text{\ref{fig:EnablingState}}$ cannot be generated by such a causal structure.
On the other hand, $\rho_{BC}$ in Fig.~\ref{fig:EnablingState} is a node which does satisfy the requirements, and thus, both structures are not equivalent.

Alternatively, this result can also be understood through monogamy of nonlocality: Access to $\rho_\text{\ref{fig:EnablingState}}$ enables maximally violating the CHSH inequality by either players $A$ and $B$ or by players $A$ and $C$, where nonlocal parties are flagged by the value of a measurement on  $\rho^\prime_D$. This result precisely replicates the example in Fig.~\ref{fig:Exog}, except that the switch setting $S$ in Fig.~\ref{fig:NonExog} is replaced by measurement on $\rho^\prime_D$ in Fig.~\ref{fig:EnablingState} in the computational basis.

An explicit realization of $\rho_\text{\ref{fig:EnablingState}}$ in Fig.~\ref{fig:EnablingState} is as follows. Let $\rho_{ABC}$ and $\rho_{BCD}$ be, respectively, projectors onto the pure states $(\ket{00}+\ket{11})_{AB_1}$ and \mbox{$(\ket{00}+\ket{11})_{B_2D}\otimes\ket{0}_C$}. Then, perform a controlled swap on qubits $B_2$ and $C$ depending on $B_1$'s computational basis value, and trace out the subsystem $B_1$.

In Sec.~\ref{sec:nonexog}, we prescribe encoding all quantum channels in the causal structure as unspecified unitaries that can be incorporated into the generating monomials. Though Fig.~\ref{fig:Exog} there illustrates classically controlled quantum channels, uncontrolled quantum channels can equally well be incorporated into the elementary monomials.
Consider applying quantum inflation to Fig.~\ref{fig:ExogMoreSubtle}. Inflation imbues every unitary with a pair of indices corresponding to which copies of the root quantum states the given unitary pertains to. For the unrestricted case in Fig.~\ref{fig:RestrictedState}, we find that misaligning the copy index pertaining to one of the root quantum states leads to a pair of operators which need not commute, namely,
\begin{align}
  U^{i,j}_{BC} \!\!\cdot U^{i,k}_{BC} \neq U^{i,k}_{BC}\!\!\cdot U^{i,j}_{BC}\quad\text{for}\quad j\neq k .
\end{align}
On the other hand, the local unitaries relevant to Fig.~\ref{fig:EnablingState} \emph{do} commute even in the presence of misaligned copy indices, namely
\begin{align}
  U^{i,j}_{B} \!\!\cdot U^{i,k}_{C} = U^{i,k}_{C}\!\!\cdot U^{i,j}_{B}\quad\text{for}\quad j\neq k .
\end{align}
One can readily see, therefore, how inflation levels \mbox{$n\geq 2$} give rise to SDPs with \emph{more constrained} moment matrices when quantum inflation is applied to Fig.~\ref{fig:RestrictedState} instead of Fig.~\ref{fig:EnablingState}. This example is a device-independent distinction afforded by the prescription of Sec.~\ref{sec:nonexog} which cannot be recovered using the sequential measurements framework of Ref.~\cite{bowles2020bounding}.

\section{Monomial sets for NPO problems}\label{app:sets}
The hierarchies of semidefinite programs that bound the solutions of NPO problems can be described in terms of sets of products of the noncommutative operators in the problem.
In this work, we use two different hierarchies that are both asymptotically complete.
The levels in the first hierarchy are known as \term{NPA levels}~\cite{npa,npa2}.
The NPA level $n$, $\mathcal{S}_n$, is associated to the set of all products of operators in the problem, of length no larger than $n$.
For example, the set $\mathcal{S}_2$ associated to the inflations of the quantum triangle scenario discussed in the main text is
\begin{equation*}
  \mathcal{S}_2=\{\id\}\cup\{H^{i,j}_p\}_{p,i,j}\cup\{H^{i,j}_p (H')^{k,l}_q\}_{p,q,i,j,k,l},
\end{equation*}
where $H_p,H'_q=E,F,G$.
On the other hand, some problems may achieve tighter results at lower levels if one instead considers \term{local levels}~\cite{LocalLevel}.
The local level $n$, $\mathcal{L}_n$, is built from the products of operators that contain at most $n$ operators of a same party.
For instance, in the quantum triangle scenario, the set $\mathcal{L}_1$ associated to its inflations is
\begin{align*}
  \mathcal{L}_1=&\{\id\}\cup\{H^{i,j}_p\}_{p,i,j}\cup\{H^{i,j}_p (H')^{k,l}_q\}_{p,q,i,j,k,l}\\
  &\cup\{E^{i,j}_a F^{k,l}_b G^{m,n}_c\}_{a,b,c,i,j,k,l,m,n},
\end{align*}
where $H_p\neq H'_q$.
While both hierarchies are asymptotically complete, they satisfy the relation $\mathcal{S}_n\subset\mathcal{L}_n\not\subset\mathcal{S}_{n+1}$ (in fact, the smallest set of the NPA hierarchy that contains $\mathcal{L}_n$ is $\mathcal{S}_{pn}$, where $p$ is the number of parties), and, thus, the use of finite levels of one or the other hierarchy may be more or less convenient depending on the specific problem to solve.

\section{Quantum key distribution in the tripartite-line scenario}\label{app:crypto}
The distribution observed by $A$, $B$, and $C$ that allows them to guarantee that a quantum eavesdropper $E$ cannot guess the outputs of the extremes $A$ and $C$ with a confidence higher than that shown in Fig.~\ref{fig:cryptoresults} is generated in a standard quantum repeater scenario, where the sources distribute maximally entangled bipartite quantum states and the repeater party $B$ performs a Bell state measurement in the standard basis.
Following Ref.~\cite{lee2018crypto}, we choose the sources to distribute singlets $\ket{\psi^-}=\left(\ket{01}-\ket{10}\right)/\sqrt{2}$ and the parties $A$ and $C$ to perform the measurements $\left(\sigma_z+\sigma_x\right)/\sqrt{2}$ for $x{=}z{=}0$ and $\left(\sigma_z-\sigma_x\right)/\sqrt{2}$ for $x{=}z{=}1$.
In such a case, the distribution observed by $A$, $B$ and $C$ is $P_{\text{ABC{\textbar}XZ}}(a,b_1b_0,c|x,z)\coloneqq$
\begin{equation*}
    \frac{2+(-1)^{a+c}\left[(-1)^{b_0}+(-1)^{b_1+x+z}\right]}{32}.
\end{equation*}
As shown in Sec.~\ref{sec:cryptography}, if $A$, $B$, and $C$ observe this distribution, they are guaranteed that an eavesdropper $E$ cannot predict the outputs of $A$ and $C$ better than randomly guessing them.

If the sources are not perfect and instead distribute noisy singlets, $\rho=v\ket{\psi^-}\bra{\psi^-}+(1-v)\id/4$, then the distribution observed by $A$, $B$, and $C$ takes the form $P_{\text{ABC{\textbar}XZ},v}(a,b_1b_0,c|x,z)\coloneqq$
\begin{align}
    v^2 &P_{\text{ABC{\textbar}XZ}}(a,b_1b_0,c|x,z) + \frac{1-v^2}{16} \notag\\
    &=\frac{2+v^2(-1)^{a+c}\left[(-1)^{b_0}+(-1)^{b_1+x+z}\right]}{32}.
    \label{eq:cryptodistribution}
\end{align}
If $A$, $B$ and $C$ observe this distribution, they can be sure only that an eavesdropper cannot guess the outputs of $A$ and $C$ with a probability higher than that depicted in Fig.~\ref{fig:cryptoresults}.
\end{document}